\documentclass[reqno]{amsart}
\usepackage{amssymb}
\usepackage{amsfonts}

\usepackage{mathrsfs}%对于花体的支持
\usepackage{caption}
\usepackage{mathtools}
\usepackage{enumitem}
\usepackage{pifont}%\ding{}172-181: 白底序号/182-191：黑底序号
\usepackage[numbers,sort&compress]{natbib}%引用多篇参考文献
\usepackage{multirow}%复杂表格
\makeatletter
\@addtoreset{equation}{section}
\makeatother
\renewcommand{\theequation}{\arabic{section}.\arabic{equation}}
\usepackage{microtype}

\usepackage{quoting}

\usepackage{mathrsfs}

\DeclareMathOperator*{\diam}{diam}

\DeclareMathOperator{\tr}{Tr}

\DeclareMathOperator{\re}{Re}

\DeclareMathOperator{\sgn}{sgn}

\newtheorem{theorem}{Theorem}[section]
\newtheorem{assumption}[theorem]{Assumption}
\newtheorem{corollary}[theorem]{Corollary}
\newtheorem{proposition}[theorem]{Proposition}
\newtheorem{lemma}[theorem]{Lemma}
\newtheorem{definition}[theorem]{Definition}
\newtheorem{remark}[theorem]{Remark}

\newcommand{\averageint}{-\kern-11pt\int}

\usepackage{hyperref}
\allowdisplaybreaks[4]%允许公式跨页，1~4表示强度，4最强

\usepackage{amsthm}

\setlist[enumerate]{leftmargin=*} % 让enumerate靠左

\begin{document}
\title[From many-body fermionic system to BCS functionals]{Derivation of BCS functionals and gap equations from quantum many-body fermionic system}
%\author{}
\date{\today}
\author[X. Chen]{Xuwen Chen}
\address{Department of Mathematics, University of Rochester, Rochester, NY 14627, USA}

\email{xuwenmath@gmail.com}

\author[J. Wu]{Jiahao Wu}
\address{School of Mathematical Sciences, Zhejiang University, Hangzhou, Zhejiang, 310058, China}
\email{wjh00043@gmail.com}

\author[Z. Zhang]{Zhifei Zhang}
\address{School of Mathematical Sciences, Peking University, Beijing, 100871, China}

\email{zfzhang@math.pku.edu.cn}

\begin{abstract}
  We consider 3D single-species spin-$1/2$ Fermi systems with general dispersion relation and general attractive interaction initially inside a finite box. Up to some critical temperature $T_c$ determined by the interaction and system size, we establish, for the expected discrete BCS gap equation, existence of positive maximal solution. We then prove the uniqueness of nonzero (not identically zero) nonnegative solution. That is, any nonzero nonnegative solution, if exist or obtained via other methods, must equal to the positive maximal solution. Building on such a well-posedness result and assuming quasi-freeness, we prove the emergence of the $SU(2)$ symmetry of the finite system. Finally, letting the system size tend to infinity, we prove the emergence of the continuous BCS functional and gap equation for which we also have the uniqueness of nonzero nonnegative solution, and the convergence of the critical temperature. In particular, our work also implies there is a positive energy (hence mass) gap for aforementioned Fermi systems for $[0,T_c)$, any nonzero nonnegative solution for the gap equations are from the many-body system it suppose to describe, and the Cooper divergence of the density of states from a finite to an infinite system.
\end{abstract}

\maketitle
\tableofcontents

\section{Introduction}

\par The Bardeen-Cooper-Schrieffer (BCS) introduced the superconducting ground state (now well-known as the BCS ground state) in their paper \cite{BCS} studying theory of superconductivity. Considering a perturbation to the free fermi gas, which is a quasi free state, they guess a new quasi free ground state. This ground state is a superposition of so-called Cooper pairs, that is, pairs of  opposite spin and momentum:
\begin{equation*}
  \vert\psi_{\mathrm{BCS}}\rangle=\prod_{k}\Big(u_k+v_k a^*_{k,\uparrow}a^*_{-k,\downarrow}\Big)\vert 0\rangle.
\end{equation*}
Here $\vert 0\rangle$ denotes the vacuum state, $v_k,u_k$ are complex coefficients to be determined, such that $\vert u_k\vert^2+\vert v_k\vert^2=1$, and $a^*$ represents the creation operator on the Fock space.

\par For a physical interpretation\footnote{See, for example, Chapter 10 of \url{https://ncatlab.org/nlab/files/Timm-Superconductivity.pdf}} of the derivation of BCS model, we can start from pressure functional of the usual many-body spin-$1/2$ fermionic system, trapping in the 3D box of length $L$, on the Fock space $\mathcal{F}$:
\begin{equation*}
  -L^3P^L_{qtm}[\Gamma]= \tr_\mathcal{F}
  \big((H-\mu\mathcal{N})\Gamma\big)-\frac{1}{\beta}S_Q[\Gamma],
\end{equation*}
where $S_Q[\Gamma]= -\tr_\mathcal{F}(\Gamma\ln\Gamma)$ is the von Neumann entropy, and the operator $H-\mu\mathcal{N}$ is given by
\begin{equation*}
  H-\mu\mathcal{N}=\sum_{k,\sigma}\big(\vert k\vert^2-\mu\big)a_{k,\sigma}^*a_{k,\sigma}
  +\frac{1}{2L^3}\sum_{\substack{k,m,p,q\\\sigma,\nu}}
  \hat{v}(k,m,p,q)a^*_{k,\sigma}a^*_{m,\nu}a_{q,\nu}a_{p,\sigma}.
\end{equation*}
 Adding the BCS pairing assumption on the classical Hamiltonian, i.e. considering the BCS Hamiltonian:
\begin{equation*}
  H_{\mathrm{BCS}}-\mu\mathcal{N}=\sum_{k,\sigma}\big(\vert k\vert^2-\mu\big)a_{k,\sigma}^*a_{k,\sigma}
  +\frac{1}{2L^3}\sum_{k,k^\prime,\sigma}\hat{v}(k,k^\prime)
  a^*_{k,\sigma}a^*_{k,-\sigma}a_{-k^\prime,-\sigma}a_{k^\prime,\sigma},
\end{equation*}
and using the mean-field approximation
\begin{equation*}
  AB\simeq\langle A\rangle B+A\langle B\rangle -\langle A\rangle\langle B\rangle
\end{equation*}
with $A= a^*_{k,\sigma}a^*_{k,-\sigma}$ and $B=a_{-k^\prime,-\sigma}a_{k^\prime,\sigma}$, we can calculate the ground state energy of the system at scale $L$ by solving the BCS gap equation
\begin{equation*}
    \Delta_{k}=-\frac{1}{2L^3}\sum_{k^\prime}\hat{v}(k,k^\prime)
    \frac{\Delta_{k^\prime}}{E_{k^\prime}}\tanh\frac{\beta}{2}E_{k^\prime}.
  \end{equation*}
Here $E_{k}=\sqrt{(k^2-\mu)^2+\vert\Delta_k\vert^2}$ and $\beta=1/T$ is the reciprocal of the temperature $T\geq0$. We can formally derive the actual BCS model by letting $L$ go to infinity, i.e. considering system in the whole 3D space.

\par In BCS superconductivity theory, the fermionic system in superconducting is equivalent to the existence of nonzero solution to the BCS gap equation. In \cite{Abohr}, a possible analogy between this superconducting state with the excitation spectra of nuclei was proposed. Also, it is meaningful to consider the BCS gap equation in the box of size $L$, since for discrete $k\in(2\pi/L)\mathbb{Z}^3$, the density of state
\begin{equation*}
  \rho(k)\propto\frac{E_k}{\sqrt{E_k^2-\vert\Delta_k\vert^2}}
\end{equation*}
possesses no singularity for almost all $L>0$. However, when $L$ goes to infinity, one has to deal with this problem. See, for example, \cite{Larkin1,Larkin2,Nambu1,Nambu2}.

\par Mathematically, the origin many-body fermionic system is of great interests, this problem dates back to the original 1957 Huang-Yang formula \cite{huangyang}, i.e. the ground state energy approximation of the many-body system in the thermodynamic limit. Recently, rigorous derivations of the Huang-Yang formula were carried out in \cite{WJH,chen2025secondorderhuangyangapproximation,2024huangyangformulalowdensityfermi,
giacomelli2025huangyangconjecturelowdensityfermi}. For the bosonic case, the most recent work about the Lee-Huang-Yang-Wu formula \cite{huangyang,LHY106,TDLEECNYANG112,Wuterm} was carried out in \cite{Wutermupper}.

\par The approximations of energy gap $\Xi=\inf_k\sqrt{(k^2-\mu)^2+\vert\Delta_k\vert^2}$, and the critical temperature $T_c$, under which the nonzero solution to the BCS gap equation exists, are also important problems among mathematicians and physicists. For radially symmetric potential, there have been many studies, like \cite{critical,Criticaltem1,Criticaltem2,BCSgeneralpair}, about the continuous version of BCS model and gap equation, i.e. system lying in the whole space rather than the box of length $L$. One can see \cite{reviewBCS} for a general review of those results. Moreover, the link of the general continuous BCS model with the Ginzburg-Landau model was given in \cite{frank2012microscopic}. Many big names are working on this problem and we will cite their specific works during our proof.

\par The well-posedness of the BCS gap equation for general $\hat{v}(k,k^\prime)$ is shown in \cite{yang1991bardeen,bryce2000uniqueness}, under certain compact condition of $\hat{v}(k,k^\prime)$ as a function of $(k,k^\prime)$. However, for radially symmetric potential studied in this paper, which is of the form $\hat{v}(k-k^\prime)$, even if $\hat{v}(k)$ is compactly supported, $\hat{v}(k-k^\prime)$ is not as a function of $(k,k^\prime)$, and this is the motivation of this paper. We first derive the well-posedness of the BCS gap equation for radially symmetric potential in Theorem \ref{main1}, and with the help of it, verify the legitimacy of Cooper pair assumption, and the approximation of the whole space in Theorems \ref{main2} and \ref{main3} respectively.

\vspace{1em}
\par In this paper, we start from the spin-$1/2$ fermionic system at the positive temperature $T>0$, and we set $\beta=1/T$. Denote the set of $2$ spin states by $\mathcal{S}_{2}=\{\uparrow,\downarrow\}$. We also use the notation $-\uparrow=\downarrow$ and vice versa.

\par The space-spin variable for single particle has the form $z=(x,\sigma)$. Let $N\in\mathbb{N}_{\geq0}$ be a non-negative integer. For $N$ fermions of spin-$1/2$ , confined to a large 3D box $\Lambda_L=[-L/2,L/2]^3$ equipped with periodic boundary condition, the wave function that describes these $N$ fermions should be in the Hilbert space $L_a^2\big((\Lambda_L\times\mathcal{S}_2)^N\big)\eqqcolon \mathcal{H}^{\wedge N}$ consisting of functions that are anti-symmetric with respect to permutations of the $N$ particles.

\par The Hilbert space for the grand canonical ensemble of fermions is given by the Fock space $\mathcal{F}=\bigoplus_{N=0}^\infty\mathcal{H}^{\wedge N}$\footnote{Sometimes in this paper, we will use $\mathcal{F}_L$ to indicate the $L$ dependence of the Fock space}, where we denote $\mathcal{H}^{\wedge 0}=\mathbb{C}$. The standard Hamiltonian is correspondingly given by $H=\bigoplus_{N=0}^\infty H_N$, where $H_0=0$, $H_1=-\Delta_{x}$ and
\begin{equation}\label{Hamilton n}
 H_{N}=\sum_{j=1}^{N}-\Delta_{{x}_j}+\sum_{1\leq i<j\leq N}v^L(\vert {x}_i-{x}_j\vert)
\end{equation}
for $N\geq2$.  Here $\vert \cdot\vert$ denotes the metric on the torus $\Lambda_L$.

\par In this paper, we consider a more generalized Hamiltonian given by
\begin{equation}\label{Hamilton n generalized}
 H_{N}=\sum_{j=1}^{N}Q(\nabla_j)+\sum_{1\leq i<j\leq N}v^L(\vert {x}_i-{x}_j\vert)
\end{equation}
where $Q$ satisfies the following assumption:
\begin{assumption}\label{assumption Q}
 $Q(ix)\in\mathbb{R}_{\geq0}$ for $x\in \mathbb{R}^3$ is a continuous radially-symmetric function such that $Q(0)<\mu$ and $Q(ix)$ grows monotonously in $\vert x\vert$ to infinity as $\vert x\vert\to\infty$. The degree $d$ of $Q$ is defined by
\begin{equation}\label{deg Q}
  C_1\leq\frac{Q(ix)}{\vert x\vert^d}\leq C_2
\end{equation}
for some $\vert x\vert>C_3$. $C_i,i=1,2,3$ are constants depends only on $Q$. We set
$0<d<\infty$.
\par Notice that $Q(0)<\mu$ and $Q(ix)$ is monotone in $\vert x\vert$ implies $Q(ir)-\mu$ has exactly one zero point for $r=\vert x\vert\in[0,\infty)$.
\end{assumption}
\noindent The operator $Q(\nabla_j)$ is defined for $\psi\in\mathcal{H}^{\wedge N}$
\begin{equation}\label{define pesudo Q}
  \frac{1}{L^{\frac{3}{2}}}\int_{\Lambda_L} \big[Q(\nabla_j)\psi\big](x_1,\dots,x_N)e^{-ikx_j}dx_j=
  \frac{Q(ik)}{L^{\frac{3}{2}}}\int_{\Lambda_L}\psi(x_1,\dots,x_N)e^{-ikx_j}dx_j
\end{equation}
for $k\in(2\pi/L)\mathbb{Z}^3$.

\par A typical example of $Q$ is the classical kinetic energy operator given in (\ref{Hamilton n}), where $Q(ix)=\vert x\vert^2$. Notice that $Q(0)=0$, thus for any $\mu>0$ fixed, $Q(ix)=\vert x\vert^2$ satisfies Assumption \ref{assumption Q}. Another intriguing example is considering BCS in relativistic quantum theory, see for example \cite{tadafumi1,tadafumi2}, where we can normalize $Q(ix)=\sqrt{\vert x\vert^2+1}$. Notice that this time $Q(0)=1$, thus $Q$ satisfies Assumption \ref{assumption Q} only for $\mu>1$.

\par  $v^L$ is the periodization of the interaction potential $v$ given by
\begin{equation}\label{v^L}
  v^L(x)=\sum_{k\in \mathbb{Z}^3}v(x+kL).
\end{equation}
Throughout this paper, we always assume $v$ satisfies following properties:
\begin{assumption}\label{assumption v}
  We require $v\in L^1(\mathbb{R}^3)$ to be radially-symmetric, and its Fourier transform satisfies $\hat{v}\leq0$ and $\hat{v}(0)<0$, here
\begin{equation}\label{fourier v}
  \hat{v}(k)=\int_{\mathbb{R}^3}v(x)e^{-ikx}dx.
\end{equation}
 Let $d$ given in (\ref{deg Q}), if $0<d\leq 3$, then we require for any $k\in\mathbb{R}^3$,
\begin{equation}\label{requirement for v 1}
  \vert\hat{v}(k)\vert\leq \frac{C}{1+\vert k\vert^{3-d+\varepsilon}},
\end{equation}
Moreover, if $0<d<3$, we require for any $x\in\mathbb{R}^3$ and $\langle x\rangle=\sqrt{1+x^2}$,
\begin{equation}\label{requirement for v 2}
  \int_{\mathbb{R}^3}\vert v(x)\vert^{\frac{3}{d}}\langle x\rangle^{\frac{3(3-d)}{d}
  (1+\varepsilon)}dx<\infty,
\end{equation}
and if $d=3$, we require there exists some $0<d^\prime<3$
\begin{equation}\label{requirement for v 3}
   \int_{\mathbb{R}^3}\vert v(x)\vert^{\frac{3}{d^\prime}}\langle x\rangle^{\frac{3(3-d^\prime)}{d^\prime}
  (1+\varepsilon)}dx<\infty,
\end{equation}
where $C,\varepsilon>0$ are universal constants.
\end{assumption}
\noindent Notice that (\ref{requirement for v 2}) implies $v\in L^1\cap L^{3/d}$ for $0<d\leq 3$. We leave the properties of periodization $v^L$ to Appendix \ref{property of v^L}.

%Moreover, we assume $v$ has scattering length $\mathfrak{a}_0$ (For a precise discussion of the scattering length one can see, for example, \cite[Appendix C]{lieb2005mathematics}).
\par For further usage, we also denote the non-interacting Hamiltonian by $\mathcal{K}=\bigoplus_{N=0}^\infty\mathcal{K}_N$, where $\mathcal{K}_0=0$ and
\begin{equation}\label{Hamilton K_n non-inter}
 \mathcal{K}_{N}=\sum_{j=1}^{N}Q(\nabla_j)
\end{equation}
for $N\geq1$. The interaction operator is given by $\mathcal{V}=\bigoplus_{N=0}^\infty\mathcal{V}_N$, where $\mathcal{V}_0=\mathcal{V}_1=0$ and
\begin{equation}\label{Interaction op V_n}
  \mathcal{V}_N=\sum_{1\leq i<j\leq N}v^L({x}_i-{x}_j)
\end{equation}
for $N\geq2$. Moreover, the particle number operator on the Fock space is defined by $\mathcal{N}=\bigoplus_{N=0}^\infty N$.

\vspace{1em}
\par An operator $\Gamma$ on $\mathcal{F}$ is a density matrix, if it is a non-negative trace class operator with $\tr_{\mathcal{F}}\Gamma=1$. We say $\Gamma$ preserves particle number if it commutes with the particle number operator $\mathcal{N}$.

\par For a density matrix $\Gamma$, the thermodynamic pressure functional for Hamiltonian $H$ is defined by
\begin{equation}\label{pressure functional interacting}
  -L^3P^L_{qtm}[\Gamma]\coloneqq \tr_\mathcal{F}
  \big((H-\mu\mathcal{N})\Gamma\big)-\frac{1}{\beta}S_Q[\Gamma],
\end{equation}
where $S_Q[\Gamma]\coloneqq -\tr_\mathcal{F}(\Gamma\ln\Gamma)$ is the von Neumann entropy, and $\mu>0$ denotes the chemical potential. We can always consider $\Gamma$ such that all the eigenfunctions of $\Gamma$ lies in the quadratic form domain of both $H$ and $\mathcal{N}$, otherwise we can simply take $\tr_\mathcal{F}
\big((H-\mu\mathcal{N})\Gamma\big)=+\infty$.

\par Following \cite{GenH_F}, a state $g$ is a complex-valued linear functional defined on the space of $\mathcal{L}(\mathcal{F})$ i.e. the bounded linear operator on the Fock space, and $g$ preserves the $C^*$ algebra, that is, $g$ satisfies $g(\mathbf{I}_{\mathcal{F}})=1$, $g(A^*A)\geq0$ (and therefore $g(A^*)=\overline{g(A)}$) for any $A\in\mathcal{L}(\mathcal{F})$. Moreover, we always consider $g$ that can be extended to operators bounded below.

\par Notice that
\begin{equation}\label{orthonormal basis H}
  f_{k,\sigma}(x,\nu)=\frac{1}{L^{\frac{3}{2}}}e^{ikx}
  \chi_{\nu=\sigma},\quad (k,\sigma)\in(2\pi/L)\mathbb{Z}^3
  \times \mathcal{S}_2,
\end{equation}
is an orthonormal basis of the Hilbert space $\mathcal{H}$. For $h=\sum_{k,\sigma}h_{k,\sigma}f_{k,\sigma}\in\mathcal{H}$, the anti-unitary involution\footnote{It is also known as the complex conjugation operation of a $C^*$ algebra, different from the usual conjugation of a complex function} is given by $h^c=\sum_{k,\sigma}\overline{h_{k,\sigma}}f_{k,\sigma}$. The generalized one-particle density matrix $M$ of $g$ is an operator on $\mathcal{H}\oplus\mathcal{H}$, defined via the quadratic form:
\begin{equation}\label{definiiton of generalized 1pdm}
  \langle M(h_1,h_2),(k_1,k_2)\rangle=g\big([a^*(h_1)+a(h_2^c)][a(k_1)+a^*(k_2^c)]\big).
\end{equation}
By CAR (see (\ref{anticommutator})), it is clear that $0\leq M\leq 1$. $M$ can be written in a $2\times2$ matrix
\begin{equation}\label{2times2}
 0\leq M=\begin{pmatrix}
      \gamma & \alpha  \\
      \alpha^* & 1-{\gamma}^c
    \end{pmatrix}\leq1
\end{equation}
where
\begin{equation}\label{gamma and alpha}
  \begin{aligned}
  \langle\gamma f_{k,\sigma},f_{m,\nu}\rangle&=g(a^*_{k,\sigma}a_{m,\nu})\eqqcolon\gamma_
  {\mathbf{k}\mathbf{m}}\\
  \langle\alpha f_{k,\sigma},f_{m,\nu}\rangle&=g(a_{k,\sigma}a_{m,\nu})\eqqcolon\alpha_
  {\mathbf{k}\mathbf{m}}
  \end{aligned}
\end{equation}
and we let $\mathbf{k}=(k,\sigma), \mathbf{m}=(m,\nu)$ for short.  The conjugation of an operator $A$ on $\mathcal{H}$ is defined through the orthonormal basis $\{f_{k,\sigma}\}$:
\begin{equation}\label{conjugate of A}
  \langle A^cf_{k,\sigma},f_{m,\nu}\rangle=\overline{\langle Af_{k,\sigma},f_{m,\nu}\rangle}.
\end{equation}
It is straightforward to check
\begin{equation}\label{property of gamma and alpha}
  \gamma^*=\gamma,\quad \alpha^T\coloneqq (\alpha^*)^c =-\alpha.
\end{equation}
Therefore, we have
\begin{equation}\label{property of gammakm and alphakm}
  \gamma_{\mathbf{k}\mathbf{m}}=\overline{\gamma}_{\mathbf{m}\mathbf{k}},\quad
  \alpha_{\mathbf{k}\mathbf{m}}=-\alpha_{\mathbf{m}\mathbf{k}}.
\end{equation}
\begin{definition}\label{def 1}
  A state is said to be with finite particle number if $g(\mathcal{N})=\tr_{\mathcal{H}}\gamma<\infty$. Also we say any $(\gamma,\alpha)$ satisfies (\ref{2times2})-(\ref{property of gammakm and alphakm}) with finite particle number is \textbf{admissible}.
\end{definition}
\noindent By (\ref{gamma and alpha}), the integral kernels of $\gamma$ and $\alpha$ can be given by
\begin{equation}\label{define alpha gamma function}
  \begin{aligned}
  \gamma(\mathbf{x};\mathbf{y})&=\sum_{\mathbf{k},\mathbf{m}}\gamma_{\mathbf{m}\mathbf{k}}
  f_{\mathbf{k}}(\mathbf{x})\overline{f_{\mathbf{m}}}(\mathbf{y}),\\
  \alpha(\mathbf{x};\mathbf{y})&=\sum_{\mathbf{k},\mathbf{m}}\alpha_{\mathbf{m}\mathbf{k}}
  f_{\mathbf{k}}(\mathbf{x})\overline{f_{\mathbf{m}}}(\mathbf{y}),
  \end{aligned}
\end{equation}
where we use the short hand notations
\begin{equation}\label{short hand}
  \begin{aligned}
  \mathbf{x}=(x,\sigma),\,\mathbf{y}=(y,\nu)&\in \Lambda_L\times\mathcal{S}_{2},\\
  \mathbf{k}=(k,\tau),\,\mathbf{m}=(m,\pi)&\in (2\pi/L)\mathbb{Z}^3\times\mathcal{S}_{2}.
  \end{aligned}
\end{equation}
By (\ref{property of gammakm and alphakm}), we can check that
\begin{equation}\label{property of kernel}
  \begin{aligned}
  \overline{\gamma}(\mathbf{y};\mathbf{x})=\gamma(\mathbf{x};\mathbf{y}),\quad
  \alpha(-y,\nu;-x,\sigma)=-\alpha(x,\sigma;y,\nu).
  \end{aligned}
\end{equation}

\par  By a quasi-free state $g$, we mean $g$ can be decomposed following Wick's theorem. That is for a set of operators $\{e_i\}_{i=1}^{2l}$, which are each either creation or annihilation operator on $\mathcal{F}$, then $g(e_1e_2\dots e_{2l-1})=0$ and
\begin{equation}\label{Wick Thm}
  g(e_1e_2\dots e_{2l})=\underset{\tau}
  \sum^\prime\sgn\sigma g(e_{\tau(1)}e_{\tau(2)})\cdots
  g(e_{\tau(2l-1)}e_{\tau(2l)}),
\end{equation}
where $\sum^\prime_{\tau}$ is the sum over permutations $\tau\in S_{2l}$ satisfying
\begin{equation}\label{permute condition}
  \begin{aligned}
  \tau(1)<\tau(3)<\cdots<\tau(2l-1),\\
  \tau(2j-1)<\tau(2j)\quad \text{for all $1\leq j\leq l$}.
  \end{aligned}
\end{equation}
In particular, for $l=2$, we have
\begin{equation}\label{Wick Thm k=2}
  g(e_1e_2e_3e_4)=g(e_1e_2)g(e_3e_4)-g(e_1e_3)g(e_2e_4)+g(e_1e_4)g(e_2e_3).
\end{equation}

\par If we additionally assume finite particle number, i.e. $\tr_{\mathcal{H}}\gamma<\infty$, then quasi-free state $g$ and generalized one-particle density matrix $M$ one-to-one corresponds with each other.
\begin{proposition}\label{1to1correspondencequasifree1-pdm}
\
\begin{enumerate}[label=(\arabic*)]
 \item For $0\leq M\leq 1$ satisfies (\ref{2times2}) and (\ref{property of gamma and alpha}), and $\tr_{\mathcal{H}}\gamma<\infty$. Then there exists a unique quasi-free state $g$ such that $g(\mathcal{N})<\infty$ and has one-particle density matrix $M$.
 \item For any quasi-free state $g$ with finite particle number $g(\mathcal{N})<\infty$, there exists a density matrix $\Gamma_g$, such that for any $A\in\mathcal{L}(\mathcal{F})$,
     \begin{equation}\label{induced density matrix}
  g(A)=\tr_\mathcal{F}A\Gamma_g.
\end{equation}
\end{enumerate}
\end{proposition}
\noindent Proposition \ref{1to1correspondencequasifree1-pdm} (2) is the byproduct of the proof of \cite[Theorem 2.3]{GenH_F}. However, Proposition \ref{1to1correspondencequasifree1-pdm} (1) is proved in \cite[Theorem 2.3]{GenH_F} under the assumption that\footnote{(\ref{assumption1to1}) has also been discussed in stack exchange page \url{https://mathoverflow.net/questions/364720/diagonalization-of-the-generalized-1-particle-density-matrix} without an answer.}
\begin{equation}\label{assumption1to1 intro}
  \dim_{\mathbb{C}}\ker\big(1/2-M\big)\equiv0\mod 2.
\end{equation}
We will complete the proof of Proposition \ref{1to1correspondencequasifree1-pdm} in Section \ref{HF pre}. We remark that (\ref{assumption1to1 intro}) is not trivial for infinite dimensional $\mathcal{H}$, and one needs an index theorem \cite{atyitah-singermod2} by Atiyah and Singer to reach this fact.

\par A density matrix $\Gamma$ satisfying $\tr_{\mathcal{F}}\mathcal{N}\Gamma<\infty$ naturally induce a state $g_\Gamma$, defined by
\begin{equation}\label{induced state}
  g_{\Gamma}(A)=\tr_\mathcal{F}A\Gamma,
\end{equation}
with finite particle number $g_{\Gamma}(\mathcal{N})<\infty$. We notice that state in the form of (\ref{induced state}) can be naturally extended to operators bounded below. On the other hand, Proposition \ref{1to1correspondencequasifree1-pdm} tells us the inverse is also true. Therefore in the following discussion, we will no longer differ a state and a density matrix.

\par By restricting the many-body pressure functional $P^L_{qtm}$ to quasi-free states, and drop the so-called exchange and the direct terms (this is also assumed and explained in many literature such as \cite{Leggett1980DiatomicMA}), we get to
\begin{equation}\label{generalized BCS}
  \begin{aligned}
  -P^{L}[g]\coloneqq&\frac{1}{L^3}\sum_{k,\sigma}\big(Q(ik)
  -\mu\big)\gamma_{\mathbf{k}\mathbf{k}}
  -\frac{1}{\beta L^3}S_{\mathrm{HF}}[g]\\
  &+\frac{1}{2L^3}
  \int_{(\Lambda_L\times\mathcal{S}_{2})^2}v^L(x-y)
  \vert\alpha(-x,\sigma;y,\nu)\vert^2dxd\sigma dyd\nu.
  \end{aligned}
\end{equation}
$-P^L$ is in fact real-valued.

\par Under (\ref{gamma and alpha}), the translation invariance and $SU(2)$ symmetry conditions are explicitly given by
\begin{equation}\label{assumption}
    \begin{aligned}
    &\gamma_{\mathbf{k}\mathbf{m}}=0,\quad \mathbf{k}\neq\mathbf{m},\\
    &\alpha_{\mathbf{k}\mathbf{m}}=0,\quad \mathbf{k}\neq-\mathbf{m}.
    \end{aligned}
\end{equation}
By (\ref{assumption}), if we denote $\gamma_{\mathbf{k}}=\gamma_{\mathbf{k}\mathbf{k}}$, $\alpha_{\mathbf{k}}=\alpha_{-\mathbf{k},\mathbf{k}}$, and
\begin{equation}\label{define alpha gamma function assumption}
  \begin{aligned}
  \gamma(\mathbf{x})=\sum_{\mathbf{k}}\gamma_{\mathbf{k}}f_{\mathbf{k}}(\mathbf{x}),\quad
  \alpha(\mathbf{x})=\sum_{\mathbf{k}}\alpha_{\mathbf{k}}f_{\mathbf{k}}(\mathbf{x}).
  \end{aligned}
\end{equation}
Then (\ref{define alpha gamma function}) can be rewritten by
\begin{equation}\label{rewritten alpha gamma function}
  \gamma(x,\sigma;y,\nu)=\frac{1}{L^{\frac{3}{2}}}\delta_{\sigma,\nu}\gamma(x-y,\sigma),\quad
  \alpha(x,\sigma;y,\nu)=\frac{1}{L^{\frac{3}{2}}}\delta_{\sigma,-\nu}\alpha(x+y,\sigma).
\end{equation}
Moreover,(\ref{assumption}) can greatly simplify the expression of entropy $S_{\mathrm{HF}}$. In fact, under (\ref{assumption}), every generalized 1-pdm $M$ is unitarily conjugated to a block matrix with each block given by
\begin{equation}\label{M(k)}
  M(\mathbf{k})=\begin{pmatrix}
      \gamma_{\mathbf{k}} & \alpha_{\mathbf{k}}  \\
      \overline{\alpha}_{\mathbf{k}} & 1-{\gamma}_{-\mathbf{k}}
    \end{pmatrix}.
\end{equation}
Then one can check that
\begin{equation}\label{entropy HF assumption}
  \tr_{\mathcal{H}^2}M\ln M=\sum_{\mathbf{k}}\tr_{\mathbb{C}^2}M(\mathbf{k})\ln M(\mathbf{k}),
\end{equation}
Therefore, we can give the definition of classical BCS pressure
\begin{equation}\label{Pressure BCS}
\begin{aligned}
  -L^3P_{\mathrm{BCS}}^L[g]\coloneqq &\sum_{k,\sigma}\big(Q(ik)-\mu\big)\gamma_{\mathbf{k}}
  +\frac{1}{\beta}\sum_{\mathbf{k}}\tr_{\mathbb{C}^2}M(\mathbf{k})\ln M(\mathbf{k})\\
  &+\frac{1}{2}\int_{\Lambda_L\times\mathcal{S}_{2}}
  v^L(x)\vert\alpha(x,\sigma)\vert^2dxd\sigma.
\end{aligned}
\end{equation}
Sometimes half potential $\frac{1}{2}v^L$ may be rewritten by $V$ instead.

\par For positive temperature $T>0$, the BCS gap equation, derived from (\ref{Pressure BCS}), is given by
\begin{equation}\label{Gap eqn intro}
    \Delta_{k,\sigma}=-\frac{1}{2L^3}\sum_{k^\prime}\hat{v}(k-k^\prime)
    \frac{\Delta_{k^\prime,\sigma}}{E_{k^\prime,\sigma}}\tanh\frac{\beta}{2}E_{k^\prime,\sigma},
  \end{equation}
where
\begin{equation}\label{define E_k intro}
  E_k\coloneqq\sqrt{\big(Q(ik)-\mu\big)^2+\vert\Delta_{k,\sigma}\vert^2}.
\end{equation}
For $T=0$, the gap equation becomes
\begin{equation}\label{Gap eqn intro T=0}
    \Delta_{k,\sigma}=-\frac{1}{2L^3}\sum_{k^\prime}\hat{v}(k-k^\prime)
    \frac{\Delta_{k^\prime,\sigma}}{E_{k^\prime,\sigma}}.
  \end{equation}
Like $\alpha_{-\mathbf{k}}=-\alpha_{\mathbf{k}}$ (see (\ref{property of gammakm and alphakm}) and (\ref{assumption})), the gap $\Delta$ should also satisfy $\Delta_{-\mathbf{k}}=-\Delta_{\mathbf{k}}$. Therefore, given that $\Delta$ solves (\ref{Gap eqn intro}) or (\ref{Gap eqn intro T=0}), for any constant phase. We can therefore omit the spin variable and simply consider
\begin{equation}\label{Gap eqn again start intro}
  \Delta_{k}=-\frac{1}{2L^3}\sum_{k^\prime}\hat{v}(k-k^\prime)
    \frac{\Delta_{k^\prime}}{E_{k^\prime}}\tanh\frac{\beta}{2}E_{k^\prime},
\end{equation}
and we can generate solution(s) to (\ref{Gap eqn intro}) or (\ref{Gap eqn intro T=0}) by defining
\begin{equation}\label{generate gap soln}
  \Delta_{k,\sigma}=c(\sigma)\Delta_{k}
\end{equation}
with $c(-\sigma)=-c(\sigma)\in\mathcal{S}^1$.

\par  Obviously, $\Delta\equiv 0$ is a solution to (\ref{Gap eqn again start intro}). On the other hand, notice that since $\hat{v}(0)<0$ and $\hat{v}$ is continuous, there exists some universal constant $r>0$ such that $\hat{v}(k)<0$ for any $\vert k\vert<r$. Therefore, as long as $L$ is large enough, such that $B_r\cap\big((2\pi/L)\mathbb{Z}^3\backslash\{0\}\big)\neq\emptyset$, one can prove by induction that any nonzero nonnegative solution of (\ref{Gap eqn again start intro}) is strictly positive.

\par Let $L$ tend to infinity, the discrete classical BCS pressure $P^L_{\mathrm{BCS}}$ given in (\ref{Pressure BCS}) formally becomes
\begin{equation}\label{pressure BCS continuous}
\begin{aligned}
  -(2\pi)^3P_{\mathrm{BCS}}[\gamma,\alpha]=&\int_{\mathbb{R}^3\times\mathcal{S}_{2}}
\big(Q(ik)-\mu\big)\gamma_{\mathbf{k}}d\mathbf{k}+\frac{1}{2}
\int_{\mathbb{R}^3\times\mathcal{S}_{2}}
v(x)\vert\check{\alpha}(x,\sigma)\vert^2dxd\sigma\\
&+\frac{1}{\beta}\int_{\mathbb{R}^3\times\mathcal{S}_{2}}
\tr_{\mathbb{C}^2}M(\mathbf{k})\ln M(\mathbf{k})d\mathbf{k},
\end{aligned}
\end{equation}
where
\begin{equation}\label{check alpha}
  \check{\alpha}(x,\sigma)=\frac{1}{(2\pi)^{\frac{3}{2}}}
  \int_{\mathbb{R}^3}
  \alpha_{k,\sigma}e^{ikx}dk,
\end{equation}
and
\begin{equation}\label{M continuous}
  M(\mathbf{k})=\begin{pmatrix}
      \gamma_{\mathbf{k}} & \alpha_{\mathbf{k}}  \\
      \overline{\alpha}_{\mathbf{k}} & 1-{\gamma}_{-\mathbf{k}}
    \end{pmatrix}
\end{equation}
for $\mathbf{k}\in\mathbb{R}^3\times\mathcal{S}_{2}$, which satisfies $0\leq M(\mathbf{k})\leq1$. We let $\mathcal{D}$ denote the domain of $P_{\mathrm{BCS}}$:
\begin{equation}\label{domain D}
  \mathcal{D}=\Big\{(\gamma,\alpha)\,\vert\,
  \int_{\mathbb{R}^3\times\mathcal{S}_{2}}(1+k^d)\gamma_{k,\sigma}dkd\sigma<\infty,
  0\leq M(\mathbf{k})\leq1
  \Big\}.
\end{equation}
We remark that $(\gamma,\alpha)\in\mathcal{D}$ implies that $\check{\alpha}\in H^{d/2}$. Corresponding to (\ref{Gap eqn intro}) and (\ref{Gap eqn intro T=0}), the BCS gap equation of (\ref{pressure BCS continuous}) is given by
\begin{equation}\label{gap eqn intro continuous}
   \Delta_{k,\sigma}=-\frac{1}{2(2\pi)^3}\int_{\mathbb{R}^3}\hat{v}(k-k^\prime)
    \frac{\Delta_{k^\prime,\sigma}}{E_{k^\prime,\sigma}}\tanh\frac{\beta}{2}E_{k^\prime,\sigma}
    dk^\prime
\end{equation}
for $T>0$, and
\begin{equation}\label{gap eqn intro continuous T=0}
   \Delta_{k,\sigma}=-\frac{1}{2(2\pi)^3}\int_{\mathbb{R}^3}\hat{v}(k-k^\prime)
    \frac{\Delta_{k^\prime,\sigma}}{E_{k^\prime,\sigma}}dk^\prime
\end{equation}
for $T=0$. We recall that $\hat{v}(k)$ is given by (\ref{fourier v}). Similarly, we can omit the spin variables and simply consider
\begin{equation}\label{gap eqn intro continuous again}
   \Delta_{k}=-\frac{1}{2(2\pi)^3}\int_{\mathbb{R}^3}\hat{v}(k-k^\prime)
    \frac{\Delta_{k^\prime}}{E_{k^\prime}}\tanh\frac{\beta}{2}E_{k^\prime}dk^\prime,
\end{equation}
and we can simply regard $\tanh(\beta E_{k^\prime}/2)\equiv 1$ when $T=0$.

 \par We leave the precise introduction of $P^{L}$ and $P^{L}_{\mathrm{BCS}}$ to Section \ref{HF pre}.

\subsection{Main Theorems}\label{main}
\
\par Theorem \ref{main1} states the well-posedness of equation (\ref{Gap eqn again start intro})
\begin{theorem}\label{main1}
  Let $v$ satisfies Assumption \ref{assumption v}, then there exists a constant $C(v,\mu,Q)$ such that for fixed $L>C(v,\mu,Q)$,
  \begin{enumerate}[label=(\arabic*)]
    \item There exists a bounded maximal nonnegative solution $\tilde{\Delta}\geq0$ of (\ref{Gap eqn again start intro}), such that for any other nonnegative solution $\Delta\geq0$, we always have $0\leq\Delta\leq\tilde{\Delta}$. Moreover, there exists a positive temperature $0<T_c=T_c(v,\mu,Q,L)<\infty$, such that for any $0\leq T<T_c$, $\tilde{\Delta}>0$, and for $T\geq T_c$, $\tilde{\Delta}\equiv0$, that is, there is no nonzero nonnegative solution for $T\geq T_c$. Moreover, for each $k\in (2\pi/L)\mathbb{Z}^3$, $\tilde{\Delta}_k$ is a continuous function in temperature $T\in[0,\infty)$.
    \item For $0\leq T<T_c$, the nonzero nonnegative solution to (\ref{Gap eqn again start intro}) is unique. That is, for $0\leq T<T_c$, (\ref{Gap eqn again start intro}) has a unique positive solution $\tilde{\Delta}$, and every nonzero nonnegative solution equals to the maximal nonnegative solution $\tilde{\Delta}$. For $T\geq T_c$, the only nonnegative solution to (\ref{Gap eqn again start intro}) is zero.
  \end{enumerate}
  Moreover, $(1)$ and $(2)$ also hold true when $L=\infty$, with the BCS gap equation now being (\ref{gap eqn intro continuous again}) rather than (\ref{Gap eqn again start intro}).
\end{theorem}

\par Theorem \ref{main1} in fact implies the uniqueness of maximizer to the classical BCS pressure functional $P^L_{\mathrm{BCS}}$.
\begin{corollary}\label{uniqueness maximizer intro}
  For $0\leq T<T_c$, up to phase changes of $\alpha^{sym}$, $P^L_{\mathrm{BCS}}$ has a unique maximizer $(\gamma^{sym},\alpha^{sym})$, and this maximizer can be chosen so that $\alpha^{sym}_{k,\uparrow}$ is positive everywhere. For $T\geq T_c$, $P^L_{\mathrm{BCS}}$ has a unique maximizer $(\gamma^{sym},0)$. In both case, the maximizer can be generated by
  \begin{align}
  \gamma^{sym}_{k,\sigma}=&\frac{1}{2}\Big(1-\frac{\xi_k}{E_{k,\sigma}}\tanh\frac{\beta}{2}
  E_{k,\sigma}\Big),\label{Gap gamma intro}\\
  \alpha^{sym}_{k,\sigma}=&\frac{\Delta_{k,\sigma}}{2E_{k,\sigma}}\tanh\frac{\beta}{2}E_{k,\sigma},
  \label{Gap alpha intro}
  \end{align}
  where
  \begin{align}
  \xi_k\coloneqq&\big(Q(ik)-\mu\big),\label{define xi_k intro coro}\\
  E_k\coloneqq&\sqrt{\xi_k^2+\vert\Delta_{k,\sigma}\vert^2}.\label{define E_k intro coro}
  \end{align}
and $\Delta_{k,\sigma}$ solves (\ref{Gap eqn intro}) when $T>0$ or (\ref{Gap eqn intro T=0}) when $T=0$ such that $\Delta_{k,\uparrow}>0$ for $0\leq T<T_c$ and $\Delta_{k,\sigma}\equiv
0$ for $T\geq T_c$.
\par Moreover, Corollary \ref{uniqueness maximizer intro} also holds true when $L=\infty$ with (\ref{Gap eqn intro}) and (\ref{Gap eqn intro T=0}) being replaced by (\ref{gap eqn intro continuous}) and (\ref{gap eqn intro continuous T=0}), and $P^L_{\mathrm{BCS}}$ being replaced by $P_{\mathrm{BCS}}$.
\end{corollary}

\par Furthermore, for temperature not greater than the critical temperature in Theorem \ref{main1}, the maximizer of the generalized BCS pressure functional $P^L$ defined in (\ref{generalized BCS}) in fact satisfies the translation invariance and $SU(2)$ symmetry condition (\ref{assumption}).

\begin{theorem}\label{main2}
  For $C(v,\mu,Q)<L<\infty$ and $0\leq T<T_c(v,\mu,Q,L)$, $P^L$ has a unique maximizer $(\gamma^{sym},\alpha^{sym})$ given in Corollary \ref{uniqueness maximizer intro}, up to a phase change of $\alpha^{sym}$, and this maximizer can be chosen so that $\alpha^{sym}_{k,\uparrow}$ is positive everywhere. In particular,
  \begin{equation}\label{main2 PBCS=P}
    \min-L^3P^L=\min-L^3P^L_{\mathrm{BCS}}.
  \end{equation}
Moreover, any nonzero nonngeative solution to the gap equation (\ref{Gap eqn intro}) or (\ref{Gap eqn intro T=0}) generates this unique maximizer (up to a phase change of $\alpha^{sym}$).
\end{theorem}

\par As $L\to\infty$, $P_{\mathrm{BCS}}$ is in fact the limit model of $P_{\mathrm{BCS}}^{L}$, and their maximum can be represented in the form of $\Delta^L$ or $\Delta$, the solutions of BCS gap equation (\ref{Gap eqn intro}) or (\ref{gap eqn intro continuous}) respectively. We have the following result.
\begin{theorem}\label{main3}
  For $C(v,\mu,Q)<L<\infty$ and $0\leq T<T_c(v,\mu,Q,L)$, let $\Delta^L$ be the unique positive solution to (\ref{Gap eqn intro}), or (\ref{Gap eqn intro T=0}) when $T=0$, such that $\Delta_{k,\uparrow}^L>0$. Then we have
  \begin{equation}\label{value min -P_BCS^L}
  \begin{aligned}
    \min-L^3P^L_{\mathrm{BCS}}=&-\frac{1}{\beta}
    \sum_{k,\sigma}\ln\big(1+e^{-\beta E_{k,\sigma}}\big)
    +\frac{1}{2}\sum_{k,\sigma}\big(\xi_k-E_{k,\sigma}\big)\\
    &+\frac{1}{4}\sum_{k,\sigma}\frac{\vert\Delta_{k,\sigma}^L\vert^2}{E_{k,\sigma}}
    \tanh\frac{\beta}{2}E_{k,\sigma}.
  \end{aligned}
  \end{equation}
  Here $\xi_{k}$ and $E_{k,\sigma}$ is given by (\ref{define xi_k intro coro}) and (\ref{define E_k intro coro}), and $k\in(2\pi/L)\mathbb{Z}^3$. (\ref{value min -P_BCS^L}) also holds true when $T\geq T_c(v,\mu,Q,L)$, with $\Delta^L$ now being identically zero.
  \par For $L=\infty$ and $0\leq T<T_c(v,\mu,Q,\infty)$, let $\Delta$ be the unique positive solution to (\ref{gap eqn intro continuous}), or (\ref{gap eqn intro continuous T=0}) when $T=0$, such that $\Delta_{k,\uparrow}>0$. Then we
  \begin{equation}\label{value min -P_BCS}
  \begin{aligned}
    \min-(2\pi)^3P_{\mathrm{BCS}}=&-\frac{1}{\beta}
    \int_{\mathbb{R}^3\times\mathcal{S}_{2}}\ln\big(1+e^{-\beta E_{k,\sigma}}\big)
    dkd\sigma
    +\frac{1}{2}\int_{\mathbb{R}^3\times\mathcal{S}_{2}}\big(\xi_k-E_{k,\sigma}\big)
    dkd\sigma\\
    &+\frac{1}{4}\int_{\mathbb{R}^3\times\mathcal{S}_{2}}
    \frac{\vert\Delta_{k,\sigma}\vert^2}{E_{k,\sigma}}
    \tanh\frac{\beta}{2}E_{k,\sigma}dkd\sigma.
  \end{aligned}
  \end{equation}
  (\ref{value min -P_BCS}) also holds true when $T\geq T_c(v,\mu,Q,\infty)$, with $\Delta$ now being identically zero.
  \par Moreover, there exists a constant $C^\prime=C^\prime(v,\mu,Q)\geq C(v,\mu,Q)$, such that when $L>C^\prime(v,\mu,Q)$, $T_c(v,\mu,Q,\infty)\leq T_c(v,\mu,Q,L)$, and
  \begin{align}
    \lim_{L\to\infty}T_c(v,\mu,Q,L)=&T_c(v,\mu,Q,\infty),\label{limit critical temperature}\\
    \lim_{L\to\infty}\min -P^L_{\mathrm{BCS}}=&\min-P_{\mathrm{BCS}}.\label{limit model}
  \end{align}
  We define for any $T\geq0$ and $L>C(v,\mu,Q)$ (if $T>T_c$ we simply choose $\Delta^L\equiv0$),
  \begin{equation}\label{define tildeDelta^L}
    \tilde{\Delta}^L_{k,\sigma}\coloneqq-\frac{1}{2L^3}\sum_{k^\prime\in(2\pi/L)\mathbb{Z}^3}
    \hat{v}(k-k^\prime)\frac{\Delta_{k^\prime,\sigma}^L}{E_{k^\prime,\sigma}}
    \tanh\frac{\beta}{2}E_{k^\prime,\sigma},\quad k\in\mathbb{R}^3.
  \end{equation}
  Here we regard $ \tanh{\beta}E_{k^\prime,\sigma}/2\equiv1$ when $T=0$. Then $\tilde{\Delta}^L_{k,\sigma}=\Delta^L_{k,\sigma}$ if $k\in(2\pi/L)\mathbb{Z}^3$. We have
  \begin{equation}\label{limit delta}
    \lim_{L\to\infty}\Vert\tilde{\Delta}^L-\Delta\Vert_{L^\infty}=0.
  \end{equation}
\end{theorem}

\begin{remark}\label{remark}
  One can hardly derive the limiting model of $P^L$ defined in (\ref{generalized BCS}) by formally letting $L$ go to infinity, since we have to impose suitable boundary condition in the limiting functional. A proposed consideration to avoid this trouble is to consider $P^L$ on a smaller domain, but still larger than that of $P^L_{\mathrm{BCS}}$, that is, to assume the system is always periodic $1$, a shorter period than the original period $L$. The limiting functional of such systems is considered in, for example, \cite{persist,frank2012microscopic}.
\end{remark}

\subsection{Organization of the paper}
\
\par In Section \ref{Preliminaries}, we will thoroughly discuss the definitions of the Fock space and different types of pressure functionals given in the introduction. Moreover, in Section \ref{HF pre} we will complete the proof of Proposition \ref{1to1correspondencequasifree1-pdm}. In Section \ref{provemain1}, we prove Theorem \ref{main1} and Corollary \ref{uniqueness maximizer intro}. In Section \ref{persist}, we prove Theorem \ref{main2}, and in Section \ref{largeL}, we prove Theorem \ref{main3}. Most of the proof is hard direct analysis, only seemed to be out of the theme is the proof of Prop \ref{1to1correspondencequasifree1-pdm}, in which an Atiyah-Singer index theorem is used.

\section{Preliminaries}\label{Preliminaries}

\subsection{Fock space}
\
\par The fermionic Fock space is given by
\begin{equation}\label{Fock space definition}
  \mathcal{F}=\bigoplus_{N=0}^\infty \mathcal{H}^{\wedge N},
\end{equation}
where $\mathcal{H}^{\wedge 0}=\mathbb{C}$,
\begin{equation}\label{H define}
  \mathcal{H}=L^2(\Lambda_L\times\mathcal{S}_2)\cong L^2(\Lambda_L;\mathbb{C}^2),
\end{equation}
and $\mathcal{H}^{\wedge N}=L^2_a\big((\Lambda_L\times\mathcal{S}_{2})^N\big)$ is the space of $N$-particle fermionic wave functions that are antisymmetric with respect to permutations of particle coordinates.

\par For $k\in(2\pi/L)\mathbb{Z}^3$ and $\sigma\in\mathcal{S}_2$, we choose the orthonormal basis of $\mathcal{H}$ by
\begin{equation}\label{orthonormal basis H fock space}
  f_{k,\sigma}(x,\nu)=\frac{1}{L^{\frac{3}{2}}}
  e^{ik\cdot x}
  \chi_{\nu=\sigma},\quad z=(x,\nu)\in\Lambda_L\times\mathcal{S}_2.
\end{equation}
For non-negative integers $n,m$, and $\psi\in L^2\big((\Lambda_L\times\mathcal{S}_2)^n\big)$, $\varphi\in L^2\big((\Lambda_L\times\mathcal{S}_2)^m\big)$, we define the antisymmetric wedge product between $\psi$ and $\varphi$
\begin{equation*}\label{wedge product}
\begin{aligned}
  \psi\wedge\varphi(z_1,\dots,z_{n+m})=& \frac{1}{\sqrt{n!m!(n+m)!}}\\
&\times\sum_{\varrho\in S_{n+m}}\sgn{\varrho}\,\psi(z_{\varrho(1)},\dots,z_{\varrho(n)})
  \varphi(z_{\varrho(n+1)},\dots,z_{\varrho(n+m)}).
\end{aligned}
\end{equation*}
Under these definitions, it is straightforward to verify $\psi\wedge\varphi\in L^2_a\big((\Lambda_L\times\mathcal{S}_2)^{n+m}\big)$, and the wedge product is associative and anti-commutative, i.e.
\begin{equation}\label{anti-commutative}
  \psi\wedge\varphi=(-1)^{nm}\varphi\wedge\psi.
\end{equation}
Under the definition of the wedge product, the notation $L_a^2\big((\Lambda_L\times\mathcal{S}_2)^N\big)= \mathcal{H}^{\wedge N}$ holds rigorously and we find an orthonormal basis of $\mathcal{H}^{\wedge N}$ given by
\begin{equation}\label{orthonormal basis H wedge N}
  \bigwedge_{i=1}^{N} f_{k_i,\sigma_i}=f_{k_1,\sigma_1}\wedge\cdots\wedge f_{k_N,\sigma_N}.
\end{equation}

\par For any $N\in\mathbb{N}_{\geq0}$, we can define the creation and annihilation operators on $\mathcal{H}^{\wedge N}$ by
\begin{equation}\label{define a*(f) & a(g)}
\left.\begin{aligned}
  (a^*(f)\psi)({z}_1,\dots,{z}_{N+1}) &= \frac{1}{\sqrt{N+1}}\sum_{i=1}^{N+1}(-1)^{i+1}f(z_i)
\psi(z_1,\dots,\hat{z_i},\dots,z_{N+1}) \\
  (a(g)\psi)(z_1,\dots,z_{N-1}) &= \sqrt{N}\int_{\Lambda_{L}\times\mathcal{S}_2}\overline{g}(z)
\psi(z,z_1,\dots,z_{N-1})dz, \quad N\geq 1
\end{aligned}\right.
\end{equation}
We additionally define $a(g)=0$ for $N=0$. It is then also reasonable to treat $a^*$ and $a$ as operators defined on the fermionic Fock space $\mathcal{F}$. We can check the following canonical anti-commutator relation (CAR):
\begin{equation}\label{anticommutator}
\begin{aligned}
  &\{a(g),a^*(f)\}=a(g)a^*(f)+a^*(f)a(g)=\langle f,g\rangle,\\
  &\{a(g),a(f)\}=\{a^*(g),a^*(f)\}=0.
\end{aligned}
\end{equation}
Using (\ref{anticommutator}), we can bound the norm of creation and annihilation operators by
\begin{equation}\label{norm bound  of creation and annihilation}
  \Vert a(g)\Vert\leq\Vert g\Vert_2,\quad\Vert a^*(f)\Vert\leq \Vert f\Vert_2.
\end{equation}
Using (\ref{orthonormal basis H}), we denote
\begin{equation}\label{define a_k,sigma a*_k,sigma}
  a_{k,\sigma}=a(f_{k,\sigma}),\quad a^*_{k,\sigma}=a^*(f_{k,\sigma}).
\end{equation}

\par Using creation and annihilation operators defined above, we can rewrite the Hamiltonian $H$ defined through (\ref{Hamilton n generalized}) by:
\begin{equation}\label{second quantization H_N}
  H=\mathcal{K}+\mathcal{V}=\sum_{k,\sigma}Q(ik) a_{k,\sigma}^*a_{k,\sigma}
  +\frac{1}{2L^3}\sum_{k,p,q,\sigma,\nu}\hat{v}(k)
  a^*_{p-k,\sigma}a^*_{q+k,\nu}a_{q,\nu}a_{p,\sigma},
\end{equation}

\subsection{Quantum Thermodymamic Pressure}\label{quantum}
\
\par Recall that an operator $\Gamma$ on $\mathcal{F}$ is a density matrix, if it is a non-negative trace class operator with $\tr_{\mathcal{F}}\Gamma=1$. We say $\Gamma$ preserves particle number if it commutes with the particle number operator $\mathcal{N}$. It is straight-forward to verify that, for a density matrix $\Gamma$ which preserves particle number, it can be written as $\Gamma=\bigoplus_{N=0}^\infty\Gamma_N$, where $\Gamma_N$ is a non-negative trace class operator on $\mathcal{H}^{\wedge N}$. Moreover, we say $\Gamma$ preserves particle number, if  it commutes with particle number operator $\mathcal{N}$. Since $\Gamma_N$ is a trace class operator, it can be characterized by an integral kernel, which we also denote it by $\Gamma_N$.

\par For a density matrix $\Gamma$ that preserves particle number, we define its normalized one-particle density matrix $\gamma$ as follows. $\gamma$ is an operator on $\mathcal{H}$ defined through the quadratic form, for $f,g\in\mathcal{H}$:
\begin{equation}\label{1pdm define}
  \langle\gamma f,g\rangle=\tr_\mathcal{F}\big(a^*(f)a(g)\Gamma\big).
\end{equation}
Notice that this definition is consistent with (\ref{gamma and alpha}). By definition, its integral kernel is given by
\begin{equation}\label{gamma 1pdm}
  \gamma(z,z^\prime)=\sum_{N=1}^\infty
  N\int_{(\Lambda_L\times\mathcal{S}_{2})^N}\Gamma_N(z,Z_{N-1};z^\prime,Z_{N-1})
  dZ_{N-1}.
\end{equation}
We can easily verify by definition that
\begin{equation}\label{tr gamma=tr NGamma}
  \tr_{\mathcal{H}}\gamma=\tr_{\mathcal{F}}(\mathcal{N}\Gamma).
\end{equation}
We will only consider $\Gamma$ such that $\tr_\mathcal{F} (\mathcal{N}\Gamma)$ is finite, i.e. $\gamma$ is a positive trace class operator on $\mathcal{H}$.

\par We also introduce the normalized two-particle density matrix $\gamma^{(2)}$ of a general density matrix $\Gamma$, which is an operator on $\mathcal{H}\otimes\mathcal{H}$ given by
\begin{equation}\label{2pdm define}
  \langle\gamma^{(2)} f_1\otimes f_2,g_1\otimes
  g_2\rangle=\tr_\mathcal{F}\big(a^*(f_1)a^*(f_2)a(g_2)a(g_1)\Gamma\big).
\end{equation}
Similar to (\ref{gamma 1pdm}), we can calculate the integral kernel of $\gamma^{(2)}$ of $\Gamma$ that preserves particle number:
\begin{equation}\label{gamma 2 2pm}
\begin{aligned}
  &\gamma^{(2)}(z_1,z_2;z^\prime_1,z^\prime_2)\\
  =&\sum_{N=2}^\infty
  N(N-1)\int_{(\Lambda_L\times\mathcal{S}_{2})^{N-2}}\Gamma_N(z_1,z_2,Z_{N-2};
  z^\prime_1,z_2^\prime,Z_{N-2})
  dZ_{N-2}.
\end{aligned}
\end{equation}

\par Let $Z_N=(z_1,\dots,z_N)\in (\Lambda_L\times\mathcal{S}_{2})^N$ and $h\in\mathbb{R}^3$, we apply the short-hand notation
\begin{equation}\label{short hand notation Z_N+h}
  Z_N+h\coloneqq(x_1+h,\sigma_1,\dots,x_N+h,\sigma_N).
\end{equation}
For $\Psi=(\psi_0,\psi_1,\dots,\psi_N,\dots)\in\mathcal{F}$, we define the translation operator $\mathcal{T}_h$ on $\mathcal{F}$ by
\begin{equation}\label{translation op}
  \mathcal{T}_h\Psi=
  (\psi_0,\psi_1(Z_1+h),\dots,\psi_N(Z_N+h),\dots).
\end{equation}
Obviously $\mathcal{T}_h^{-1}=\mathcal{T}_h^*=\mathcal{T}_{-h}$. An operator $\Gamma$ on $\mathcal{F}$ is said to be translation-invariant, if $\mathcal{T}_{h}^*\Gamma\mathcal{T}_h=\Gamma$. For a density matrix $\Gamma$ which preserves particle number and possesses translation-invariance, its one-particle density matrix can be written in the form
\begin{equation}\label{1-pdm translation invariance}
  \gamma(z,z^\prime)=\gamma(x-x^\prime,\sigma,\sigma^\prime).
\end{equation}
In particular, we denote $\gamma(x,\sigma,\sigma)=\gamma(x,\sigma)$ for short. We then use (\ref{1-pdm translation invariance}) to define for $k\in(2\pi/L)\mathbb{Z}^3$ and $\nu\in\mathcal{S}_{2}$
\begin{equation}\label{gamma_k,sigma}
\begin{aligned}
  \hat{\gamma}(k,\nu)&\coloneqq \frac{1}{L^3}
  \int_{(\Lambda_L\times\mathcal{S}_{2})^2}
  \gamma(z,z^\prime)e^{-ikx/L}e^{ikx^\prime/L}\chi_{\sigma=\sigma^\prime=\nu}dzdz^\prime\\
  &=\int_{\Lambda_L}\gamma(x,\nu)e^{-ikx/L}dx,
\end{aligned}
\end{equation}
which satisfies $0\leq\hat{\gamma}(k,\nu)\leq1$ since $0\leq\gamma\leq1$ by (\ref{1pdm define}) and CAR.

\par Recall that for a density matrix $\Gamma$, the thermodynamic pressure functional for Hamiltonian $H$ is defined by
\begin{equation}\label{pressure functional interacting sec}
  -L^3P^L_{qtm}[\Gamma]\coloneqq \tr_\mathcal{F}
 \big((H-\mu\mathcal{N})\Gamma\big)-\frac{1}{\beta}S_Q[\Gamma],
\end{equation}
where $S_Q[\Gamma]\coloneqq -\tr_\mathcal{F}(\Gamma\ln\Gamma)$ is the von Neumann entropy, and $\mu>0$ denotes the chemical potential. Notice that, using (\ref{1pdm define}) and (\ref{2pdm define}), we can rewrite
\begin{equation}\label{pressure functional rewrt}
  \tr_{\mathcal{F}}(\mathcal{K}-\mu\mathcal{N})\Gamma=\tr_{\mathcal{H}}(\mathcal{K}-\mu\mathcal{N})
  \gamma,\quad
  \tr_{\mathcal{F}}\mathcal{V}\Gamma=
  \frac{1}{2}\tr_{\mathcal{H}\otimes\mathcal{H}}\mathcal{V}\gamma^{(2)},
\end{equation}
where on the right-hand side of each equality, we treat respectively $(\mathcal{K}-\mu\mathcal{N})$ an operator on $\mathcal{H}$ and $\mathcal{V}$ an operator on $\mathcal{H}\otimes\mathcal{H}$.

\par Let $P^L_{qtm}(\beta,\mu)$ denote the supremum of the thermodynamic pressure functional $P^L_{qtm}[\Gamma]$ over all density matrices. $P^L_{qtm}(\beta,\mu)$ is uniquely attained by the Gibbs density matrix $G$ (see, for example, \cite[Theorem 2.8]{GenH_F})
\begin{equation}\label{Gibbs state interacting}
  G=Z^{-1}e^{-\beta(H-\mu\mathcal{N})},\quad Z=\tr_\mathcal{F} e^{-\beta(H-\mu\mathcal{N})}.
\end{equation}
One can, on the other hand, consider the free energy for system with definite particle number. In physics terminology, the free energy and the thermodynamic pressure can inter-convert via Legendre transformation. However, as mentioned above, the thermodynamic pressure is more convenient for direct analysis, according to \cite[8.6]{huang2008statistical}.

\par The above discussion can be applied correspondingly to the non-interacting Hamiltonian $\mathcal{K}$. The Gibbs density matrix $G_0$ to the non-interacting pressure functional $P^L_{qtm,0}[\Gamma]$ is given by
\begin{equation}\label{Gibbs state non-inter}
  G_0=Z_0^{-1}e^{-\beta(\mathcal{K}-\mu\mathcal{N})},\quad Z_0=\tr_\mathcal{F} e^{-\beta(\mathcal{K}-\mu\mathcal{N})}.
\end{equation}
$G_0$ is translation-invariant and preserves particle and spin numbers. Moreover, by \cite{wickformula}, the state
\begin{equation}\label{define g_0}
  g_0(A)\coloneqq\tr_{\mathcal{F}}AG_0
\end{equation}
is a quasi-free state. By direct calculation, we have
\begin{equation}\label{P^L_{qtm}_0[G_0]}
  P^L_{qtm,0}(\beta,\mu)=P^L_{qtm,0}[G_0]=\frac{1}{L^3\beta}\ln Z_0.
\end{equation}
Let $\gamma_0$ be the one-particle density matrix to $G_0$. From \cite[(8.65)]{huang2008statistical}, we can calculate the diagonal Fourier coefficients (\ref{gamma_k,sigma}) to $\gamma_0$ by
\begin{equation}\label{gamma_0}
  \hat{\gamma}_0(k,\sigma)=\frac{1}{1+\mathfrak{z}^{-1}e^{\beta Q(ik)}},
\end{equation}
where $\mathfrak{z}=e^{\beta\mu}$ is the fugacity. From \cite[(8.63)]{huang2008statistical}, we can write the limit pressure $P_0(\beta,\mu)$ by
\begin{equation}\label{pressure non_interacting}
  P_0(\beta,\mu)=\frac{2}{(2\pi)^3\beta}\int_{\mathbb{R}^3}
  \ln\big(1+\mathfrak{z}e^{-\beta Q(ix)}\big)dx.
\end{equation}
We can calculate directly the corresponding density $\rho_0=\partial P_0/\partial\mu$
\begin{equation}\label{density non interacting}
  \rho_0(\beta,\mu)=\frac{2}{(2\pi)^3}\int_{\mathbb{R}^3}
  \frac{dx}{1+\mathfrak{z}^{-1}e^{\beta Q(ix)}}.
\end{equation}
%or we can write it by (see \cite[25.12]{NIST})
%\begin{equation}\label{density non intercating Fermi Dirac dis}
%  \rho_{0}(\beta,\mu)=-\frac{2}{(4\pi\beta)^{\frac{3}{2}}}L_{\frac{3}{2}}(-\mathfrak{z})
%  =\frac{2}{(4\pi\beta)^{\frac{3}{2}}}F_{\frac{1}{2}}(\beta\mu),
%\end{equation}
%where $L_s$ is the polylogarithm, and $F_s$ is the Fermi-Dirac integral. When $\beta\mu\gg1$ (see for example \cite{dingle1973asymptotic}), we have
%\begin{equation}\label{asymptotic rho_0}
 % \rho_{0}(\beta,\mu)=\frac{2}{6\pi^2\beta^{\frac{3}{2}}}
%  \Big((\beta\mu)^{\frac{3}{2}}+O\big((\beta\mu)^{-\frac{1}{2}}\big)\Big).
%\end{equation}

\subsection{BCS Pressure}\label{HF pre}
\
\par We first complete the proof of Proposition \ref{1to1correspondencequasifree1-pdm}.

\begin{proof}[Proof of Proposition \ref{1to1correspondencequasifree1-pdm}]
\
\par Proposition \ref{1to1correspondencequasifree1-pdm} is proved in \cite[Theorem 2.3]{GenH_F} under the assumption that
\begin{equation}\label{assumption1to1}
  \dim_{\mathbb{C}}\ker\big(1/2-M\big)\equiv0\mod 2.
\end{equation}
We now prove (\ref{assumption1to1}) under the requirements in Proposition \ref{1to1correspondencequasifree1-pdm}.
  \par Let $J$ be the anti-linear complex conjugation on $\mathcal{H}$ as defined in Section \ref{HF pre}, i.e. $Jf=f^{c}$. Notice that the anti-linear conjugation of an operator $T$ on $\mathcal{H}$ is exactly $T^{c}=JTJ$. Also let
\begin{equation}\label{convolution 2x2}
 \mathcal{J}=\begin{pmatrix}
      0 & J  \\
      J & 0
    \end{pmatrix},
\end{equation}
such that $\mathcal{J}^2=\mathcal{I}$ on $\mathcal{H}\oplus\mathcal{H}$. Then it is easy to verify for $M$ given in Proposition \ref{1to1correspondencequasifree1-pdm},
\begin{equation}\label{conjuagte with J1}
  \mathcal{J}\big(M-1/2\big)\mathcal{J}=\big(1/2-M\big).
\end{equation}
If we let $A=i\big(M-1/2\big)$, then we have $A^*=-A$ and $\mathcal{J}A\mathcal{J}=A$. Notice that these imply $A$ can be treated as a real skew-adjoint operator on the real space
\begin{equation}\label{real subspace H}
  \big(\mathcal{H}\oplus\mathcal{H}\big)_{\mathrm{R}}\coloneqq \big\{(f,g)\in\mathcal{H}\oplus\mathcal{H}\,\vert\, \mathcal{J}(f,g)=(f,g)\big\}.
\end{equation}
For simplicity, we let
\begin{equation}\label{simpnotationappendixA}
  A_0=\begin{pmatrix}
      -i/2 & 0  \\
      0& i/2
    \end{pmatrix},\quad
   B=\begin{pmatrix}
      \gamma & \alpha  \\
      \alpha^* & -\bar{\gamma}
    \end{pmatrix}
\end{equation}
By the condition $\tr_{\mathcal{H}}\gamma<\infty$, we know $\tr B<\infty$ and hence $B$ is compact. Since $A=A_0+iB$, A is a real skew-adjoint Fredholm operator.

\par For $t\in[0,1]$, let $A_t=A_0+t(A-A_0)$. It is straight-forward to check the path $A_t$ connects $A_0$ and $A$ and $A_t$ is a real skew-adjoint Fredholm operator for any $t$. From the index theorem for skew-symmetry operator by Atiyah and Singer in \cite{atyitah-singermod2},
\begin{equation}\label{indextheroymod2}
  \dim_{\mathbb{C}}\ker A\equiv\dim_{\mathbb{C}}\ker A_0\equiv 0\mod 2.
\end{equation}
Then (\ref{assumption1to1}) follows immediately.
\end{proof}

%\par We can also say a density matrix is quasi-free, if its induced state (\ref{induced state}) is quasi-free. On the other hand, by the proof of \cite[Theorem 2.3]{GenH_F} and Proposition \ref{1to1correspondencequasifree1-pdm}, we know that for a quasi-free state $g$ with finite particle number $g(\mathcal{N})<\infty$, there exists a density matrix $\Gamma_g$, such that for any $A\in\mathcal{L}(\mathcal{F})$:

%Therefore, we can say $\Gamma_g$ is quasi-free. Moreover, by (\ref{induced density matrix}), $\Gamma_g$ is uniquely determined by $g$.
\vspace{1em}

\par As in \cite{GenH_F}, the Hartree-Fock pressure functional of the Hamilton $H$ is defined over all quasi-free state $g$ with finite particle number:
\begin{equation}\label{HF pressure}
  -L^3P_{\mathrm{HF}}^L[g]\coloneqq \mathcal{E}_{\mathrm{HF}}[g]-\frac{1}{\beta}S_{\mathrm{HF}}[g],
\end{equation}
where
\begin{equation}\label{HF pressure energy and entropy}
  \begin{aligned}
  \mathcal{E}_{\mathrm{HF}}[g]\coloneqq g(H-\mu\mathcal{N}),\quad
  S_{\mathrm{HF}}[g]=-\tr_{\mathcal{H}\oplus\mathcal{H}}M\ln M.
  \end{aligned}
\end{equation}
Notice that by (\ref{2times2}) and (\ref{property of gamma and alpha}) $UMU^*=1-M^c$ with
\begin{equation}\label{Unitary U}
  U=\begin{pmatrix}
      0 & \mathbf{I}_{\mathcal{H}} \\
      \mathbf{I}_{\mathcal{H}} & 0
    \end{pmatrix},
\end{equation}
Hence
\begin{equation}\label{entropy equivalent definition}
   S_{\mathrm{HF}}[g]=-\frac{1}{2}\tr_{\mathcal{H}\oplus\mathcal{H}}\big\{M\ln M
   +(1-M)\ln(1-M)\big\}.
\end{equation}
From (\ref{induced density matrix}) in Proposition \ref{1to1correspondencequasifree1-pdm}, we can rewrite
\begin{equation}\label{re HF energy}
  \mathcal{E}_{\mathrm{HF}}[g]=\tr_\mathcal{F}(\mathcal{H}-\mu\mathcal{N})\Gamma_g.
\end{equation}
Also, by the construction of $\Gamma_g$ in \cite[Theorem 2.3]{GenH_F}, it is straight-forward to check that
\begin{equation}\label{re HF entropy}
  S_{\mathrm{HF}}[g]=S_{Q}[\Gamma_g].
\end{equation}
Therefore, by definition (\ref{pressure functional interacting}), we know that
\begin{equation}\label{pressure equivalent}
  P^L_{qtm}[\Gamma_g]=P_{\mathrm{HF}}^L[g].
\end{equation}

\par By Wick's theorem (\ref{Wick Thm k=2}) and the second quantization of $H$ in (\ref{second quantization H_N}), we can expand $g(\mathcal{H}-\mu\mathcal{N})$ by
\begin{equation}\label{expansion g(K-mu N)}
  g(\mathcal{K}-\mu\mathcal{N})=\sum_{k,\sigma}\big(Q(ik)
  -\mu\big)\gamma_{\mathbf{k}\mathbf{k}},
\end{equation}
and
\begin{equation}\label{expasion g(V)}
  \begin{aligned}
  g(\mathcal{V})=\frac{1}{2L^3}\sum_{\substack{k,p,q\\\sigma,\nu}}\hat{v}_k\big(
  \overline{\alpha}_{\mathbf{q+k},\mathbf{p-k}}\alpha_{\mathbf{q}\mathbf{p}}-
  \gamma_{\mathbf{p-k},\mathbf{q}}\gamma_{\mathbf{q+k},\mathbf{p}}+
  \gamma_{\mathbf{p-k},\mathbf{p}}\gamma_{\mathbf{q+k},\mathbf{q}}\big).
  \end{aligned}
\end{equation}
Plugging (\ref{define alpha gamma function}) into (\ref{expasion g(V)}), we get to
\begin{equation}\label{H-F-B functional}
  \begin{aligned}
  g(\mathcal{V})=\frac{1}{2}\int_{(\Lambda_L\times\mathcal{S}_{2})^2}v^L(x-y)
  \Big\{&
  \vert\alpha(-x,\sigma;y,\nu)\vert^2+\vert\gamma(x,\sigma;y,\nu)\vert^2\\
  &-\gamma(x,\sigma;x,\sigma)\gamma(y,\nu;y,\nu)\Big\}dxdyd\sigma d\nu.
  \end{aligned}
\end{equation}
Therefore, the Hartree-Fock pressure can be rewritten by
\begin{equation}\label{HF pressure rewritten general}
  \begin{aligned}
  -P_{\mathrm{HF}}^{L}[g]=\frac{1}{L^3}\sum_{k,\sigma}\big(Q(ik)
  -\mu\big)\gamma_{\mathbf{k}\mathbf{k}}+\frac{1}{L^3}g(\mathcal{V})
  -\frac{1}{\beta L^3}S_{\mathrm{HF}}[g]
  \end{aligned}
\end{equation}
where $g(\mathcal{V})$ is given in (\ref{H-F-B functional}).

\par The second and the third terms on the right-hand side of (\ref{H-F-B functional}) is called the exchange term and the direct term respectively. If we assume the absence of the exchange and the direct terms (this is also assumed and explained in many literature such as \cite{Leggett1980DiatomicMA}), we reach the generalized BCS type pressure given by
\begin{equation}\label{generalized BCS sec}
  \begin{aligned}
  -P^{L}[g]=&\frac{1}{L^3}\sum_{k,\sigma}\big(Q(ik)
  -\mu\big)\gamma_{\mathbf{k}\mathbf{k}}
  -\frac{1}{\beta L^3}S_{\mathrm{HF}}[g]\\
  &+\frac{1}{2L^3}
  \int_{(\Lambda_L\times\mathcal{S}_{2})^2}v^L(x-y)
  \vert\alpha(-x,\sigma;y,\nu)\vert^2.
  \end{aligned}
\end{equation}
Notice that $-P^L$ is real-valued. The first and the second terms are obviously real. For the third term, notice that
\begin{equation}\label{real-valued}
  \frac{1}{2L^3}
  \int_{(\Lambda_L\times\mathcal{S}_{2})^2}v^L(x-y)
  \vert\alpha(-x,\sigma;y,\nu)\vert^2=\frac{1}{2L^3}\sum_{\substack{k,p,q\\\sigma,\nu}}
  \hat{v}(k)
  \overline{\alpha}_{\mathbf{q+k},\mathbf{p-k}}\alpha_{\mathbf{q}\mathbf{p}}.
\end{equation}
One can check the right-hand side of (\ref{real-valued}) is in fact real, using the change of variables $\tilde{k}=-k,\tilde{p}=p-k,\tilde{q}=q+k$ and the fact that $\hat{v}\leq0$ and $\hat{v}(-k)=\hat{v}(k)$.

\par The classical BCS pressure (on the torus) is derived by further assuming (\ref{assumption}), i.e.
\begin{equation}\label{assumption sec}
    \begin{aligned}
    &\gamma_{\mathbf{k}\mathbf{m}}=0,\quad \mathbf{k}\neq\mathbf{m},\\
    &\alpha_{\mathbf{k}\mathbf{m}}=0,\quad \mathbf{k}\neq-\mathbf{m}.
    \end{aligned}
\end{equation}
Recall that we let $\gamma_{\mathbf{k}}=\gamma_{\mathbf{k}\mathbf{k}}$, $\alpha_{\mathbf{k}}=\alpha_{-\mathbf{k},\mathbf{k}}$. Plugging (\ref{rewritten alpha gamma function}) into (\ref{H-F-B functional}), we have
\begin{equation}\label{H-F-B functional assumption}
  g(\mathcal{V})=\frac{1}{2}\int_{\Lambda_L\times\mathcal{S}_{2}}
  v^L(x)\Big\{\vert\alpha(x,\sigma)\vert^2-
  \vert\gamma(x,\sigma)\vert^2\Big\}dxd\sigma+\frac{\hat{v}(0)}{2}\Big(
  \sum_{\sigma}\gamma(0,\sigma)\Big)^2.
\end{equation}
Similarly we drop the exchange term and the direct term in (\ref{H-F-B functional assumption}), we reach the classical BCS pressure functional:
\begin{equation}\label{Pressure BCS sec}
\begin{aligned}
  -L^3P_{\mathrm{BCS}}^L[g]= &\sum_{k,\sigma}\big(Q(ik)-\mu\big)\gamma_{\mathbf{k}}
  +\frac{1}{\beta}\sum_{\mathbf{k}}\tr_{\mathbb{C}^2}M(\mathbf{k})\ln M(\mathbf{k})\\
  &+\frac{1}{2}\int_{\Lambda_L\times\mathcal{S}_{2}}
  v^L(x)\vert\alpha(x,\sigma)\vert^2dxd\sigma.
\end{aligned}
\end{equation}

\section{Well-posedness of Discrete BCS Gap Equation}\label{provemain1}

\par We first proof Theorem \ref{main1} and Corollary \ref{uniqueness maximizer intro} for $0<L<\infty$, the $L=\infty$ case will be explained at the end of Section \ref{corollaries}. For fixed $0<L<\infty$, the discrete version of BCS gap equation:
\begin{equation}\label{Gap eqn again start}
  \Delta_{k}=-\frac{1}{2L^3}\sum_{k^\prime}\hat{v}(k-k^\prime)
    \frac{\Delta_{k^\prime}}{E_{k^\prime}}\tanh\frac{\beta}{2}E_{k^\prime},
\end{equation}
where
\begin{equation}\label{define E_k start}
  E_k\coloneqq\sqrt{\big(Q(ik)-\mu\big)^2+\vert\Delta_k\vert^2},
\end{equation}
and $k\in(2\pi/L)\mathbb{Z}^3$. Notice that for any $\Delta$ satisfies (\ref{Gap eqn again start}) and any phase $\theta\in[0,2\pi)$, any phase change $e^{i\theta}\Delta$ also satisfies (\ref{Gap eqn again start}).

\par Recall that, $\Delta\equiv 0$ is a solution to (\ref{Gap eqn again start}), and since $\hat{v}(0)<0$ and $\hat{v}$ is continuous, there exists some universal constant $r>0$ such that $\hat{v}(k)<0$ for any $\vert k\vert<r$. Therefore, as long as $L$ is large enough, such that $B_r\cap\big((2\pi/L)\mathbb{Z}^3\backslash\{0\}\big)\neq\emptyset$, one can prove by induction that any nonzero nonnegative solution of (\ref{Gap eqn again start}) is strictly positive.

\par Equation (\ref{Gap eqn again start}) also implies for any uniformly bounded solution $\Delta$ must decay at the order of some power of $\vert k\vert^{-1}$, with respect to the degree $d$ of $Q$. Thus we have the following lemma:
\begin{lemma}\label{decay Delta}
  Let uniformly bounded $\Delta$ be a solution to the discrete BCS gap equation (\ref{Gap eqn again start}) for some fixed $L>0$ and temperature $T\geq0$. Recall that $Q$ satisfies Assumption \ref{assumption Q}, $v$ satisfies Assumption \ref{assumption v} and $d$, $\varepsilon$ are given is (\ref{deg Q}) and (\ref{requirement for v 1}) respectively. Then for any $s>0$, there exist constants $C(v,\mu,Q)>0$ and $C(\mu,Q)>2$, such that for $0<d\leq 3$ $\vert k\vert>C(\mu,Q)$,
  \begin{equation}\label{decay Delta eq}
    \vert\Delta_{k}\vert\leq\frac{C(v,\mu,Q)}{\vert k\vert^{3-d+\varepsilon}},
  \end{equation}
  and for $d>3$, $\vert k\vert>C(\mu,Q)$,
    \begin{equation}\label{decay Delta eq2}
    \vert\Delta_{k}\vert\leq C(v,\mu,Q)\max\Big\{\sup_{\vert\xi\vert>\vert k\vert/2}\hat{v}(\xi),\vert k\vert^{3-d}\Big\}.
  \end{equation}
  In particular, $C(v,\mu,Q)$ is independent of $L$.
\end{lemma}
\begin{proof}
  \par The proof is a combination of standard decay estimate of convolution and the bootstrap argument. If $0<d\leq3$, from (\ref{requirement for v 1}), we have
  \begin{equation}\label{decay requirement for v 1}
  \vert\hat{v}(k)\vert\leq \frac{C(v)}{1+\vert k\vert^{3-d+\varepsilon}},
\end{equation}
for some $\varepsilon>0$. Also, since $\Delta$ is uniformly bounded, by Assumption (\ref{assumption Q}), we have for $\vert k\vert>C(\mu,Q)>2$,
\begin{equation}\label{decat tanh}
  \Big\vert\frac{\Delta_{k}}{E_{k}}\tanh\frac{\beta}{2}E_{k}\Big\vert\leq
  \frac{C(\mu,Q)}{1+\vert k\vert^{d}}.
\end{equation}
\par For $\vert k\vert>C(\mu,Q)$, we divide the sum on the right-hand side of (\ref{Gap eqn again start}) into six parts, and estimate them respectively:
\begin{flushleft}
  \textbf{Case 1: $\vert k^\prime\vert\geq2\vert k\vert$.}
\end{flushleft}
In this case we have $\vert k-k^\prime\vert\geq\vert k^\prime\vert/2$. Therefore
\begin{equation}\label{case 1 lemma 2.1}
  \begin{aligned}
  \Big\vert\frac{1}{2L^3}\sum_{\text{Case 1}}\hat{v}(k-k^\prime)
    \frac{\Delta_{k^\prime}}{E_{k^\prime}}\tanh\frac{\beta}{2}E_{k^\prime}\Big\vert\leq&
    C(v,\mu,Q)\sum_{\vert k^\prime\vert\geq2\vert k\vert}\frac{1}{\vert k^\prime\vert^{3+\varepsilon}}\Big(\frac{2\pi}{L}\Big)^3\\
    \leq&\frac{C(v,\mu,Q)}{\vert k\vert^\varepsilon}.
  \end{aligned}
\end{equation}

\begin{flushleft}
  \textbf{Case 2: $\vert k-k^\prime\vert\geq2\vert k\vert$, $\vert k^\prime\vert<2\vert k\vert$.}
\end{flushleft}
In this case we have $\vert k^\prime\vert\geq\vert k-k^\prime\vert/2$. Therefore
\begin{equation}\label{case 2 lemma 2.1}
  \begin{aligned}
  \Big\vert\frac{1}{2L^3}\sum_{\text{Case 2}}\hat{v}(k-k^\prime)
    \frac{\Delta_{k^\prime}}{E_{k^\prime}}\tanh\frac{\beta}{2}E_{k^\prime}\Big\vert\leq&
    C(v,\mu,Q)\sum_{\vert k-k^\prime\vert\geq2\vert k\vert}\frac{1}{\vert k-k^\prime\vert^{3+\varepsilon}}\Big(\frac{2\pi}{L}\Big)^3\\
    \leq&\frac{C(v,\mu,Q)}{\vert k\vert^\varepsilon}.
  \end{aligned}
\end{equation}

\begin{flushleft}
  \textbf{Case 3: $\vert k\vert/2\leq\vert k-k^\prime\vert<2\vert k\vert$, $\vert k\vert/2\leq\vert k^\prime\vert<2\vert k\vert$.}
\end{flushleft}
In this case we have
\begin{equation}\label{case 3 lemma 2.1}
  \begin{aligned}
  \Big\vert\frac{1}{2L^3}\sum_{\text{Case 3}}\hat{v}(k-k^\prime)
    \frac{\Delta_{k^\prime}}{E_{k^\prime}}\tanh\frac{\beta}{2}E_{k^\prime}\Big\vert\leq&
    C(v,\mu,Q)\sum_{\vert k^\prime\vert<2\vert k\vert}\frac{1}{\vert k\vert^{3+\varepsilon}}\Big(\frac{2\pi}{L}\Big)^3\\
    \leq&\frac{C(v,\mu,Q)}{\vert k\vert^\varepsilon}.
  \end{aligned}
\end{equation}

\begin{flushleft}
  \textbf{Case 4: $\vert k\vert/2\leq\vert k-k^\prime\vert<2\vert k\vert$, $\vert k^\prime\vert<\vert k\vert/2$.}
\end{flushleft}
In this case we have $\vert k-k^\prime\vert\geq\vert k\vert/2$. Therefore, we have
\begin{equation}\label{case 4 lemma 2.1}
  \begin{aligned}
  &\Big\vert\frac{1}{2L^3}\sum_{\text{Case 4}}\hat{v}(k-k^\prime)
    \frac{\Delta_{k^\prime}}{E_{k^\prime}}\tanh\frac{\beta}{2}E_{k^\prime}\Big\vert\\
    \leq&
    \frac{C(v,\mu,Q)}{\vert k\vert^{3-d+\varepsilon}}
    \Big\{\sum_{\vert k^\prime\vert<1}\Big(\frac{2\pi}{L}\Big)^3+
    \sum_{1\leq\vert k^\prime\vert<\vert k\vert/2}\frac{1}{\vert k^\prime\vert^{d}}\Big(\frac{2\pi}{L}\Big)^3\Big\}\\
    \leq&\frac{C(v,\mu,Q)}{\vert k\vert^\varepsilon}.
  \end{aligned}
\end{equation}

\begin{flushleft}
  \textbf{Case 5: $\vert k-k^\prime\vert<\vert k\vert/2$, $\vert k\vert/2\leq\vert k^\prime\vert<2\vert k\vert$.}
\end{flushleft}
In this case we have $\vert k^\prime\vert\geq\vert k\vert/2$. Therefore, we have
\begin{equation}\label{case 5 lemma 2.1}
  \begin{aligned}
  &\Big\vert\frac{1}{2L^3}\sum_{\text{Case 5}}\hat{v}(k-k^\prime)
    \frac{\Delta_{k^\prime}}{E_{k^\prime}}\tanh\frac{\beta}{2}E_{k^\prime}\Big\vert\\
    \leq&
    \frac{C(v,\mu,Q)}{\vert k\vert^{d}}
    \Big\{\sum_{\vert k-k^\prime\vert<1}\Big(\frac{2\pi}{L}\Big)^3+
    \sum_{1\leq\vert k-k^\prime\vert<\vert k\vert/2}\frac{1}{\vert k^\prime\vert^{3-d+\varepsilon}}\Big(\frac{2\pi}{L}\Big)^3\Big\}\\
    \leq&\frac{C(v,\mu,Q)}{\vert k\vert^d}\Big(1+\vert k\vert^{d-\varepsilon}\Big)
    \leq\frac{C(v,\mu,Q)}{\vert k\vert^s},
  \end{aligned}
\end{equation}
where $s=\min\{d,\varepsilon\}$.

\begin{flushleft}
  \textbf{Case 6: $\vert k-k^\prime\vert<\vert k\vert/2$, $\vert k^\prime\vert<\vert k\vert/2$.}
\end{flushleft}
This case is illegal since in case 6,
\begin{equation*}
  \vert k-k^\prime\vert+\vert k^\prime\vert<\vert k\vert\leq
  \vert k-k^\prime\vert+\vert k^\prime\vert.
\end{equation*}

\par Combining (\ref{case 1 lemma 2.1})-(\ref{case 5 lemma 2.1}), we have for $\vert k\vert>C(\mu,Q)$ and $0<d\leq3$,

\begin{equation}\label{decay Delta 1}
  \vert\Delta_{k}\vert\leq\frac{C(v,\mu,Q)}{\vert k\vert^{s}},
\end{equation}
where $s=\min\{d,\varepsilon\}$. Plugging (\ref{decay Delta 1}) back into (\ref{decat tanh}) we have for $\vert k\vert>C(\mu,Q)$,
\begin{equation}\label{decat tanh2}
  \Big\vert\frac{\Delta_{k}}{E_{k}}\tanh\frac{\beta}{2}E_{k}\Big\vert\leq
  \frac{C(v,\mu,L)}{1+\vert k\vert^{d+s}}.
\end{equation}
Then a bootstrap argument yields the (\ref{decay Delta eq}). We remark that the bootstrap argument will stop after finite steps, since at $n-th$ step, (\ref{case 4 lemma 2.1}) becomes
\begin{equation}\label{case 4 lemma 2.1 bootstrap}
  \begin{aligned}
  &\Big\vert\frac{1}{2L^3}\sum_{\text{Case 4}}\hat{v}(k-k^\prime)
    \frac{\Delta_{k^\prime}}{E_{k^\prime}}\tanh\frac{\beta}{2}E_{k^\prime}\Big\vert\\
    \leq&
    \frac{C(v,\mu,Q)}{\vert k\vert^{3-d+\varepsilon}}
    \Big\{\sum_{\vert k^\prime\vert<1}\Big(\frac{2\pi}{L}\Big)^3+
    \sum_{1\leq\vert k^\prime\vert<\vert k\vert/2}\frac{1}{\vert k^\prime\vert^{d+(n-1)s}}\Big(\frac{2\pi}{L}\Big)^3\Big\}\\
    \leq&\frac{C(v,\mu,Q)}{\vert k\vert^{3-d+\varepsilon}}\Big(1+\vert k\vert^
    {3-d-(n-1)s}\Big),
  \end{aligned}
\end{equation}
and (\ref{case 4 lemma 2.1 bootstrap}) is bounded by $\vert k\vert^{d-3-\varepsilon}$ when $n$ is large enough so that $3-d-(n-1)s<0$.

\par For $d>3$, since $v\in L^1$, obviously
\begin{equation}\label{decay Delta 3}
  \vert \hat{v}(k)\vert\leq C(v)
\end{equation}
On the other hand, we still have (\ref{decat tanh}). Then we can get to (\ref{decay Delta eq2}) via a similar argument (without bootstrap).
\end{proof}

\subsection{Existence of maximizer without (\ref{assumption})}
\
\begin{proposition}\label{existenceofminimizer}
  For fixed $L>0$, $P^L$ defined in (\ref{generalized BCS}) has at least one maximizer in its domain.
\end{proposition}
\begin{proof}
  \par Notice that, we do not have (\ref{assumption}) nor the simplification it provides here, so we have to deal with the entropy $S_{\mathrm{HF}}[g]$ to its full extend.

  \par First notice that the quantum thermodymamic pressure for an non-interacting system is apparently bounded below, and its Gibbs state is quasi-free (see \cite{wickformula}). Hence
  \begin{equation}\label{noninteracting bounded below}
   C_1(Q,L)\coloneqq \inf_{\text
   {$g$ quasi-free}}\Big\{\frac{1}{4}\sum_{k,\sigma}\big(
   Q(ik)-\mu\big)\gamma_{\mathbf{k}\mathbf{k}}-\frac{1}{\beta}S_{\mathrm{HF}}[g]\Big\}
   >-\infty.
  \end{equation}
  Therefore for any quasi-free state $g$ with finite particle number,
  \begin{equation}\label{generalizedBCS bounded below1}
  \begin{aligned}
    -L^3P^L[g]\geq& C_1(L)+\frac{3}{4}\sum_{k,\sigma}\big(
   Q(ik)-\mu\big)\gamma_{\mathbf{k}\mathbf{k}}\\
   &+\frac{1}{2}\int_{(\Lambda_L\times\mathcal{S}_
   {2})^2}v^L(x-y)\vert\alpha(-x,\sigma;y,\nu)\vert^2.
  \end{aligned}
  \end{equation}
  For $0<d<3$, since $v^L\in L^{3/d}(\Lambda_L)$ according to Proposition \ref{Lp v^L}, by Sobolev embedding $H^{d/2}(\Lambda_L)\hookrightarrow L^{6/(3-d)}(\Lambda_L)$, together with (\ref{deg Q}), we can deduce (see for example \cite[Chapter 11]{lieb2001analysis}) for any $\varphi\in H^{d/2}(\Lambda_L)$ and $y\in\Lambda_L$,
  \begin{equation}\label{Sobolev imbedding}
    \int_{\Lambda_L}\Big\{\frac{1}{4}\vert Q(ik)^{1/2}\varphi\vert^2
    +\frac{1}{2} v^L(x-y)\vert\varphi\vert^2\Big\}dx\geq
    -C_2(Q,L)\int_{\Lambda_L}\vert\varphi\vert^2dx.
  \end{equation}
  For $d>3$, since $v^L\in L^{1}(\Lambda_L)$ according to Proposition \ref{Lp v^L}, and by Sobolev embedding, we have $H^{d/2}(\Lambda_L)\hookrightarrow L^{\infty}(\Lambda_L)$, and for $d=3$, we have $H^{d^\prime/2}(\Lambda_L)\hookrightarrow L^{6/(3-d^\prime)}(\Lambda_L)$ where $d^\prime$ is given in (\ref{requirement for v 3}). Therefore, we can also deduce (\ref{Sobolev imbedding}) for $d\geq3$.

  \par Also, notice that the generalized 1-pdm of $g$ satisfies $0\leq M\leq 1$, which implies $M(1-M)\geq0$ and hence $\alpha\alpha^*\leq \gamma-\gamma^2\leq\gamma\leq1$. Using (\ref{gamma and alpha}), we can rewrite this inequality by
  \begin{equation}\label{inequality M(1-M)=0}
    \sum_{\mathbf{m}}\vert\alpha_{\mathbf{k}\mathbf{m}}\vert^2\leq
    \gamma_{\mathbf{k}\mathbf{k}}-\sum_{\mathbf{m}}\vert\gamma_{\mathbf{k}\mathbf{m}}\vert^2\leq
    \gamma_{\mathbf{k}\mathbf{k}}\leq1.
  \end{equation}
  Applying (\ref{deg Q}), (\ref{Sobolev imbedding}) and (\ref{inequality M(1-M)=0} to (\ref{generalizedBCS bounded below1}), we reach
  \begin{equation}\label{generalizedBCS bounded below2}
    -L^3P^L[g]\geq C_1(L)+\frac{1}{4}\sum_{\mathbf{k}}\big(1+\vert k\vert^d\big)\gamma_{\mathbf{k}\mathbf{k}}
    +\sum_{\mathbf{k}}\Big(\frac{1}{4}\vert k\vert^d-\frac{1}{4}-\frac{3}{4}\mu-C_2(L)\Big)
    \gamma_{\mathbf{k}\mathbf{k}}.
  \end{equation}
  Since $0\leq\gamma_{\mathbf{k}\mathbf{k}}\leq1$, the last term on the right-hand side of (\ref{generalizedBCS bounded below2}) is bounded below, hence
  \begin{equation}\label{generalizedBCS bounded below3}
    -L^3P^L[g]\geq -C_3(Q,L)+\frac{1}{4}\sum_{\mathbf{k}}\big(1+\vert k\vert^d\big)
    \gamma_{\mathbf{k}\mathbf{k}}.
  \end{equation}
   Therefore, $-L^3P^L[g]$ is bounded from below. Notice that by inequality (\ref{generalizedBCS bounded below3}), we can w.l.g. restrict the domain of $\gamma$ on
    \begin{equation}\label{condition gamma}
      \sum_{\mathbf{k}}\big(1+\vert k\vert^d\big)\gamma_{\mathbf{k}\mathbf{k}}<\infty.
    \end{equation}
    Also notice that by (\ref{inequality M(1-M)=0}) and (\ref{property of gammakm and alphakm}),
  \begin{equation}\label{generalizedBCS bounded below4}
    \Vert\gamma\Vert^2_{H^{d/2}_{\mathbf{x}}L^2_{\mathbf{y}}}+
    \Vert\alpha\Vert^2_{H^{d/2}_{\mathbf{x}}L^2_{\mathbf{y}}}
    \leq C_4\sum_{\mathbf{k}}(1+k^2)\gamma_{\mathbf{k}\mathbf{k}}.
  \end{equation}
Viewing $\mathbf{y}\in \Lambda_L\times\mathcal{2}$ as a parameter, i.e. taking
\begin{equation}\label{parameter y}
  \begin{aligned}
  &\gamma_{\mathbf{y}}(\mathbf{x})=\gamma(\mathbf{x},\mathbf{y}),\\
  &\alpha_{\mathbf{y}}(\mathbf{x})=\alpha(\mathbf{x},\mathbf{y}),
  \end{aligned}
\end{equation}
for $\gamma,\alpha$ defined in (\ref{define alpha gamma function}), then under (\ref{condition gamma}) we have $\gamma_{\mathbf{y}},\alpha_{\mathbf{y}}$ are in $H^1$. From (\ref{generalizedBCS bounded below3}), we can find a minimizing sequence $(\gamma^{(n)},\alpha^{(n)})$ such that
\begin{equation}\label{uniform bound gamma alpha}
  \Vert\gamma^{(n)}\Vert^2_{H^{d/2}_{\mathbf{x}}L^2_{\mathbf{y}}}+
    \Vert\alpha^{(n)}\Vert^2_{H^{d/2}_{\mathbf{x}}L^2_{\mathbf{y}}}
    \leq C_4\sum_{\mathbf{k}}\big(1+\vert k\vert^d\big)\gamma^{(n)}_{\mathbf{k}\mathbf{k}}\leq C_5(Q,L).
\end{equation}
We are going to show that, up to a subsequence,
\begin{align}
  \lim_{n\to\infty}\int_{(\Lambda_L\times\mathcal{S}_{2})^2}
  v^L(x-y)\vert\alpha^{(n)}(-x,\sigma;y,\nu)\vert^2=&
  \int_{(\Lambda_L\times\mathcal{S}_{2})^2}
  v^L(x-y)\vert\tilde{\alpha}(-x,\sigma;y,\nu)\vert^2,\label{limit l.s.m1}\\
  \liminf_{n\to\infty}\sum_{\mathbf{k}}\big(Q(ik)-\mu\big)\gamma^{(n)}_{\mathbf{k}\mathbf{k}}
  -\frac{1}{\beta}S_{\mathrm{HF}}[g^{(n)}]\geq&
  \sum_{\mathbf{k}}\big(Q(ik)-\mu\big)\tilde{\gamma}_{\mathbf{k}\mathbf{k}}
  -\frac{1}{\beta}S_{\mathrm{HF}}[\tilde{g}],\label{limit l.s.m2}
\end{align}
for some $(\tilde{\gamma},\tilde{\alpha})$ satisfy (\ref{2times2}), (\ref{property of gamma and alpha}) and (\ref{condition gamma}), and hence, by Proposition \ref{1to1correspondencequasifree1-pdm}, induces the expected maximizer $\tilde{g}$ of $P^L$.

  \par For (\ref{limit l.s.m1}), since $\alpha^{(n)}$ is uniformly bounded in $H^{d/2}_{\mathbf{x}}L^2_{\mathbf{y}}$ (for fixed $L$), then up to a subsequence they converge weakly in $H^{d/2}_{\mathbf{x}}L^2_{\mathbf{y}}$ to some $\tilde{\alpha}$. For $0<d<3$, $H^{d/2}_{\mathbf{x}}L^2_{\mathbf{y}}\hookrightarrow L^2_{\mathbf{y}}L^{6/(3-d)}_{\mathbf{x}}$, by (\ref{uniform bound gamma alpha}) and property of weak convergence, $\alpha^{(n)}$ and $\tilde{\alpha}$ are uniformly bounded in $L^2_{\mathbf{y}}L^{6/(3-d)}_{\mathbf{x}}$. Also from the compact Sobolev embedding $H^{d/2}_{\mathbf{x}}L^2_{\mathbf{y}}\Subset L^2_{\mathbf{y}}L^{p}_{\mathbf{x}}$ for $1\leq p< 6/(3-d)$, we have $\alpha^{(n)}\to\tilde{\alpha}$ in $L^2_{\mathbf{y}}L^{p}_{\mathbf{x}}$. On the other hand, we take for $B>0$,
  \begin{equation}\label{partition of v^L}
    v^L_B=\min\{\vert v^L\vert,B\}\sgn v^L,
  \end{equation}
  such that $v^L_B\in L^{\infty}$ and $v^L_B\to v^L\, a.e.$ as $B\to\infty$. Since $v^L\in L^{3/d}$ by Proposition \ref{Lp v^L}, we have
  \begin{equation}\label{est l.s.c 1}
  \begin{aligned}
    &\Big\vert\int_{(\Lambda_L\times\mathcal{S}_{2})^2}
    \big(v^L(x-y)-v^L_B(x-y)\big)\big\vert\alpha^{(n)}(-x,\sigma;y,\nu)\big\vert^2\Big\vert\\
    \leq&
    C_6(L)\Vert v^L-v^L_B\Vert_{L^{3/2}}\sup_n\Vert\alpha^{(n)}\Vert^2
    _{L^2_{\mathbf{y}}L^{6/(3-d)}_{\mathbf{x}}}\\
    \leq& C_7(L)\Vert v^L-v^L_B\Vert_{L^{3/2}}
  \end{aligned}
  \end{equation}
  For fixed $L$, the last term tends to $0$ as $B\to\infty$, independent of $n$, by dominated convergence theorem. The same estimate also holds if one replaces $\alpha^{(n)}$ by $\tilde{\alpha}$ in (\ref{est l.s.c 1}), therefore,
  \begin{equation}\label{est l.s.c 2}
    \lim_{B\to\infty}\sup_{n}\int_{(\Lambda_L\times\mathcal{S}_{2})^2}
    \big\vert v^L(x-y)-v^L_B(x-y)\big\vert
    \big(\vert\alpha^{(n)}\vert^2+\vert\tilde{\alpha}\vert^2\big)
    =0.
  \end{equation}
   On the other hand, notice that
   \begin{equation}\label{est l.s.c 3}
     \int_{\Lambda_L\times\mathcal{S}_{2}}
     \Big\vert\vert\alpha^{(n)}_{\mathbf{y}}\vert^2-\vert\tilde{\alpha}_\mathbf{y}
     \vert^2\Big\vert^kd\mathbf{x}\leq
     \Vert \alpha^{(n)}_\mathbf{y}-\tilde{\alpha}_\mathbf{y}\Vert_{L^{2k}_\mathbf{x}}^k\Vert \alpha^{(n)}_\mathbf{y}+\tilde{\alpha}_\mathbf{y}\Vert_{L^{2k}_\mathbf{x}}^k.
   \end{equation}
   Therefore, for $1\leq k\leq 3$,
   \begin{equation}\label{est l.s.c 4}
   \begin{aligned}
    \Big\vert \int v_B^L(x-y)\big(\vert\alpha^{(n)}\vert^2-\vert\tilde{\alpha}\vert^2\big)
    \Big\vert&\leq C_8(L,B)\Vert \alpha^{(n)}-\tilde{\alpha}\Vert
    _{L^1_{\mathbf{y}}L^{2k}_{\mathbf{x}}}\\
    &\leq C_9(L,B)\Vert \alpha^{(n)}-\tilde{\alpha}\Vert
    _{L^2_{\mathbf{y}}L^{2k}_{\mathbf{x}}}
   \end{aligned}
   \end{equation}
   As long as we choose $1\leq k< 3/(3-d)$, the right-hide side of (\ref{est l.s.c 4}) will converge to $0$ as $n\to\infty$ as $B$ fixed. Therefore, combining (\ref{est l.s.c 2}) and (\ref{est l.s.c 4}) we reach (\ref{limit l.s.m1}) for $0<d<3$.

   \par For $d=3$ and $d>3$, the proofs are similar once we make use of compact Sobolev embedding:
   \begin{equation}\label{Sobolev embedding 2}
     \begin{aligned}
     &H^{3/2}_{\mathbf{x}}L^2_{\mathbf{y}}\Subset L^2_{\mathbf{y}}L^p_{\mathbf{x}},\quad
     \text{for all $1\leq p<\infty$ when $d=3$,}\\
     &H^{d/2}_{\mathbf{x}}L^2_{\mathbf{y}}\Subset L^2_{\mathbf{y}}L^{\infty}_{\mathbf{x}}
     ,\quad\text{when $d>3$}.
     \end{aligned}
   \end{equation}

   \par For (\ref{limit l.s.m2}), notice that we do not have the translation and $SU(2)$ invariance (\ref{assumption}) in the first place, and we have to deal with the implicity of the entropy $S_{\mathrm{HF}}[g]$. First, by (\ref{uniform bound gamma alpha}), we have, up to a subsequence, $\alpha^{(n)}\to\tilde{\alpha}$, $\gamma^{(n)}\to\tilde{\gamma}$ in $L^2_{\mathbf{x},\mathbf{y}}$ when $n\to\infty$. This implies $\alpha^{(n)}$ and $\gamma^{(n)}$, regarded as operators on $L^2\big(\Lambda_L\times\mathcal{S}_{2}\big)$, converge to $\tilde{\alpha}$ and $\tilde{\gamma}$ respectively in Hilbert-Schmidt norm. Hence it is straight-forward to verify $(\tilde{\gamma},\tilde{\alpha})$ satisfies (\ref{2times2}) and (\ref{property of gamma and alpha}). Also the Hilbert-Schmidt norm convergence implies
   \begin{equation}\label{gamma^n_kk to gammatilde_kk}
     \tilde{\gamma}_{\mathbf{k}\mathbf{k}}
     =\langle\tilde{\gamma}f_{\mathbf{k}},f_{\mathbf{k}}\rangle=\lim_{n\to\infty}
     \langle\gamma^{(n)}f_{\mathbf{k}},f_{\mathbf{k}}\rangle
     =\lim_{n\to\infty}\gamma^{(n)}_{\mathbf{k}\mathbf{k}}.
   \end{equation}
   Since $0\leq \gamma^{(n)}\leq1$, by Fatou's lemma, we have
   \begin{equation}\label{gamma L1bound}
     \sum_{\mathbf{k}}(1+k^2)\tilde{\gamma}_{\mathbf{k}\mathbf{k}}\leq C_{5}(Q,L).
   \end{equation}

  \par We then would like to show inequality (\ref{limit l.s.m2}), using the convergence in Hilbert-Schmidt norm. It is equivalent to show
  \begin{equation}\label{define P^L_0}
    -L^3P_0^L[g]\coloneqq \sum_{\mathbf{k}}\big(Q(ik)-\mu\big)\gamma_{\mathbf{k}\mathbf{k}}
  -\frac{1}{\beta}S_{\mathrm{HF}}[g]
  \end{equation}
   is lower semi-continuous in Hilbert-Schmidt norm. From (\ref{induced density matrix})-(\ref{pressure equivalent}), we know that
   \begin{equation}\label{equvalent free gas pressure}
     -L^3P_0^L[g]=-L^3P^L_{qtm,0}[\Gamma_g].
   \end{equation}
    From (\ref{Gibbs state non-inter}), $-L^3P_0^L[g]$ attains its minimum at the Gibbs state $G_0$, and by (\ref{gamma_0}) we know that the generalized 1-pdm of $G_0$ is given by
    \begin{equation}\label{generalized 1-pdm gibbs_0}
      M_0=\begin{pmatrix}
            \big(1+e^{\beta Q(\nabla)-\beta\mu}\big)^{-1} & 0 \\
            0 & \big(1+e^{-\beta Q(\nabla)+\beta\mu}\big)^{-1}
          \end{pmatrix}.
    \end{equation}
    We let $g_0$ denote the corresponding quasi-free state such that $g_0(A)=\tr_{\mathcal{F}}AG_0$. Thus it is equivalent to prove that, the relative entropy
    \begin{equation}\label{relative entropy}
      \frac{1}{\beta} S_{\mathrm{HF}}(M,M_0)\coloneqq -L^3P_{0}^L[g]+L^3P_{0}^L[g_0]
    \end{equation}
   is lower semi-continuous in Hilbert-Schmidt norm of $(\gamma,\alpha)$ (from Proposition \ref{1to1correspondencequasifree1-pdm} we know admissible $(\gamma,\alpha)$ 1-to-1 corresponds to $\Gamma_g$ and hence quasi-free state $g$). Notice that
   \begin{equation}\label{relative entropy0}
     \begin{aligned}
     \sum_{\mathbf{k}}\big(Q(ik)-\mu\big)\gamma_{\mathbf{k}\mathbf{k}}=\tr_{\mathcal{H}}\big(
     Q(\nabla)-\mu\big)\gamma.
     \end{aligned}
   \end{equation}
   Also by (\ref{gamma_0}), we have
   \begin{equation}\label{relative entropy-1}
     \frac{1}{\beta}\ln\frac{1-\gamma_0}{\gamma_0}= Q(\nabla)-\mu,
   \end{equation}
   where $\gamma_0=\big(1+e^{\beta Q(\nabla)-\beta\mu}\big)^{-1}$ is the upper-left block of $M_0$ in (\ref{generalized 1-pdm gibbs_0}). Since the family of eigenvectors of $\gamma_0$ is $\{f_{k,\sigma}\}$, the standard orthonormal basis of $\mathcal{H}$, then from (\ref{conjugate of A}), together with (\ref{relative entropy0}) and (\ref{relative entropy-1}, we can check that
   \begin{equation}\label{relative entropy-2}
     \sum_{\mathbf{k}}\big(Q(ik)-\mu\big)\gamma_{\mathbf{k}\mathbf{k}}=
     \frac{1}{\beta}\big(\tr_{\mathcal{H}}\gamma^c\ln(1-\gamma_0)-\tr_{\mathcal{H}}
     \gamma\ln\gamma_0\big).
   \end{equation}
   Plugging (\ref{entropy equivalent definition}) and (\ref{relative entropy-2}) into (\ref{relative entropy}), we have
   \begin{equation}\label{relative entropy 2}
   \begin{aligned}
     &\frac{1}{\beta} S_{\mathrm{HF}}(M,M_0)\\
     =&\frac{1}{2\beta}\Big\{\tr_{\mathcal{H}\oplus\mathcal{H}}
     M\big(\ln M-\ln M_0\big)+\tr_
     {\mathcal{\mathcal{H}\oplus\mathcal{H}}}(1-M)\big(\ln(1-M)-\ln(1-M_0)\big)\Big\}.
   \end{aligned}
   \end{equation}
   %Let $\{\lambda_i\}$ be the family of eigenvalues of $\Gamma_g$ in decreasing order, and $\{\psi_i\}$ be its corresponding eigenvectors which form an orthonormal basis of $\mathcal{F}$. Accordingly we let $\{\mu_i\}$ be the family of eigenvalues of $G_0$ in decreasing order and the orthonormal basis $\{\phi_i\}$ be its corresponding eigenvectors. Assume also
   %\begin{equation}\label{unitary trans orthonormal basis}
   %  \psi_i=\sum_j a_{ij}\phi_j,
   %\end{equation}
   %then obviously
   %\begin{equation}\label{unitary trans orthonormal basis1}
   %  \sum_{j}\vert a_{ij}\vert^2=1.
   %\end{equation}
   %Now we expand (\ref{relative entropy 2}) using notations above and get to
   Now we introduce the Umegaki-Lindblad type relative entropy from \cite{umegaki1954conditional,lindblad1973entropy}:
   \begin{equation}\label{U_L relative entropy}
     D(M \vert M_0)\coloneqq \tr_{\mathcal{H}\oplus\mathcal{H}}
     M\big(\ln M-\ln M_0\big)-M+M_0\geq0.
   \end{equation}
   Hence,
   \begin{equation}\label{U-L r U}
     S_{\mathrm{HF}}(M,M_0)=\frac{1}{2}\Big\{D(M\vert M_0)+D(1-M\vert1-M_0)\Big\}
   \end{equation}
   The Umegaki-Lindblad type relative entropy is lower semi-continuous in the weak*-topology in trace class space with respect to $M$, i.e. topology generated by point-wise convergence of functional
    \begin{equation}\label{functional M}
      \tr_{\mathcal{H}\oplus\mathcal{H}}AM,\quad A\in\mathcal{C}(\mathcal{H}\oplus\mathcal{H}),
    \end{equation}
    where $\mathcal{C}(\mathcal{H}\oplus\mathcal{H})$ is the compact operator space, see for example \cite[Lemma 4]{LSCrelativeentropy}. On the other hand, \cite[Lemma 4]{Lindblad} states that
   \begin{equation}\label{relative entropy 3}
     D(M \vert M_0)=\sup_{P}D(PMP\vert PM_0P),
   \end{equation}
   where the supremum is taken over all finite dimensional projections $P$. Once we show that $PM^{(n)}P$ tends to $P\tilde{M}P$ in the weak*-topology of $\mathcal{L}_{1}(\mathcal{H}\oplus\mathcal{H})$ for arbitrary finite dimensional projection $P$, then by (\ref{relative entropy 3})
   \begin{equation}\label{relative entropy 4}
   \begin{aligned}
     \liminf_{n\to\infty}D(M^{(n)}\vert M_0)\geq
     \liminf_{n\to\infty}D(PM^{(n)}P\vert PM_0P)
     \geq D(P\tilde{M}P\vert PM_0P),
   \end{aligned}
   \end{equation}
   and we reach the lower semi-continuity of the relative entropy (\ref{relative entropy}) in Hilbert-Schmidt norm of $(\gamma,\alpha)$ by taking supremum over all $P$ on the right-hand side of (\ref{relative entropy 4}).

  \par To show $PM^{(n)}P$ tends to $P\tilde{M}P$ in the weak*-topology of $\mathcal{L}_{1}(\mathcal{H}\oplus\mathcal{H})$, we first notice that $(\gamma^{(n)},\alpha^{(n)})\to(\tilde{\gamma},\tilde{\alpha})$ in Hilbert-Schmidt norm immediately implies $M^{(n)}\to\tilde{M}$ in Hilbert-Schmidt norm. Since $PM^{(n)}P$ and $P\tilde{M}_gP$ are in fact acting on finite dimensional space, then convergence in weak*-topology of $\mathcal{L}_{1}(\mathcal{H}\oplus\mathcal{H})$ follows naturally. The same property of $P(1-M^{(n)})P$ can be deduced analogously, and therefore we have finished the proof.

\end{proof}

\subsection{Deduction of discrete BCS gap equation}\label{discrete sec}

\begin{proposition}\label{discrete Gap eqn}
  Let the quasi-free state $g$ satisfying (\ref{condition gamma}) be a minimizer of $-L^3P_{\mathrm{BCS}}^L$ for some fixed $L>0$, then for any $\mathbf{k}=(k,\sigma)\in(2\pi/L)\mathbb{Z}^3\times\mathcal{S}_{2}$, we have
  \begin{align}
  \gamma_{k,\sigma}=&\frac{1}{2}\Big(1-\frac{\xi_k}{E_{k,\sigma}}\tanh\frac{\beta}{2}
  E_{k,\sigma}\Big),\label{Gap gamma}\\
  \alpha_{k,\sigma}=&\frac{\Delta_{k,\sigma}}{2E_{k,\sigma}}\tanh\frac{\beta}{2}E_{k,\sigma},
  \label{Gap alpha}
  \end{align}
  where
  \begin{align}
  \Delta_{k,\sigma}&\coloneqq-\frac{1}{L^3}\sum_{k^\prime}\hat{v}(k-k^\prime)
  \alpha_{k^\prime,\sigma},\label{define Delta}\\
  \xi_k&\coloneqq\big(Q(ik)-\mu\big),\label{define xi}\\
  E_{k,\sigma}&\coloneqq\sqrt{\xi_k^2+\vert\Delta_{k,\sigma}\vert^2}.\label{define E_k}
  \end{align}
  Moreover, $\Delta_{k,\sigma}$ satisfies the discrete BCS gap equation:
  \begin{equation}\label{Gap eqn}
    \Delta_{k,\sigma}=-\frac{1}{2L^3}\sum_{k^\prime}\hat{v}(k-k^\prime)
    \frac{\Delta_{k^\prime,\sigma}}{E_{k^\prime,\sigma}}\tanh\frac{\beta}{2}E_{k^\prime,\sigma}.
  \end{equation}
  In particular, (\ref{Gap gamma}), (\ref{Gap alpha}) and (\ref{Gap eqn}) still hold true when temperature $T=0$, with all the $\tanh\frac{\beta}{2}E$ replaced by $1$.
\end{proposition}
\begin{proof}
  \par According to Proposition \ref{existenceofminimizer} and its Corollary \ref{corollary existence}, the minimizer $g$ actually exists. \cite[Lemma 1]{BCSgeneralpair} deduced the BCS gap equation of the continuous version of BCS pressure, and can be generalized to the discrete version. First we notice under (\ref{assumption}), (\ref{inequality M(1-M)=0}) can be simplified to
  \begin{equation}\label{inequality M(1-M) simplified}
    \vert\alpha_{\mathbf{k}}\vert^2\leq\gamma_{\mathbf{k}}(1-\gamma_{\mathbf{k}}).
  \end{equation}
  On the other hand , using (\ref{Wick Thm k=2}) to expand $g(a^*_{\mathbf{k}}a_{-\mathbf{k}}a_{-\mathbf{k}}^*a_{\mathbf{k}})\geq0$, we have
  \begin{equation}\label{inequality M(1-M) vari}
    \vert\alpha_{\mathbf{k}}\vert^2\leq\gamma_{\mathbf{k}}(1-\gamma_{-\mathbf{k}}).
  \end{equation}
  Since under (\ref{assumption}), every generalized 1-pdm $M$ is unitarily conjugated to a block matrix with each block given by (\ref{M(k)}). Taking $\tilde{\alpha}_{\mathbf{k}}=\alpha_{\mathbf{k}}$, $\tilde{\gamma}_{\mathbf{k}}=\frac{1}{2}(\gamma_{\mathbf{k}}+\gamma_{-\mathbf{k}})$, then using (\ref{M(k)}), (\ref{inequality M(1-M) simplified}) and (\ref{inequality M(1-M) vari}), we have $0\leq\tilde{M}\leq1$. It is also easy to see $\tr_{\mathcal{H}}\tilde{\gamma}<\infty$, then by Proposition \ref{1to1correspondencequasifree1-pdm}, $\tilde{M}$ uniquely induces a quasi-free state $\tilde{g}$. Using (\ref{entropy HF assumption}), it is elementary to check that
  \begin{equation}\label{symmetric gamma_k=gamma_-k}
    -\frac{1}{\beta}S_{\mathrm{HF}}[g]\geq-\frac{1}{\beta}S_{\mathrm{HF}}[\tilde{g}],
  \end{equation}
  hence $-L^3P_{\mathrm{BCS}}^L$ decreases as we replace $g$ with $\tilde{g}$. Therefore we can w.l.g. assume $\gamma_{\mathbf{k}}=\gamma_{-\mathbf{k}}$ in the first place.

  \par If inequality (\ref{inequality M(1-M) vari}) becomes an equality for some $\mathbf{k}$, then the two eigenvalues of $M(\mathbf{k})$ are $1$ and $0$. We immediately find that we can lower $-L^3P_{\mathrm{BCS}}^L$ by modifying $\gamma_{\mathbf{k}}$ or $\alpha_{\mathbf{k}}$ so that two eigenvalues of $M(\mathbf{k})$ become closer to 1/2. Since the derivative of $f(x)=x\ln x+(1-x)\ln(1-x)$ diverges as $x$ being closer to $0$ or $1$ and $f$ decreases as $x$ being closer to $1/2$ from both sides, $-L^3P_{\mathrm{BCS}}^L$ will actually be lowered. Hence we can furthermore w.l.g. assume (\ref{inequality M(1-M) vari}) is a strict inequality for any $\mathbf{k}$.

  \par Therefore, for temperature $T>0$, and $\vert t\vert$ be small enough, $\theta\in[0,2\pi)$, and some $\mathbf{k}$ be fixed, we can define two new admissible pair $(\gamma^{(i)},\alpha^{(i)})$, whose only differences with $(\gamma,\alpha)$ are
  \begin{equation}\label{test state gamma alpha}
    \gamma_{\mathbf{k}}^{(1)}=\gamma_{\mathbf{k}}+t,\quad\alpha^{(2)}_{\mathbf{k}}=
    \alpha_{\mathbf{k}}+te^{i\theta}.
  \end{equation}
  Since $(\gamma,\alpha)$ minimizes $-L^3P_{\mathrm{BCS}}^L$, direct calculation gives
  \begin{equation}\label{variation 1}
   0= \frac{d}{dt}\big(-L^3P_{\mathrm{BCS}}^L[g^{(1)}]\big)\Big\vert_{t=0}
   =\xi_k+\frac{2\gamma_{\mathbf{k}}-1}{\beta\sqrt{D}}\ln\frac{1+\sqrt{D}}{1-\sqrt{D}},
  \end{equation}
  and
  \begin{equation}\label{variation 2}
     0= \frac{d}{dt}\big(-L^3P_{\mathrm{BCS}}^L[g^{(2)}]\big)\Big\vert_{t=0}
     =\re\Big\{e^{-i\theta}\big(-\Delta_{\mathbf{k}}+\frac{2\alpha_{\mathbf{k}}}
     {\beta\sqrt{D}}\ln\frac{1+\sqrt{D}}{1-\sqrt{D}}\big)\Big\},
  \end{equation}
  for any $\theta\in[0,2\pi)$, where
  \begin{equation}\label{Define D}
    D=(2\gamma_{\mathbf{k}}-1)^2+4\vert\alpha_{\mathbf{k}}\vert^2.
  \end{equation}
  Rearranging (\ref{variation 1}) and (\ref{variation 2}) we reach (\ref{Gap gamma})-(\ref{Gap eqn}).

  \par For the zero temperature case, the entropy term vanishes and elementary analysis concludes that $-L^3P_{\mathrm{BCS}}^L$ is minimized at the boundary, i.e. $\vert\alpha_{\mathbf{k}}\vert^2=\gamma_{\mathbf{k}}(1-\gamma_{\mathbf{k}})$. Then elementary calculus gives (\ref{Gap gamma})-(\ref{Gap eqn}) for $T=0$.
\end{proof}

%\begin{remark}
%  \par By (\ref{assumption}), we deduce $\Delta_{-\mathbf{k}}=-\Delta_{\mathbf{k}}$ for any $\Delta$ defined by (\ref{define Delta}) and satisfying (\ref{Gap eqn}). Let $c(\sigma)\in\mathbb{S}^1\subset\mathbb{C}$ denote a phase. We assume $c(\sigma)=c(-\sigma)$. For any aforementioned $\Delta$, $\tilde{\Delta}=c(\sigma)\Delta$ also solves (\ref{Gap eqn}), and satisfies $\tilde{\Delta}_{-\mathbf{k}}=-\tilde{\Delta}_{\mathbf{k}}$. In this paper, we call this kind of phase change of $\Delta$, or equivalently of $\alpha$ due to (\ref{Gap alpha}), an \textbf{admissible phase change}.
%\end{remark}

\par Let $(\gamma,\alpha)$ minimizes $-L^3P_{\mathrm{BCS}}^L$ and $\alpha_{\mathbf{k}}=\vert
\alpha_{\mathbf{k}}\vert e^{i\phi_{\mathbf{k}}}$, since $\hat{v}\leq0$, it is easy to check
\begin{equation}\label{alpha positve}
  \begin{aligned}
  g(\mathcal{V})=&\frac{1}{2L^3}\sum_{k,k^\prime,\sigma}\hat{v}(k-k^\prime)
  \alpha_{k^\prime,\sigma}\overline{\alpha_{k,\sigma}}=
  \frac{1}{2L^3}\sum_{k,k^\prime,\sigma}\hat{v}(k-k^\prime)
  \vert\alpha_{k^\prime,\sigma}\vert\vert{\alpha_{k,\sigma}}\vert e^{i(\phi_{k^\prime,\sigma}-\phi_{k,\sigma})}\\
  =&\frac{1}{2L^3}\sum_{k,k^\prime,\sigma}\hat{v}(k-k^\prime)
  \vert\alpha_{k^\prime,\sigma}\vert\vert{\alpha_{k,\sigma}}\vert
  \cos(\phi_{k^\prime,\sigma}-\phi_{k,\sigma})\\
  \geq& \frac{1}{2L^3}\sum_{k,k^\prime,\sigma}\hat{v}(k-k^\prime)
  \vert\alpha_{k^\prime,\sigma}\vert\vert{\alpha_{k,\sigma}}\vert.
  \end{aligned}
\end{equation}
Equality holds if and only if for each spin $\sigma$ and some fixed $\theta_\sigma$, $\phi_{k,\sigma}\equiv\theta_{\sigma}$ for all $k\in (2\pi/L)\mathbb{Z}^3$. Since $(\gamma,\alpha)$ minimizes $-L^3P_{\mathrm{BCS}}^L$, we can w.l.g. consider $\alpha_{k,\uparrow}\geq0$, which implies $\alpha_{-k,\downarrow}=-\alpha_{k,\uparrow}\leq0$. Using (\ref{Gap alpha}) and (\ref{define Delta}), that is equivalent to consider $\Delta_{k,\uparrow}\geq0$. Since we also have $\Delta_{-k,\downarrow}=-\Delta_{k,\uparrow}$, in the following discussion, we can omit the spin subscript and focus on the nonnegative solution of the BCS gap equation (\ref{Gap eqn again start}), i.e.
\begin{equation}\label{Gap eqn again}
  \Delta_{k}=-\frac{1}{2L^3}\sum_{k^\prime}\hat{v}(k-k^\prime)
    \frac{\Delta_{k^\prime}}{E_{k^\prime}}\tanh\frac{\beta}{2}E_{k^\prime}.
\end{equation}

\par Notice that $\Delta\equiv0$ is a solution to (\ref{Gap eqn}). The zero solution can induce a pair $(\gamma_0,\alpha_0)$ through (\ref{Gap gamma}) and (\ref{Gap alpha}). Direct calculation yields $\gamma_0=\big(1+e^{-\beta(Q(\nabla)+\mu)}\big)$ given in (\ref{gamma_0}) and $\alpha_0\equiv0$, and they actually induce a quasi-free state $g_0$ such that $g_0(A)=\tr_{\mathcal{F}}AG_0$, where $G_0$ is the free gas Gibbs state given in (\ref{Gibbs state non-inter}).

\par On the other hand, if (\ref{Gap eqn again start}) has a unique positive solution $\Delta>0$, we can check that $\Delta_k=\Delta_{-k}$. After multiplying $\Delta$ by a phase $c(\sigma)\in\mathbb{S}^1\subset\mathbb{C}$, which satisfies $c(-\sigma)=-c(\sigma)$, we construct a solution to (\ref{Gap eqn}), also denoted as $\Delta$, such that $\Delta_k=\Delta_{-k}$. Using (\ref{Gap gamma}) and (\ref{Gap alpha}), we can construct $(\gamma,\alpha)$ through this $\Delta$, and they satisfy $\gamma_{\mathbf{k}}=\gamma_{-\mathbf{k}}$ and $\alpha_{\mathbf{k}}=-\alpha_{-\mathbf{k}}\neq0$.

\vspace{1em}
\par Next, a natural question is that can we reconstruct the minimizer of $-L^3P^L_{\mathrm{BCS}}$ by solving the gap equation (\ref{Gap eqn}). Let $(\gamma,\alpha)$ satisfies (\ref{Gap gamma}) and (\ref{Gap alpha}), direct calculation yields
\begin{equation}\label{chech alpha gamma 1}
  \gamma_{\mathbf{k}}(1-\gamma_{\mathbf{k}})-\vert\alpha_{\mathbf{k}}\vert^2
  =\frac{1}{4}\big(1-\tanh^2\frac{\beta}{2}E_{\mathbf{k}}\big)>0.
\end{equation}
That is, the pair $(\gamma,\alpha)$ satisfies (\ref{2times2})-(\ref{property of gammakm and alphakm}) and (\ref{assumption}). To apply Proposition \ref{1to1correspondencequasifree1-pdm}, we have to estimate $\tr_{\mathcal{H}}\gamma$. We have the following lemma:
\begin{lemma}\label{check tr gamma lemma}
  Let uniformly bounded $\Delta>0$ be the unique positive solution of discrete BCS gap equation (\ref{Gap eqn again start}) for some fixed $L>0$ and temperature $T\geq0$. By multiplying a phase $c(\sigma)\in\mathbb{S}^1\subset\mathbb{C}$, which satisfies $c(-\sigma)=-c(\sigma)$, we can construct a solution, also denoted by $\Delta$, solves (\ref{Gap eqn}), and therefore induce $(\gamma,\alpha)$ through (\ref{Gap gamma}) and (\ref{Gap alpha}). Then we have
  \begin{equation}\label{chech tr gamma}
    \sum_{\mathbf{k}}(1+k^d)\gamma_{\mathbf{k}}<\infty.
  \end{equation}
\end{lemma}
\begin{proof}
  \par Since $\gamma_0=\big(1+e^{\beta Q(\nabla)-\beta\mu}\big)$, we have
  \begin{equation}\label{gamma_0 k}
    \gamma_{0,\mathbf{k}}=\frac{1}{1+e^{\beta Q(ik)-\beta\mu}}.
  \end{equation}
  Using (\ref{deg Q}), it is easy to check
  \begin{equation}\label{check tr gamma_0}
    \sum_{\mathbf{k}}(1+k^d)\gamma_{0,\mathbf{k}}<\infty.
  \end{equation}
  On the other hand, using (\ref{Gap gamma}), if we let
  \begin{equation}\label{check tr gamma1}
    f_{\mathbf{k}}(\Delta_{\mathbf{k}})=\frac{\tanh\frac{\beta}{2}
    \sqrt{\xi_k^2+\vert\Delta_{\mathbf{k}}\vert^2}}{\sqrt{\xi_k^2+\vert\Delta_{\mathbf{k}}\vert^2}},
  \end{equation}
  then
  \begin{equation}\label{check tr gamma2}
    \gamma_{\mathbf{k}}-\gamma_{0,\mathbf{k}}=
    \frac{\xi_k}{2}\big(f_{\mathbf{k}}(0)-f_{\mathbf{k}}(\Delta_{\mathbf{k}})\big).
  \end{equation}
  We can calculate directly
  \begin{equation}\label{check tr gamma3}
    \frac{\partial f_{\mathbf{k}}}{\partial\vert\Delta_{\mathbf{k}}\vert}=
    \frac{\vert\Delta_{\mathbf{k}}\vert\big(\beta\sqrt{\xi_k^2+\vert\Delta_{\mathbf{k}}\vert^2}-
    \sinh\beta\sqrt{\xi_k^2+\vert\Delta_{\mathbf{k}}\vert^2}\big)}{2
    \big({\xi_k^2+\vert\Delta_{\mathbf{k}}\vert^2}\big)^{\frac{3}{2}}
    \cosh^2\frac{\beta}{2}\sqrt{\xi_k^2+\vert\Delta_{\mathbf{k}}\vert^2}}
  \end{equation}
  for $T>0$ and
  \begin{equation}\label{check tr gamma3 T=0}
     \frac{\partial f_{\mathbf{k}}}{\partial\vert\Delta_{\mathbf{k}}\vert}=-
    \frac{\vert\Delta_{\mathbf{k}}\vert}{
    \big({\xi_k^2+\vert\Delta_{\mathbf{k}}\vert^2}\big)^{\frac{3}{2}}
    }
  \end{equation}
  for $T=0$. Therefore, we can bound
  \begin{equation}\label{check tf gamma4}
    \vert\gamma_{\mathbf{k}}-\gamma_{0,\mathbf{k}}\vert\leq \frac{C\vert\Delta_{\mathbf{k}}\vert^2}{\vert\xi_k\vert^2}.
  \end{equation}
  Then (\ref{chech tr gamma}) follows by combining (\ref{check tr gamma_0}), (\ref{check tf gamma4}) and Lemma \ref{decay Delta}.
\end{proof}

\par Now the remaining question is, whether the energy $-L^3P^L_{\mathrm{BCS}}$ at $(\gamma,\alpha)$ induced by the unique positive solution is actually lower than free gas Gibbs state $(\gamma_0,0)$. We have the following proposition.
\begin{proposition}\label{relation sln and minimizer}
   Let $g_0(A)=\tr_{\mathcal{F}}AG_0$ be the quasi-free state corresponding to free gas Gibbs state $G_0$ given in (\ref{Gibbs state non-inter}). Also let $M_0$ denote its generalized 1-pdm and $(\gamma_0,0)$ be two matrix elements of $M_0$. Let uniformly bounded $\Delta>0$ be the unique positive solution of discrete BCS gap equation (\ref{Gap eqn again start}) for some fixed $L>0$ and temperature $T\geq0$. By multiplying a phase $c(\sigma)\in\mathbb{S}^1\subset\mathbb{C}$, which satisfies $c(-\sigma)=-c(\sigma)$, we can construct a solution, also denoted by $\Delta$, solves (\ref{Gap eqn}), and therefore induce $(\gamma,\alpha)$ through (\ref{Gap gamma}) and (\ref{Gap alpha}). According to Lemma \ref{check tr gamma lemma} and Proposition \ref{1to1correspondencequasifree1-pdm}, $(\gamma,\alpha)$ uniquely induces a quasi-free state $g$. We have
   \begin{equation}\label{lower than free gas}
    \min-L^3P^L_{\mathrm{BCS}}= -L^3P^L_{\mathrm{BCS}}[g]< -L^3P^L_{\mathrm{BCS}}[g_0].
   \end{equation}
\end{proposition}

\begin{proof}
  \par This proposition is a generalization of \cite[Theorem 1]{BCSgeneralpair} to discrete BCS gap equation.

  \par We stick to the notations in the proof of Lemma \ref{check tr gamma lemma}. We take a family of admissible pair $(\tilde{\gamma},\tilde{\alpha})$ with
  \begin{equation}\label{test admissible pair}
    \tilde{\gamma}=\gamma_0, \quad
    \tilde{\alpha}_{\mathbf{k}}=t\alpha_{\mathbf{k}}\chi_{\vert k\vert<R}
  \end{equation}
  for some finite $R>0$. Since $\gamma_{0,\mathbf{k}}>0$, when $\vert t\vert$ is small enough, $(\tilde{\gamma},\tilde{\alpha})$ is actually admissible. We let $\tilde{g}$ be its corresponding quasi-free state by Proposition \ref{1to1correspondencequasifree1-pdm}. Direct calculation yields
  \begin{equation}\label{second dereivative}
  \begin{aligned}
    \frac{d^2}{dt^2}\big(-L^3P^L_{\mathrm{BCS}}[\tilde{g}]\big)\Big\vert_{t=0}
    =&\frac{1}{L^3}\sum_{k,k^\prime,\sigma}\hat{v}(k-k^\prime)
  \alpha_{k^\prime,\sigma}\overline{\alpha_{k,\sigma}}\chi_{\vert k\vert<R}
  \chi_{\vert k^\prime\vert<R}\\
  &+\sum_{k,\sigma}\frac{2\vert\alpha_{k,\sigma}\vert^2\chi_{\vert k\vert<R}}{\sqrt{D}}
  \ln\frac{1+\sqrt{D}}{1-\sqrt{D}},
  \end{aligned}
  \end{equation}
  where
  \begin{equation}\label{relation sln and minimizer1}
    D=(2\gamma_{0,\mathbf{k}}-1)^2.
  \end{equation}
  However, using (\ref{Gap gamma}), (\ref{Gap alpha}), (\ref{define Delta}) and (\ref{gamma_0 k}), we can rearrange
  \begin{equation}\label{relation sln and minimizer2}
    \begin{aligned}
    &\frac{1}{L^3}\sum_{k,k^\prime,\sigma}\hat{v}(k-k^\prime)
  \alpha_{k^\prime,\sigma}\overline{\alpha_{k,\sigma}}
  +\sum_{k,\sigma}\frac{2\vert\alpha_{k,\sigma}\vert^2}{\sqrt{D}}
  \ln\frac{1+\sqrt{D}}{1-\sqrt{D}}\\
  =&\sum_{\mathbf{k}}2\xi_{k}\vert\alpha_{\mathbf{k}}\vert^2
  \Big(\frac{1}{2\gamma_{0,\mathbf{k}}-1}-\frac{1}{2\gamma_{\mathbf{k}}-1}\Big),
    \end{aligned}
  \end{equation}
  and (\ref{check tr gamma2})-(\ref{check tr gamma3 T=0}) imply the right-hand side of (\ref{relation sln and minimizer2}) is strictly negative. Since we have Lemma \ref{decay Delta}, when $R>0$ large enough, by dominated convergence theorem, we also have the right-hand side of (\ref{second dereivative}) is strictly negative. This fact implies the minimizer $(\gamma^{min},\alpha^{min})$ of $-L^3P^L_{\mathrm{BCS}}$ can not be $(\gamma_0,0)$, and thus $\alpha^{min}_{\mathbf{k}}$ is not identically zero. Therefore, by Proposition \ref{discrete Gap eqn}, and the uniqueness assumption in this proposition, any minimizer $(\gamma^{min},\alpha^{min})$ equals to $(\gamma,\alpha)$ up to phase changes of $\alpha$.

\end{proof}

\subsection{Existence of the maximal solution}\label{existence maximal sln sec}
\

%Notice that since $\hat{v}(0)<0$ and $\hat{v}$ is continuous, there exists some universal constant $r>0$ such that $\hat{v}(k)<0$ for any $\vert k\vert<r$. Therefore, as long as $L$ large enough, such that $B_r\cap\big((2\pi/L)\mathbb{Z}^3\backslash\{0\}\big)\neq\emptyset$, one can prove by induction that any nonzero nonnegative solution of (\ref{Gap eqn again}) is strictly positive.

\par Proposition \ref{discrete Gap eqn} has already implied the existence of nonnegative solution of (\ref{Gap eqn again}). In the next proposition, we are going to prove the existence of the maximal solution, which is part of $(1)$ of Theorem \ref{main1}. It is crucial in proving the uniqueness of nonnegative solution of (\ref{Gap eqn again}), and the emergence of translation and $SU(2)$ symmetry in Section \ref{persist}.

\par Some technique used in this section comes from \cite{yang1991bardeen,bryce2000uniqueness}, where a compactly supported $\hat{v}(k,k^\prime)$ was required. However, in our setting, even if $\hat{v}(k)$ is compactly supported, $\hat{v}(k-k^\prime)$ is not as a function of $(k,k^\prime)$.

  \begin{proposition}\label{existence of maximal sln}
    For fixed $L>C(v,\mu,Q)$, there exists a bounded maximal nonnegative solution $\tilde{\Delta}\geq0$ of (\ref{Gap eqn again}), such that for any other nonnegative solution $\Delta\geq0$, we always have for some universal constant (even independent of $L$), $C\leq\Delta\leq\tilde{\Delta}$. Moreover, there exists a positive temperature $T_c=T_c(v,\mu,Q,L)>0$, such that for any $0\leq T<T_c$, $\tilde{\Delta}>0$, and for $T>T_c$, $\tilde{\Delta}\equiv0$. Moreover, for each $k\in (2\pi/L)\mathbb{Z}^3$, $\tilde{\Delta}_k$ is a left continuous function in temperature $T\in[0,\infty)$.
  \end{proposition}

\par To prove Proposition \ref{existence of maximal sln}, we need the following lemmas. For simplicity, we denote
\begin{equation}\label{op gap eqn}
  \mathcal{M}^L[\Delta]_k\coloneqq-\frac{1}{2L^3}\sum_{k^\prime}\hat{v}(k-k^\prime)
    \frac{\Delta_{k^\prime}}{E_{k^\prime}}\tanh\frac{\beta}{2}E_{k^\prime},
\end{equation}
so that (\ref{Gap eqn again}) now becomes $\Delta=\mathcal{M}^L[\Delta]$. The corresponding continuous version is defined by
\begin{equation}\label{op gap eqn continuous}
  \mathcal{M}[\Delta]_k\coloneqq-\frac{1}{2(2\pi)^3}\int_{\mathbb{R}^3}\hat{v}(k-k^\prime)
    \frac{\Delta_{k^\prime}}{E_{k^\prime}}\tanh\frac{\beta}{2}E_{k^\prime}dk^\prime.
\end{equation}

\begin{lemma}\label{comparison principle}
  Let $0\leq\Delta^{(2)}\leq\Delta^{(1)}$ be two bounded function of $k\in(2\pi/L)\mathbb{Z}^3$ for some fixed $L$ and temperature $T$, then $0\leq\mathcal{M}^L[\Delta^{(2)}]\leq\mathcal{M}^L[\Delta^{(1)}]$.
\end{lemma}
\begin{proof}
  \par Let
  \begin{equation}\label{function f g}
    f_k(\Delta)=\frac{\Delta_k}{2\sqrt{\xi_k^2+\Delta_k^2}},\quad
    g_k(\Delta)=\tanh\frac{\beta}{2}\sqrt{\xi_k^2+\Delta_k^2}.
  \end{equation}
  Both of them are monotone increasing for $\Delta\geq0$. Since $\hat{v}\leq0$ and
  \begin{equation}\label{op gap rewrite}
    \mathcal{M}^L[\Delta]_k=-\frac{1}{L^3}\sum_{k^\prime}
    \hat{v}(k-k^\prime)f_{k^\prime}(\Delta)g_{k^\prime}(\Delta),
  \end{equation}
  the comparison principle follows immediately.
\end{proof}

\begin{lemma}\label{existence of subsolution}
  For fixed $L,\beta>C(v,\mu,Q)$, there exists a bounded nonzero subsolution $\Delta_0\geq0$ of (\ref{Gap eqn again}) such that $\Delta_0<\mathcal{M}^L[\Delta_0]$.
\end{lemma}
\begin{proof}
  \par Since $\hat{v}(0)<0$ and $\hat{v}$ is continuous, there is some universal constant $r_1>0$ such that
  \begin{equation}\label{inf hatv m}
    \inf_{\vert k\vert\leq r_1}\big(-\hat{v}\big)=m>0.
  \end{equation}
  Also according to Assumption \ref{assumption Q}, we let $r_0>0$ be the unique zero point of $Q(ir)-\mu$. We then construct a subsolution support on the annulus
   \begin{equation}\label{annulus A}
      A^L=\Big\{k\in(2\pi/L)\mathbb{Z}^3\,\vert\,r_0-\frac{r_1}{2}\leq
   \vert k\vert\leq r_0+\frac{r_1}{2}\Big\},
   \end{equation}
   and
   \begin{equation}\label{annulus A continuous}
      A=\Big\{k\in\mathbb{R}^3\,\vert\,r_0-\frac{r_1}{2}\leq
   \vert k\vert\leq r_0+\frac{r_1}{2}\Big\},
   \end{equation}
   Notice that $A$ can be equally divided into $n(r_1)$ parts such that for each part $A_i$, $\diam A_i\leq r_1$, and
   \begin{equation}\label{annulus A_i}
     \int_{A_i}f(x)dx=\frac{1}{n({r_1})}\int_{A}f(x)dx
   \end{equation}
   for all radially symmetric function $f$. For any $k\in A_i\cap(2\pi/L)\mathbb{Z}^3\subset A^L$ and $\varepsilon>0$,
   \begin{equation}\label{sub est 1}
     -\frac{1}{L^3}\sum_{ k^\prime\in A^L}\frac{\hat{v}(k-k^\prime)}
     {\sqrt{\big(Q(ik^\prime)-\mu\big)^2+\varepsilon^2}}\geq
     \frac{m}{L^3}\sum_{k^\prime\in A_i}\frac{1}
     {\sqrt{\big(Q(ik^\prime)-\mu\big)^2+\varepsilon^2}}.
   \end{equation}
   Notice $Q(ik^\prime)$ is a monotone increasing function in $\vert k^\prime\vert$, thus for fixed $\varepsilon>0$ and large $L>C(\varepsilon)$, the right-hand side of (\ref{sub est 1}) can be bounded from below by
   \begin{equation}\label{sub est 2}
     -\frac{m}{L^3}\sum_{k^\prime\in A_i}\frac{1}
     {\sqrt{\big(Q(ik^\prime)-\mu\big)^2+\varepsilon^2}}\geq \frac{m}{2(2\pi)^3n(r_1)}\int_{A}
     \frac{dk^\prime}
     {\sqrt{\big(Q(ik^\prime)-\mu\big)^2+\varepsilon^2}}
   \end{equation}
   for any part $A_i\subset A$. Notice that the right-hand side of (\ref{sub est 2}) tends to infinity as $\varepsilon\to0$, we then choose $\varepsilon_0=\varepsilon_0(m,r_1,\mu,Q)$ small enough such that
   \begin{equation}\label{sub est  3}
     \frac{m}{2(2\pi)^3n(r_1)}\int_{A}
     \frac{dk^\prime}
     {\sqrt{\big(Q(ik^\prime)-\mu\big)^2+\varepsilon_0^2}}\geq 10.
   \end{equation}
   On the other hand, for $\beta>C(\varepsilon_0)$ large enough,
   \begin{equation}\label{sub est 4}
     \tanh\frac{\beta}{2}\sqrt{\big(Q(ik^\prime)-\mu\big)^2+\varepsilon_0^2}>\frac{1}{5}.
   \end{equation}
   Therefore, for all $k\in A^L$, as long as $L,\beta>C(v,\mu)$,
   \begin{equation}\label{sub est 5}
      -\frac{1}{L^3}\sum_{ k^\prime\in A^L}\hat{v}(k-k^\prime)
      \frac{\tanh\frac{\beta}{2}\sqrt{\big(Q(ik^\prime)-\mu\big)^2+\varepsilon_0^2}}
     {\sqrt{\big(Q(ik^\prime)-\mu\big)^2+\varepsilon_0^2}}\geq2>1.
   \end{equation}
   We then define for $k\in(2\pi/L)\mathbb{Z}^3$
   \begin{equation}\label{sub sln}
     \Delta_0(k)=\left\{\begin{aligned}
     \varepsilon_0,\quad k\in A^L,\\
     0,\quad k\notin A^L.
     \end{aligned}\right.
   \end{equation}
   It is easy to verify $\Delta_0<\mathcal{M}^L[\Delta_0]$ using (\ref{sub est 5}).
\end{proof}

\begin{lemma}\label{existence of supersolution}
  For fixed $L>C(v,\mu,Q)$ and any temperature $0\leq T<\infty$, there exists a bounded nonzero supersolution $\Delta^0\equiv\delta_0>0$ of (\ref{Gap eqn again}) such that $\Delta^0>\mathcal{M}^L[\Delta^0]$.
\end{lemma}

\begin{proof}
  \par In the case $0<d\leq3$, for any $k\in(2\pi/L)\mathbb{Z}^3$, $\delta>0$ and $0\leq T<\infty$, using (\ref{deg Q}) and (\ref{requirement for v 1}), we have
  \begin{equation}\label{sup est 1}
    \begin{aligned}
    0\leq&-\frac{1}{L^3}\sum_{k^\prime}\hat{v}(k-k^\prime)
      \frac{\tanh\frac{\beta}{2}\sqrt{\big(Q(ik^\prime)-\mu\big)^2+\delta^2}}
     {\sqrt{\big(Q(ik^\prime)-\mu\big)^2+\delta^2}}\\
     \leq&\Big(\frac{1}{L^3}\sum_{k^\prime}\frac{C}{1+\vert k-k^\prime\vert^{p(3-d+\varepsilon)}}\Big)^{\frac{1}{p}}
     \Big(\frac{1}{L^3}\sum_{k^\prime}
     \frac{1}{\big((\vert k^\prime\vert^d-\mu)^2+\delta^2\big)^{p^\prime/2}}
     \Big)^{\frac{1}{p^\prime}},
    \end{aligned}
  \end{equation}
  where $p=(3+\varepsilon)/(3-d+\varepsilon)<3/(3-d)$ and $p^{\prime}=p/(p-1)>3/d$. Notice that for fixed $\delta$, when $L>C(\delta)$ large enough, the right-hand side of (\ref{sup est 1}) can be bounded by
  \begin{equation}\label{sup est 2}
    \begin{aligned}
    C\Big(\int_{\mathbb{R}^3}\frac{dk^\prime}{1+\vert k^\prime\vert^{p(3-d+\varepsilon)}}
    \Big)^{\frac{1}{p}}
    \Big(\int_{\mathbb{R}^3}\frac{dk^\prime}{\big((\vert k^\prime\vert^d-\mu)^2+\delta^2\big)^{p^\prime/2}}
    \Big)^{\frac{1}{p^\prime}}<\infty.
    \end{aligned}
  \end{equation}
  The second integral of (\ref{sup est 2}) tends to $0$ as $\delta\to\infty$. Hence we can choose a fixed $\delta_0=\delta_0(v,\mu,Q)$ large enough so that (\ref{sup est 2}) is bounded by, say, $1/4$. Then we define
  \begin{equation}\label{sup sln}
     \Delta^0(k)=\delta_0,
   \end{equation}
   which is a desired supersulotion.
   \par In the case $d>3$, it is much easier since $\vert k\vert^{-d}$ itself is integrable at infinity, we thus omit further details.
 \end{proof}

 \begin{lemma}\label{coro super}
   For $L>C(v,\mu,Q)$ given in Lemma \ref{existence of supersolution}, any nonnegative subsolution $\Delta$ of (\ref{Gap eqn again}) is bounded by $\delta_0$ given in (\ref{sup sln}).
 \end{lemma}

 \begin{proof}
   \par Otherwise $\bar{\Delta}\coloneqq\sup\Delta(k)>\delta_0$, then from Lemma \ref{comparison principle} the proof of Lemma \ref{existence of supersolution} we know $\Delta\leq\mathcal{M}^L[\Delta]\leq\mathcal{M}^L[\bar{\Delta}]\leq\bar{\Delta}/4$, which leads to a contradiction if we take supremum with respect to $k\in(2\pi/L)\mathbb{Z}^3$ at each side of the inequality.
 \end{proof}

\begin{lemma}\label{zero sln}
  For fixed $L,T>C(v,\mu,Q)$, then the only solution of (\ref{Gap eqn again}) is zero.
\end{lemma}
\begin{proof}
  \par Assume $\Delta$ is a solution (not necessarily nonnegative) of (\ref{Gap eqn again}), let $\bar{\Delta}\coloneqq\sup\vert\Delta(k)\vert$. By virtue of (\ref{op gap eqn}), we have
  \begin{equation}\label{zero est 1}
    \vert\Delta_k\vert\leq \mathcal{M}^L[\bar{\Delta}]_k.
  \end{equation}
  Notice that (\ref{zero est 1}) implies $\bar{\Delta}\leq\delta_0$ when $L>C(v,\mu,Q)$ given in Lemma \ref{existence of supersolution}.

  \par We estimate $\mathcal{M}^L[\bar{\Delta}]_k$ by dividing it into two parts by cutting the momentum $k^\prime$ at some fixed level $\lambda$ to be decided. For the lower momentum part, first notice that
  \begin{equation}\label{zero est 2}
    \frac{\tanh\frac{\beta}{2}\sqrt{\big(Q(ik)-\mu\big)^2+\bar{\Delta}^2}}
    {\sqrt{\big(Q(ik)-\mu\big)^2+\bar{\Delta}^2}}\leq\frac{\beta}{2}.
  \end{equation}
  Since $\hat{v}$ is bounded, we have for any $L>C(\lambda)$,
  \begin{equation}\label{zero est 3}
    \begin{aligned}
    -\frac{1}{L^3}\sum_{\vert k^\prime\vert\leq\lambda}\hat{v}(k-k^\prime)
    \frac{\tanh\frac{\beta}{2}\sqrt{\big(Q(ik^\prime)-\mu\big)^2+\bar{\Delta}^2}}
    {\sqrt{\big(Q(ik^\prime)-\mu\big)^2+\bar{\Delta}^2}}
    \leq&\sum_{\vert k^\prime\vert\leq\lambda}\frac{C\beta}{L^3}\\
    \leq& C\beta\lambda^3.
    \end{aligned}
  \end{equation}
  For the higher momentum part, we bound it in a way similar to (\ref{sup est 1}). For $0<d\leq3$ (case $d>3$ can be analyzed similarly and we omit further details), we have
  \begin{equation}\label{zero est 4}
    \begin{aligned}
    &-\frac{1}{L^3}\sum_{\vert k^\prime\vert>\lambda}\hat{v}(k-k^\prime)
    \frac{\tanh\frac{\beta}{2}\sqrt{\big(Q(ik^\prime)-\mu\big)^2+\bar{\Delta}^2}}
    {\sqrt{\big(Q(ik^\prime)-\mu\big)^2+\bar{\Delta}^2}}\\
    \leq& \Big(\frac{1}{L^3}\sum_{k^\prime}\frac{C}{1+\vert k-k^\prime\vert^{p(3-d+\varepsilon)}}\Big)^{\frac{1}{p}}
     \Big(\frac{1}{L^3}\sum_{\vert k^\prime\vert>\lambda}
     \frac{1}{\big((\vert k^\prime\vert^d-\mu)^2+\bar{\Delta}^2\big)^{p^\prime/2}}
     \Big)^{\frac{1}{p^\prime}},
    \end{aligned}
  \end{equation}
  where $p=(3+\varepsilon)/(3-d+\varepsilon)<3/(3-d)$ and $p^{\prime}=p/(p-1)>3/d$. As long as $\lambda>C(\mu)$ and $L>C(v,\mu,\lambda)$, (\ref{zero est 4}) can be bounded by
  \begin{equation}\label{zero est 5}
    \begin{aligned}
    C\Big(\int_{\mathbb{R}^3}\frac{dk^\prime}{1+\vert k^\prime\vert^{p(1+\varepsilon)}}
    \Big)^{\frac{1}{p}}
    \Big(\int_{\vert k^\prime\vert>\lambda/2}\frac{dk^\prime}{\vert k^\prime\vert^{p^\prime d}}
    \Big)^{\frac{1}{p^\prime}}\leq C\lambda^{\frac{3-p^\prime d}{p^\prime}}.
    \end{aligned}
  \end{equation}
  We then choose $\lambda_0=\lambda_0(v,\mu,Q)$ large enough such that (\ref{zero est 5}) is less than $1/4$. Using (\ref{zero est 3}), for $\beta<C(\lambda_0)$ small enough, it can also be bounded by $1/4$. Therefore, for $L,T>C(v,\mu)$, we get to
  \begin{equation}\label{zero est  6}
    \vert\Delta_k\vert\leq\mathcal{M}^L[\bar{\Delta}]_k\leq\frac{1}{2}\bar{\Delta},
  \end{equation}
  which contradicts to the definition of $\bar{\Delta}$ unless $\Delta\equiv0$.
\end{proof}

\vspace{1em}

\begin{proof}[Proof of Proposition \ref{existence of maximal sln}]
  \par From Lemma \ref{existence of supersolution}, we can define $\Delta^n$ by induction
  \begin{equation}\label{define Delta^n}
    \Delta^{n+1}=\mathcal{M}^L[\Delta^n],
  \end{equation}
  where $\Delta^0\equiv\delta_0$. By Lemma \ref{comparison principle}, $0\leq\Delta^{n+1}\leq\Delta^n\leq\Delta^0=\delta_0$. Therefore, there exists a bounded nonnegative $\Delta^{\infty}$ such that $\Delta^n$ converges point-wisely to $\Delta^{\infty}$. From (\ref{define Delta^n}) we know that $\Delta^\infty$ is a solution of (\ref{Gap eqn again}). If $\Delta\geq0$ is another solution of (\ref{Gap eqn again}), from Lemma \ref{coro super}, $\Delta\leq\Delta^0$. Since $\Delta=\mathcal{M}^L[\Delta]$, using Lemma \ref{comparison principle} we deduce $\Delta\leq\Delta^n$ and hence $\Delta\leq\Delta^\infty$. Therefore, $\Delta^\infty$ is the desired maximal solution.

  \par If $\beta>C(v,\mu)$ as in Lemma \ref{existence of subsolution} large enough, we can take $\Delta_0$ be a nonzero subsolution of (\ref{Gap eqn again}). Using Lemmas \ref{comparison principle} and \ref{coro super}, we have $0\leq\Delta_0\leq\Delta^n$ and hence $\Delta_0\leq\Delta^\infty$. Since $\Delta_0$ is nonzero, by the observation near (\ref{Gap eqn again}), when $L>C(v)$ large enough, $\Delta^\infty$ must be positive.

  \par Now for fixed $L>C(v,\mu)$ large enough, let $\Delta^n=\Delta^n[T]$ and $\Delta^\infty=\Delta^\infty[T]$ to indicate its dependence of temperature $T$. We define
  \begin{equation}\label{T_c define}
    T_c\coloneqq\sup\big\{T\,\vert\,\Delta^\infty[T]>0\, \text{at temperature $T$}\big\}.
  \end{equation}
  Then the above discussion implies $T_c>0$. On the other hand, for any $0\leq T<T_c$, by definition, there must exist another temperature $T<T^\prime<T_c$ such that at $T^\prime$, $\Delta^\infty[T^\prime]>0$. However,
  \begin{equation}\label{proof 3.3 est 1}
    \begin{aligned}
    \Delta^\infty[T^\prime]_k=&-\frac{1}{L^3}\sum_{k^\prime}\hat{v}(k-k^\prime)
    \frac{\Delta^\infty[T^\prime]_{k^\prime}}{2\sqrt{\xi_{k^\prime}^2+\Delta^\infty[T^\prime]
    _{k^\prime}^2}}\tanh\frac{1}{2T^\prime}\sqrt{\xi_{k^\prime}^2+\Delta^\infty[T^\prime]
    _{k^\prime}^2}\\
    \leq&-\frac{1}{L^3}\sum_{k^\prime}\hat{v}(k-k^\prime)
    \frac{\Delta^\infty[T^\prime]_{k^\prime}}{2\sqrt{\xi_{k^\prime}^2+\Delta^\infty[T^\prime]
    _{k^\prime}^2}}\tanh\frac{1}{2T}\sqrt{\xi_{k^\prime}^2+\Delta^\infty[T^\prime]
    _{k^\prime}^2},
    \end{aligned}
  \end{equation}
  which means $\Delta^\infty[T^\prime]$ is an positive subsolution of (\ref{Gap eqn again}) at temperature $T$. Then using the argument above we deduce immediately $\Delta^\infty[T]\geq\Delta^\infty[T^\prime]>0$.

  \par For $0\leq T_1<T_2$,% by definition, $\Delta^0[T_1]=\Delta^0[T_2]=\delta_0$, and
  %\begin{equation}\label{comparison Delta^1 T_1and T_2}
   % \begin{aligned}
   % \Delta^1[T_1]=&-\frac{1}{2L^3}\sum_{k^\prime}\hat{v}(k-k^\prime)
   % \frac{\Delta^0[T_1]_{k^\prime}}{\sqrt{\xi_{k^\prime}^2+\Delta^0[T_1]
   % _{k^\prime}^2}}\tanh\frac{1}{2T_1}\sqrt{\xi_{k^\prime}^2+\Delta^0[T_1]
   % _{k^\prime}^2}\\
   % >&-\frac{1}{2L^3}\sum_{k^\prime}\hat{v}(k-k^\prime)
   % \frac{\Delta^0[T_2]_{k^\prime}}{\sqrt{\xi_{k^\prime}^2+\Delta^0[T_2]
    %_{k^\prime}^2}}\tanh\frac{1}{2T_2}\sqrt{\xi_{k^\prime}^2+\Delta^0[T_2]
   % _{k^\prime}^2}=\Delta^1[T_2].
   % \end{aligned}
  %\end{equation}
  %Using Lemma \ref{comparison principle}, and argue by induction, we have $\Delta^n[T_1]>\Delta^n[T_2]$ and thus
   we have $\delta_0\geq\Delta^\infty[T_1]\geq\Delta^\infty[T_2]\geq0$. Now for arbitrary $k\in(2\pi/L)\mathbb{Z}^3$, and $T\nearrow T_0$, there must exist a function $\tilde{\Delta}[T_0]$ such that
  \begin{equation}\label{limit TtoT_0}
    \lim_{T\to T_0}\Delta^\infty[T]_k=\tilde{\Delta}[T_0]_k.
  \end{equation}
  By dominated convergence theorem, we know this $\tilde{\Delta}[T_0]$ solves the Gap equation (\ref{Gap eqn again}), thus $\Delta^\infty[T_0]\geq\tilde{\Delta}[T_0]$. On the other hand, since $\Delta^\infty[T]_k\geq\Delta^\infty[T_0]_k$, we must have $\tilde{\Delta}[T_0]\geq\Delta^\infty[T_0]$. Therefore, $\tilde{\Delta}[T_0]=\Delta^\infty[T_0]$, and $\Delta^\infty[T]_k$ is left continuous in $T$.

  \end{proof}

\subsection{Uniqueness of positive solution}\label{uniqueness sec}
\begin{proof}[Proof of uniqueness of positive solution for $L$ greater than some threshold]
\
\par In this section, we are going to prove part of $(2)$ of Theorem \ref{main1}, that is, when $0\leq T<T_c$. The $T>T_c$ part is obvious due to $(1)$ of Theorem \ref{main1}, or in other words, Proposition \ref{existence of maximal sln}.

  \par Using Proposition \ref{existence of maximal sln}, we are going to prove any nonzero nonnegative solution of (\ref{Gap eqn again}) is in fact the maximal solution for $0\leq T<T_c$. Here we require $L$ large enough so that any nonzero nonnegative solution is everywhere positive, we then assume $\Delta^{(i)}$ for $i=1,2$ are two positive solutions and $\Delta^{(1)}$ is the maximal solution, such that $\Delta^{(1)}\geq\Delta^{(2)}>0$.

   \par Using (\ref{Gap alpha}), we can define $\alpha^{(i)}$. By definition we can check $\alpha^{(1)}_k\geq\alpha^{(2)}_k>0$. Also, let
  \begin{equation}\label{K^delta_T(k)}
    K^{\Delta}_{T}(k)\coloneqq\frac{E_k}{\tanh\frac{\beta}{2}E_k},
  \end{equation}
  and the multiplier $K^{\Delta}_T$ on $L^2(\Lambda_L)$ given by
  \begin{equation}\label{K^delta_T}
   \frac{1}{L^{\frac{3}{2}}}\int_{\Lambda_L} \big(K^{\Delta}_Tf\big)(x)e^{-ipx}dx
   =\frac{K^{\Delta}_{T}(k)}{L^{\frac{3}{2}}}\int_{\Lambda_L}f(x)e^{-ipx}dx
  \end{equation}
  for all $k\in(2\pi/L)\mathbb{Z}^3$. By definition, $K^\Delta_T$ is self-adjoint and $K_T^\Delta\geq 2T$. Notice that $\Delta^{(1)}\geq\Delta^{(2)}\geq0$ implies $K^{\Delta^{(1)}}_{T}(k)\geq K^{\Delta^{(2)}}_{T}(k)$ and hence $K^{\Delta^{(1)}}_{T}\geq K^{\Delta^{(2)}}_{T}$.

  \par We also let
  \begin{equation}\label{define V}
   Vf=\frac{1}{2}v^Lf.
  \end{equation}
   $V$ is also self-adjoint since $\hat{v}\leq0$ and $v$ is radially symmetric. By definition, we have
  \begin{equation}\label{gap eqn op form}
    \big(K^{\Delta^{(i)}}_{T}+V\big)\alpha^{(i)}=0.
  \end{equation}
  Therefore,
  \begin{equation}\label{gap compare}
    0\geq\big\langle(K^{\Delta^{(2)}}_{T}+V\big)\alpha^{(1)},\alpha^{(1)}\big\rangle.
  \end{equation}
  We claim that
  \begin{equation}\label{K^delta_T+V>=0}
    K^{\Delta^{(i)}}_{T}+V\geq0
  \end{equation}
  and $0$ is its non-degenerate eigenvalue.

  \par Assuming the claim (\ref{K^delta_T+V>=0}), combining (\ref{gap compare}) and (\ref{K^delta_T+V>=0}), we have $\alpha^{(2)}=c\alpha^{(1)}$ for some $0\leq c\leq1$ and hence $\Delta^{(2)}=c\Delta^{(1)}$ since by (\ref{Gap alpha}) and (\ref{Gap eqn again})
  \begin{equation}\label{gap delta again}
    \Delta_{k}^{(i)}\coloneqq-\frac{1}{L^3}\sum_{k^\prime}\hat{v}(k-k^\prime)
  \alpha^{(i)}_{k^\prime}.
  \end{equation}
  However, plugging $\Delta^{(2)}=c\Delta^{(1)}$ back into (\ref{Gap eqn again}) we find
  \begin{equation}\label{gap compare 2}
    \Delta^{(1)}_{k}=-\frac{1}{2L^3}\sum_{k^\prime}\hat{v}(k-k^\prime)
    \frac{\Delta^{(1)}_{k^\prime}}{K^{\Delta^{(2)}}_T(k^\prime)}
    \geq-\frac{1}{2L^3}\sum_{k^\prime}\hat{v}(k-k^\prime)
    \frac{\Delta^{(1)}_{k^\prime}}{K^{\Delta^{(1)}}_T(k^\prime)}=\Delta^{(1)}_k.
  \end{equation}
  The inequality in the middle becomes an equality if and only if $c=1$, hence we have finished the proof of uniqueness.

  \par We now prove claim (\ref{K^delta_T+V>=0}). For simplicity we omit the superscript of $\Delta$. First observe that $K^{\Delta}_T(k)\geq C\vert k^d-\mu\vert$, then using technique similar to (\ref{Sobolev imbedding}) we know that $K^{\Delta}_{T}+V$ is bounded from below and $V$ is relatively compact with respect to $K^{\Delta}_{T}$. By Weyl's theorem,
  \begin{equation}\label{spectrum ess}
    \sigma_{ess}\big(K^{\Delta}_{T}+V\big)=
    \sigma_{ess}\big(K^{\Delta}_{T}\big)=\emptyset.%\subset[2T,\infty).
  \end{equation}
  Therefore, from (\ref{gap eqn op form}) we can assume $K^{\Delta}_{T}+V$ has a lowest isolated eigenvalue $\lambda\leq0$ and $\alpha_0$ as one of its corresponding normalized eigenfunction. Define $\tilde{\alpha}_0$ so that all of its Fourier coefficients are given by taking absolute value of the Fourier coefficients of $\alpha_0$ at corresponding momentum. Using observation (\ref{alpha positve}), we can check that
  \begin{equation}\label{alpha_0 positive}
    \langle\big(K^{\Delta}_{T}+V\big)\tilde{\alpha}_0,\tilde{\alpha}_0\rangle
    \leq \langle\big(K^{\Delta}_{T}+V\big){\alpha}_0,{\alpha}_0\rangle=\lambda.
  \end{equation}
  Since $\tilde{\alpha}_0$ is normalized, it is also an eigenfunction associated with $\lambda$. However, $\alpha$ is an eigenfunction associated with $0$, but
  \begin{equation}\label{not orthgonal alpha}
    \langle\alpha,\tilde{\alpha}_0\rangle=\sum_{k}\alpha_k\cdot\vert\alpha_{0,k}\vert>0.
  \end{equation}
  (\ref{not orthgonal alpha}) holds if and only if $\lambda=0$ is non-degenerate.
\end{proof}

\par We state the result (\ref{K^delta_T+V>=0}) in the following lemma, which is useful in Section \ref{persist}.
\begin{lemma}\label{K^delta_T+V>=0 lemma}
  Let $\Delta>0$ be the positive solution of discrete BCS gap equation (\ref{Gap eqn again}) for some fixed $L>0$ and temperature $T\geq0$. Two self-adjoint operators $K_T^\Delta$ and $V$ are defined in (\ref{K^delta_T}) and (\ref{define V}) respectively. Then $0$ is the non-degenerate ground state eigenvalue of $K_T^\Delta+V$, with $\alpha$ induced through (\ref{Gap alpha}) being its corresponding eigenfunction.
\end{lemma}

\subsection{Corollaries}\label{corollaries}

\begin{corollary}\label{corollary existence}
  For fixed $L>0$, $P^L_{\mathrm{BCS}}$ defined in (\ref{Pressure BCS}) has at least one maximizer in its domains.
\end{corollary}
\begin{proof}
   \par Given the symmetry assumption provided by (\ref{assumption}), \cite[Proposition 2]{BCSgeneralpair} has proved the existence of maximizer of the continuous version of BCS pressure. For the discrete version $P^L_{\mathrm{BCS}}$, the argument is simpler than  Proposition \ref{existenceofminimizer}, since we have (\ref{assumption}), and we omit further details.
\end{proof}

\begin{corollary}\label{coro minimizer}
  Let the quasi-free state $g$ satisfying (\ref{condition gamma}) be a minimizer of $-L^3P_{\mathrm{BCS}}^L$ for some fixed $L>0$, and $M$ be its corresponding generalized 1-pdm with the form (\ref{2times2}). $M$ is unitarily conjugated to a block matrix with each block $M(\mathbf{k})$ given by (\ref{M(k)}) for $\mathbf{k}=(k,\sigma)\in(2\pi/L)\mathbb{Z}^3\times
  \mathcal{S}_{2}$. Further, let
  \begin{equation}\label{1st quantization of minimizer}
    H_{\Delta}(\mathbf{k})\coloneqq\begin{pmatrix}
                                     Q(ik)-\mu & -\Delta_{k,\sigma} \\
                                     -\bar{\Delta}_{k,\sigma} & \mu-Q(ik)
                                   \end{pmatrix},
  \end{equation}
  with $\Delta$ given in (\ref{define Delta}), then
  \begin{equation}\label{M(k)=1/1+exp(H/T)}
    M(\mathbf{k})=\frac{1}{1+e^{\beta H_{\Delta}(\mathbf{k})}}.
  \end{equation}
  Moreover, if we let $H_{\Delta}$ be an self-adjoint operator on $\mathcal{H}\oplus\mathcal{H}$ given by
  \begin{equation}\label{H_Delta}
    H_{\Delta}\coloneqq\begin{pmatrix}
                                     Q(\nabla_x)-\mu & T_\Delta \\
                                     T_\Delta^* & -Q(\nabla_y)+\mu
                                   \end{pmatrix},
  \end{equation}
  where $T_\Delta$ is defined on $\mathcal{H}$ by
  \begin{equation}\label{T_Delta}
    T_\Delta f=-\sum_{k,\sigma}c_{-k,-\sigma}\Delta_{k,\sigma}f_{k,\sigma},\quad\text{for any }
    f=\sum_{k,\sigma}c_{k,\sigma}f_{k,\sigma},
  \end{equation}
  then
  \begin{equation}\label{M=1/1+exp(H/T)}
    M=\frac{1}{1+e^{\beta H_{\Delta}}}.
  \end{equation}
\end{corollary}
\begin{proof}
  \par Notice that
  \begin{equation}\label{H_Delta^2}
     H_{\Delta}(\mathbf{k})^2=\begin{pmatrix}
                                E_{k,\sigma}^2 & 0 \\
                                0 & E_{k,\sigma}^2
                              \end{pmatrix},
  \end{equation}
  then (\ref{M(k)=1/1+exp(H/T)}) is a direct result of (\ref{Gap gamma}) and (\ref{Gap alpha}) after elementary analysis.

  \par For the proof of (\ref{M=1/1+exp(H/T)}), first we notice that the Laplacian $\Delta_x$ commutes $T_\Delta$, and for $f=\sum_{\mathbf{k}}
  c_{\mathbf{k}}f_{\mathbf{k}}\in\mathcal{H}$,
  \begin{equation}\label{TT*=T*T}
    T_\Delta T^*_{\Delta}f=T^*_{\Delta}T_\Delta f=\sum_{k,\sigma}
    c_{k,\sigma}\vert\Delta_{k,\sigma}\vert^2
    f_{k,\sigma}.
  \end{equation}
  Also notice that, by definition (\ref{define Delta}), $\Delta_{-k,-\sigma}=-\Delta_{k,\sigma}$ since $\alpha_{-k,-\sigma}=-\alpha_{k,\sigma}$ and $v$ is radially-symmetric, a similar analysis in proving (\ref{M(k)=1/1+exp(H/T)}) can be applied here to deduce (\ref{M=1/1+exp(H/T)}).
\end{proof}

\par The next corollary gives the lower bound to the unique positive solution $\Delta$ to the gap equation (\ref{Gap eqn again start}) when temperature $T$ is small enough.
\begin{corollary}\label{lower bound gap Delta}
  For $L>C(v,\mu,Q)$ and $0\leq T<\min\{T_c(v,\mu,Q,L),C(v,\mu,Q)^{-1}\}$, let $\Delta$ be the unique positive solution to (\ref{Gap eqn again start}). Also let $A^L$ be as defined in (\ref{annulus A}). Then for all $k\in A^L$, $\Delta_k$ is bounded from below by $\varepsilon(v,\mu,Q)>0$ given in Lemma \ref{existence of subsolution}.
\end{corollary}
\begin{proof}
  \par Let $\Delta_0$ defined in (\ref{sub sln}) be the nonzero nonnegative subsolution to (\ref{Gap eqn again start}). From the proof of Proposition \ref{existence of maximal sln} at the end of Section \ref{existence maximal sln sec}, we have
  \begin{equation}\label{dsyufh}
   \mathcal{M}^L[\Delta_0]\leq \Delta.
  \end{equation}
  However, (\ref{sub est 5}) implies for any $k\in A^L$ with $A^L$ defined in (\ref{annulus A}),
  \begin{equation}\label{fuigvf}
    \mathcal{M}^L[\Delta_0]_k>\varepsilon_0>0.
  \end{equation}
  Therefore, for any $k\in A$,
  \begin{equation}\label{fduddgheri}
    \Delta_k>\varepsilon_0>0.
  \end{equation}
\end{proof}

\par The next corollary consider the solution to the gap equation at the critical temperature $T_c$ and finish the final part of the proof of Theorem \ref{main1}.
\begin{corollary}\label{sln critical}
  For $L>C(v,\mu,Q)$ and $T=T_c(v,\mu,Q,L)$, then (\ref{Gap eqn again start}) has only zero solution.
\end{corollary}
\begin{proof}
  \par We argue by contradiction. Assume the contrary, then the maximal solution ${\Delta}$ at $T_c$ must be positive. Following the argument in Section \ref{uniqueness sec}, ${\Delta}$ is the unique positive solution at $T_c$. Then the proof of Lemma \ref{relation sln and minimizer} gives
  \begin{equation}\label{sln critical1}
    \begin{aligned}
    &\frac{1}{L^3}\sum_{k,k^\prime,\sigma}\hat{v}(k-k^\prime)
  \alpha_{k^\prime,\sigma}\overline{\alpha_{k,\sigma}}
  +\sum_{k,\sigma}\frac{2\vert\alpha_{k,\sigma}\vert^2}{\sqrt{D}}
  \ln\frac{1+\sqrt{D}}{1-\sqrt{D}}\\
  =&\sum_{\mathbf{k}}2\xi_{k}\vert\alpha_{\mathbf{k}}\vert^2
  \Big(\frac{1}{2\gamma_{0,\mathbf{k}}-1}-\frac{1}{2\gamma_{\mathbf{k}}-1}\Big)<0,
    \end{aligned}
  \end{equation}
  where
  \begin{equation}\label{sln critical2}
    \sqrt{D}=\vert2\gamma_{0,\mathbf{k}}-1\vert=\tanh\frac{1}{2T_c}\big\vert Q(ik)-\mu\big\vert.
  \end{equation}
  Since we have Lemma \ref{decay Delta}, by dominated convergence theorem, expression (\ref{sln critical1}) remains negative when the temperature $T$ varies in a small neighbour of $T_c$. Using the proof of Lemma \ref{relation sln and minimizer} again, we find that in this small neighbour of $T_c$, the minimizer $(\gamma^{min},\alpha^{min})$ must satisfies the fact that $\alpha^{min}_{\mathbf{k}}$ is not identically zero. In particular, for $T>T_c$ in this small neighbour, (\ref{Gap eqn again start}) has a nonzero solution, which contradicts $(2)$ of Theorem \ref{main1}.
\end{proof}

\begin{corollary}\label{uniqueness maximizer}
  For $0\leq T<T_c$, up to phase changes of $\alpha^{sym}$, $P^L_{\mathrm{BCS}}$ has a unique maximizer $(\gamma^{sym},\alpha^{sym})$, and this maximizer can be chosen so that $\alpha^{sym}_{k,\uparrow}$ is positive everywhere. For $T\geq T_c$, $P^L_{\mathrm{BCS}}$ has a unique maximizer $(\gamma^{sym},0)$.
\end{corollary}

\begin{proof}
  The continuous version is proved in \cite{critical} for $T=0$ and later in \cite{henheik2025universal} for general temperature, without proving the uniqueness of the gap equation. For the discrete version, this is a direct corollary of Theorem \ref{main1} and Propositions \ref{discrete Gap eqn} and \ref{relation sln and minimizer}.
\end{proof}

\begin{remark}\label{remark continuous}
  All the propositions, lemmas and corollaries in Section \ref{provemain1} can be correspondingly extended to the case $L=\infty$. In fact, some have already been proved, in like \cite{BCSgeneralpair,yang1991bardeen,critical,henheik2025universal}. The only difference is that (\ref{spectrum ess}) becomes
  \begin{equation}\label{spectrum ess continuous}
    \sigma_{ess}\big(K^{\Delta}_{T}+V\big)=
    \sigma_{ess}\big(K^{\Delta}_{T}\big)\subset[2T,\infty).
  \end{equation}
\end{remark}

\subsection{Conclusion}\label{conclusion}

\par Theorem \ref{main1} is a combination of Proposition \ref{existence of maximal sln}, Section \ref{uniqueness sec} and Corollary \ref{sln critical}. Corollary \ref{uniqueness maximizer intro} is a combination of Corollary \ref{uniqueness maximizer} and Propositions \ref{discrete Gap eqn} and \ref{relation sln and minimizer}.

\section{Translation Invariance and $SU(2)$ Symmetry}\label{persist}

\par In this section, we let $g$ be any admissible quasi-free state with its generalized 1-pdm $M$ having the form (\ref{2times2}). For simplicity, we might as well w.l.g. require $\gamma$ satisfies (\ref{condition gamma}). Recalling (\ref{generalized BCS})
\begin{equation}\label{generalized BCS again}
  \begin{aligned}
  -L^3P^{L}[g]=&\sum_{k,\sigma}\big(Q(ik)-\mu\big)\gamma_{\mathbf{k}\mathbf{k}}
  -\frac{1}{\beta}S_{\mathrm{HF}}[g]\\
  &+\frac{1}{2}
  \int_{(\Lambda_L\times\mathcal{S}_{2})^2}v^L(x-y)
  \vert\alpha(-x,\sigma;y,\nu)\vert^2.
  \end{aligned}
\end{equation}
In particular, the last term is real-valued. Also recall that (\ref{entropy equivalent definition})
\begin{equation}\label{entropy equivalent def again}
  S_{\mathrm{HF}}[g]=-\frac{1}{2}\tr_{\mathcal{H}\oplus\mathcal{H}}\big\{M\ln M
   +(1-M)\ln(1-M)\big\}
\end{equation}

\par On the other hand, for $L>C(v,\mu,Q)$ and $0\leq T<T_c$, using Theorem \ref{main1} (in fact Corollary \ref{uniqueness maximizer} is enough), Proposition \ref{discrete Gap eqn}, and the discussion around (\ref{alpha positve}), we choose $(\gamma^{sym},\alpha^{sym})$ (or equivalently using Proposition \ref{1to1correspondencequasifree1-pdm}, the quasi-free state $g^{sym}$ satisfies (\ref{assumption})) be the unique minimizer (up to phase changes of $\alpha^{sym}$) of $-L^3P_{\mathrm{BCS}}^L$ such that $\alpha^{sym}_{k,\uparrow}>0$. Notice that $(\gamma^{sym},\alpha^{sym})$ satisfies the translation invariance and $SU(2)$ symmetry condition (\ref{assumption}), and in particular,
\begin{equation}\label{alpha^sym}
  \alpha^{sym}(x,\sigma;y,\nu)=\frac{1}{L^{\frac{3}{2}}}\delta_{\sigma,-\nu}\alpha^{sym}
  (x+y,\sigma).
\end{equation}

\begin{proof}
\par Now we calculate the difference
\begin{equation}\label{difference 1}
  \begin{aligned}
  &-L^3P^{L}[g]+L^3P^{L}_{\mathrm{BCS}}[g^{sym}]=-L^3P^{L}[g]+L^3P^{L}[g^{sym}]\\
  =&
  \tr_{\mathcal{H}}\big(Q(\nabla)-\mu\big)(\gamma-\gamma^{sym})\\
  &+\frac{1}{2\beta}\tr_{\mathcal{H}\oplus\mathcal{H}}\big\{M\ln M
   +(1-M)\ln(1-M)\big\}\\
   &-\frac{1}{2\beta}\tr_{\mathcal{H}\oplus\mathcal{H}}\big\{M^{sym}\ln M^{sym}
   +(1-M^{sym})\ln(1-M^{sym})\big\}\\
   &+\frac{1}{2}\re
  \int_{(\Lambda_L\times\mathcal{S}_{2})^2}v^L(x-y)\Big(
  \vert\alpha(-x,\sigma;y,\nu)\vert^2-\vert\alpha^{sym}(-x,\sigma;y,\nu)\vert^2\Big).
  \end{aligned}
\end{equation}
Using an observation in \cite[Section 4]{persist}, we rewrite the last term of (\ref{difference 1}) by
\begin{equation}\label{difference 2}
  \begin{aligned}
  &\frac{1}{2}\int_{(\Lambda_L\times\mathcal{S}_{2})^2}v^L(x-y)
  \big\vert\alpha(-x,\sigma;y,\nu)-\alpha^{sym}(-x,\sigma;y,\nu)\big\vert^2\\
&-\int_{(\Lambda_L\times\mathcal{S}_{2})^2}v^L(x-y)
  \big\vert\alpha^{sym}(-x,\sigma;y,\nu)\big\vert^2\\
  &+\frac{1}{2}\re\int_{(\Lambda_L\times\mathcal{S}_{2})^2}v^L(x-y)\Big\{
  \alpha\overline{\alpha^{sym}}+\overline{\alpha}\alpha^{sym}\Big\}(-x,\sigma;y,\nu)
  \end{aligned}
\end{equation}
We let the trace of the diagonal entries of an operator $A$ on $\mathcal{H}\oplus\mathcal{H}$ (see for example \cite{frank2012microscopic,persist}):
\begin{equation}\label{Tr_0}
  \tr_0A=\tr_{\mathcal{H}\oplus\mathcal{H}}\big(P_0AP_0+Q_0AQ_0\big)
\end{equation}
where
\begin{equation}\label{P_0, Q_0}
  P_0=\begin{pmatrix}
        1 & 0 \\
        0 & 0
      \end{pmatrix},\quad
  Q_0=\begin{pmatrix}
        0 & 0 \\
        0 & 1
      \end{pmatrix}.
\end{equation}
Notice that since $M\ln M,(1-M)\ln(1-M)\leq 0$, we have
\begin{equation}\label{tr_0 equality}
  \tr_0\big\{M\ln M
   +(1-M)\ln(1-M)\big\}=
   \tr_{\mathcal{H}\oplus\mathcal{H}}\big\{M\ln M
   +(1-M)\ln(1-M)\big\}.
\end{equation}
The same equality holds with $M$ being replaced by $M^{sym}$. Recall the definition of $H_\Delta$ in (\ref{H_Delta}), one can check, after a tedious computation, that
\begin{equation}\label{difference 3}
  \begin{aligned}
  &\tr_{\mathcal{H}}\big(Q(\nabla)-\mu\big)(\gamma-\gamma^{sym})\\
  &-\int_{(\Lambda_L\times\mathcal{S}_{2})^2}v^L(x-y)
  \big\vert\alpha^{sym}(-x,\sigma;y,\nu)\big\vert^2\\
  &+\frac{1}{2}\re\int_{(\Lambda_L\times\mathcal{S}_{2})^2}v^L(x-y)\Big\{
  \alpha\overline{\alpha^{sym}}+\overline{\alpha}\alpha^{sym}\Big\}(-x,\sigma;y,\nu)\\
  =&\frac{1}{2}\tr_0H_\Delta\big(M-M^{sym}\big).
  \end{aligned}
\end{equation}
On the other hand, (\ref{M=1/1+exp(H/T)}) implies
\begin{equation}\label{idk}
  \frac{1}{\beta}\ln\frac{1-M^{sym}}{M^{sym}}=H_{\Delta}.
\end{equation}
Combining (\ref{difference 1})-(\ref{idk}) we get to
\begin{equation}\label{relative entropy sym}
  \begin{aligned}
  &-L^3P^{L}[g]+L^3P^{L}_{\mathrm{BCS}}[g^{sym}]=-L^3P^{L}[g]+L^3P^{L}[g^{sym}]\\
  =&\frac{1}{2}\int_{(\Lambda_L\times\mathcal{S}_{2})^2}v^L(x-y)
  \big\vert\alpha(-x,\sigma;y,\nu)-\alpha^{sym}(-x,\sigma;y,\nu)\big\vert^2\\
  &+\frac{1}{\beta}S_0(M,M^{sym}),
  \end{aligned}
\end{equation}
where
\begin{equation}\label{relative entropy sym 1}
  \begin{aligned}
  &\frac{1}{\beta}S_0(M,M^{sym})\\
  \coloneqq&\frac{1}{2\beta}\Big\{
  \tr_0M\big(\ln M-\ln M^{sym}\big)+\tr_0(1-M)\big(
  \ln(1-M)-\ln(1-M^{sym})\big)\Big\}.
  \end{aligned}
\end{equation}
Using a generalization of Klein's inequality in \cite[Lemma 5]{frank2012microscopic} (see also \cite[Proposition 4.2]{persist}), (\ref{relative entropy sym 1}) can be bounded from below by
\begin{equation}\label{Klein's inequality gen}
  \frac{1}{\beta}S_0(M,M^{sym})\geq \frac{1}{2}\tr_{\mathcal{H}\oplus\mathcal{H}}
  \Big\{\big(M-M^{sym}\big)
  \frac{H_{\Delta}}{\tanh\frac{\beta}{2}H_{\Delta}}\big(M-M^{sym}\big)\Big\}.
\end{equation}
From the observation in the proof of Corollary \ref{coro minimizer}, we have
\begin{equation}\label{Klein 2}
  \frac{H_{\Delta}}{\tanh\frac{\beta}{2}H_{\Delta}}=\begin{pmatrix}
                                                      K^{\Delta}_T & 0 \\
                                                      0 & K^{\Delta}_T
                                                    \end{pmatrix}
\end{equation}
where $K^\Delta_T$ is given in (\ref{K^delta_T(k)}) and (\ref{K^delta_T}) but with the additional spin variables, i.e.
  \begin{equation}\label{K^delta_T(k) sigma}
    K^{\Delta}_{T}(k,\sigma)\coloneqq\frac{E_{k,\sigma}}{\tanh\frac{\beta}{2}E_{k,\sigma}},
  \end{equation}
  and $K^{\Delta}_T$ on $\mathcal{H}$ given by
  \begin{equation}\label{K^delta_T sigma}
   \frac{1}{L^{\frac{3}{2}}}\int_{\Lambda_L} \big(K^{\Delta}_Tf\big)(x,\sigma)e^{-ipx}dx
   =\frac{K^{\Delta}_{T}(k,\sigma)}{L^{\frac{3}{2}}}\int_{\Lambda_L}f(x,\sigma)e^{-ipx}dx
  \end{equation}
  for all $(k,\sigma)\in(2\pi/L)\mathbb{Z}^3\times\mathcal{S}_{2}$. Notice that by definition $K_T^\Delta\geq 2T$.

  \par Plugging (\ref{Klein's inequality gen}) and (\ref{Klein 2}) into (\ref{relative entropy sym}), we get to
  \begin{equation}\label{difference 4}
    \begin{aligned}
    &-L^3P^{L}[g]+L^3P^{L}_{\mathrm{BCS}}[g^{sym}]=-L^3P^{L}[g]+L^3P^{L}[g^{sym}]\\
    \geq&\frac{1}{2}\int_{\Lambda_L\times\mathcal{S}_{2}}
    \Big\langle\big(K^\Delta_T+V_x\big)(\alpha-\alpha^{sym}),
    (\alpha-\alpha^{sym})\Big\rangle_{y,\nu}dyd\nu\\
    &+\frac{1}{2}\int_{\Lambda_L\times\mathcal{S}_{2}}
    \Big\langle\big(K^\Delta_T+V_y\big)(\alpha-\alpha^{sym}),
    (\alpha-\alpha^{sym})\Big\rangle_{x,\sigma}dxd\sigma\\
    &+\tr_{\mathcal{H}}K^\Delta_T(\gamma-\gamma^{sym})^2,
    \end{aligned}
  \end{equation}
  where
  \begin{equation}\label{define V_y}
    V_yf(x,\sigma)\coloneqq\frac{1}{2}v^L(x+y)f(x,\sigma)
  \end{equation}
  is an self-adjoint operator on $\mathcal{H}$ (due to $\hat{v}\leq0$ and $v$ is radially symmetric), and $V_x$ is defined analogously.

  \par Since $\mathcal{H}\cong L^2(\Lambda_L;\mathbb{C}^{2})\eqqcolon\bigoplus_{\sigma}L^2(\Lambda_L)_{\sigma}$, it is easy to see from the definition that $K_T^\Delta$ and $V_y$ both act separately on  each $L^2(\Lambda_L)_{\sigma}$. Therefore we can write
  \begin{equation}\label{separate}
    K^\Delta_T+V_y=\bigoplus_{\sigma\in\mathcal{S}_{2}}
    \big(K^\Delta_T+V_y\big)_{\sigma}.
  \end{equation}
   From Lemma \ref{K^delta_T+V>=0 lemma}, one can easily check that for fixed $\sigma\in\mathcal{S}_{2}$, $0$ is the non-degenerated ground state of the self-adjoint operator $(K^\Delta_T+V_y)_\sigma$, whose corresponding eigenfunction is $\alpha^{sym}(\cdot+y,\sigma)$ given in (\ref{alpha^sym}). The same result also holds for $K_T^\Delta+V_x$. From this observation, we have
   \begin{equation}\label{difference 5}
     \begin{aligned}
    &-L^3P^{L}[g]+L^3P^L_{\mathrm{BCS}}[g^{sym}]=-L^3P^{L}[g]+L^3P^L[g^{sym}]\\
    \geq&\frac{1}{2}\int_{\Lambda_L\times\mathcal{S}_{2}}
    \Big\langle\big(K^\Delta_T+V_y\big)\alpha,
    \alpha\Big\rangle_{x,\sigma}dyd\nu
    +\frac{1}{2}\int_{\Lambda_L\times\mathcal{S}_{2}}
    \Big\langle\big(K^\Delta_T+V_x\big)\alpha,
    \alpha\Big\rangle_{y,\nu}dxd\sigma\\
    &+\tr_{\mathcal{H}}K^\Delta_T(\gamma-\gamma^{sym})^2\\
    \geq&0.
    \end{aligned}
   \end{equation}

   \par When $0<T<T_c$, using the discussion around (\ref{separate}), and the fact that $K^T_\Delta\geq2T>0$, we know that $-L^3P^{L}[g]+L^3P_{\mathrm{BCS}}[g^{sym}]=0$ if and only if $\gamma=\gamma^{sym}$ and there exists some complex constants $c_1(y,\sigma,\nu)$ and $c_2(x,\sigma,\nu)$ such that
   \begin{equation}\label{difference alpha}
     \alpha(x,\sigma;y,\nu)=c_1(y,\sigma,\nu)\alpha^{sym}(x+y,\sigma)
     =c_2(x,\sigma,\nu)\alpha^{sym}(x+y,\nu).
   \end{equation}
   Notice that the difference between $\alpha^{sym}(x+y,\nu)$ and $\alpha^{sym}(x+y,\sigma)$ is a complex constant factor due to Corollary \ref{uniqueness maximizer}, thus there exists some complex constant $c(\sigma,\nu)$,
   \begin{equation}\label{difference alpha 2}
     \alpha(x,\sigma;y,\nu)=c(\sigma,\nu)\alpha^{sym}(x+y,\sigma).
   \end{equation}
   Since $\alpha^{sym}_{k,\sigma}=\alpha^{sym}_{-k,\sigma}$ (see Section \ref{discrete sec}), using (\ref{difference alpha 2}) and (\ref{property of gammakm and alphakm}), we must have $c(\sigma,\sigma)=0$. Therefore, we conclude that $(\gamma,\alpha)$ satisfies (\ref{assumption}). From (\ref{assumption}) and Corollary \ref{uniqueness maximizer}, we conclude that
\begin{equation}\label{difference alpha 3}
     \alpha(x,\sigma;y,\nu)=c\alpha^{sym}(x,\sigma;y,\nu)
     =\frac{c}{L^{\frac{3}{2}}}\delta_{\sigma,-\nu}\alpha^{sym}
  (x+y,\sigma).
   \end{equation}
   for some complex constant $\vert c\vert=1$. That is, the minimizer of $-L^3P^L$ and $-L^3P^L_{\mathrm{BCS}}$ are the same, up to phase changes of $\alpha$. Therefore, Theorem \ref{main2} is concluded for $0<T<T_c$.

\par Finally, for the case $T=0$, we only need to notice additionally
\begin{equation}\label{K_0>0}
  K^{\Delta}_{0}(k,\sigma)=\sqrt{\big(Q(ik)-\mu\big)^2+\vert\Delta_{k,\sigma}\vert^2}.
\end{equation}
From Corollary \ref{lower bound gap Delta}, we also have for fixed $L>C(v,\mu,Q)$, $K^{\Delta}_{0}>0$. The rest of the proof follows similarly to the $0<T<T_c$ case.

\end{proof}

\section{Large $L$ limit}\label{largeL}

\subsection{The minimal energy computation}
\
\begin{proof}[Proof of (\ref{value min -P_BCS^L}) and (\ref{value min -P_BCS})]
\
\par Using (\ref{Pressure BCS}) and Corollary \ref{uniqueness maximizer intro}, we have
\begin{equation}\label{cal discrete kinetic}
  \begin{aligned}
  \sum_{k,\sigma}\big(Q(ik)-\mu\big)\gamma^{sym}_{k,\sigma}=
  \frac{1}{2}\Big(\xi_k-\frac{\xi_k^2}{E_{k,\sigma}}
  \tanh\frac{\beta}{2}E_{k,\sigma}\Big),
  \end{aligned}
\end{equation}
and
\begin{equation}\label{cal discrete potential}
  \begin{aligned}
  \frac{1}{2}\int_{\Lambda_L\times\mathcal{S}_{2}}v^L(x)\vert\alpha^{sym}(x,\sigma)
  \vert^2dxd\sigma=&
  \frac{1}{2L^3}\sum_{k,k^\prime,\sigma}\hat{v}(k-k^\prime)\alpha^{sym}_{k^\prime,\sigma}
  \alpha^{sym}_{k,\sigma}\\
  =&-\frac{1}{4}\sum_{k,\sigma}\frac{\vert\Delta_{k,\sigma}\vert^2}{E_{k,\sigma}}
  \tanh\frac{\beta}{2}E_{k,\sigma}.
  \end{aligned}
\end{equation}
Moreover, the two eigenvalues of $M(\mathbf{k})$ given in (\ref{M(k)}) is $\lambda_{k,\sigma}$ and $1-\lambda_{k,\sigma}$, where
\begin{equation}\label{eigenvalues M(k)}
  \lambda_{k,\sigma}=\frac{1}{2}\Big(1+\tanh\frac{\beta}{2}E_{k,\sigma}\Big).
\end{equation}
Combining (\ref{cal discrete kinetic}) and (\ref{cal discrete potential}), we have
\begin{equation}\label{cal discrete together}
  \begin{aligned}
  &\sum_{k,\sigma}\big(Q(ik)-\mu\big)\gamma^{sym}_{k,\sigma}
  +\frac{1}{2}\int_{\Lambda_L\times\mathcal{S}_{2}}v^L(x)\vert\alpha^{sym}(x,\sigma)
  \vert^2\\
  =&\sum_{k,\sigma}\frac{E_{k,\sigma}}{2}\Big(1-\tanh\frac{\beta}{2}E_{k,\sigma}\Big)
  +\frac{1}{2}\sum_{k,\sigma}\big(\xi_k-E_{k,\sigma}\big)+\frac{1}{4}
  \sum_{k,\sigma}\frac{\vert\Delta_{k,\sigma}\vert^2}{E_{k,\sigma}}
  \tanh\frac{\beta}{2}E_{k,\sigma}.
  \end{aligned}
\end{equation}
On the other hand, using (\ref{eigenvalues M(k)}), we have
\begin{equation}\label{cal discrete together2}
  \begin{aligned}
  &\sum_{k,\sigma}\frac{E_{k,\sigma}}{2}\Big(1-\tanh\frac{\beta}{2}E_{k,\sigma}\Big)
  +\frac{1}{\beta}\sum_{k,\sigma}\tr_{\mathbb{C}^2}M(\mathbf{k})\ln M(\mathbf{k})\\
  =&-\frac{1}{\beta}\sum_{k,\sigma}\ln\big(1+e^{-\beta E_{k,\sigma}}\big).
  \end{aligned}
\end{equation}
Combining (\ref{cal discrete together}) and (\ref{cal discrete together2}) we have (\ref{value min -P_BCS^L}). We remark that all the terms in (\ref{cal discrete kinetic})-(\ref{cal discrete together2}) are summable by Lemmas \ref{decay Delta} and (\ref{check tr gamma lemma}). (\ref{value min -P_BCS}) can be calculated similarly using (\ref{pressure BCS continuous}) and Corollary \ref{uniqueness maximizer intro}.

\end{proof}

\subsection{Derivation of the continuous BCS functional}
\
\begin{proof}[Proof of (\ref{limit model})]
\
\par For $0\leq T<T_c(v,\mu,Q,\infty)$, by Corollary \ref{uniqueness maximizer intro}, we can choose the unique maximizer, for $\mathbf{k}=(k,\sigma)\in\mathbb{R}^3\times\mathcal{S}_{2}$,
 \begin{align}
  \gamma_{k,\sigma}=&\frac{1}{2}\Big(1-\frac{\xi_k}{E_{k,\sigma}}\tanh\frac{\beta}{2}
  E_{k,\sigma}\Big),\label{Gap gamma limit}\\
  \alpha_{k,\sigma}=&\frac{\Delta_{k,\sigma}}{2E_{k,\sigma}}\tanh\frac{\beta}{2}E_{k,\sigma},
  \label{Gap alpha limit}
  \end{align}
  where
  \begin{align}
  \Delta_{k,\sigma}&=-\frac{1}{(2\pi)^3}\int_{\mathbb{R}^3}\hat{v}(k-k^\prime)
  \alpha_{k^\prime,\sigma}dk^\prime,\label{define Delta limit}\\
  \xi_k&=\big(Q(ik)-\mu\big),\label{define xi limit}\\
  E_{k,\sigma}&=\sqrt{\xi_k^2+\vert\Delta_{k,\sigma}\vert^2},\label{define E_k limit}
  \end{align}
  and $\Delta_{k,\uparrow}>0$, $\Delta_{k,\downarrow}<0$ satisfies equation (\ref{gap eqn intro continuous}), or (\ref{gap eqn intro continuous T=0}) when $T=0$. It is straight-forward to check that $\Delta$ is bounded continuous, since $Q$ satisfies Assumption \ref{assumption Q}, $v$ satisfies Assumption \ref{assumption v}, and
  \begin{equation*}
    \frac{\Delta_{k,\sigma}}{2E_{k,\sigma}}\tanh\frac{\beta}{2}E_{k,\sigma}
  \end{equation*}
is bounded. Hence $\gamma_{k,\sigma}$ and $\alpha_{k,\sigma}$ are also continuous in $k$. We also let
\begin{equation}\label{noninteract gamma continuous}
  \gamma_{0,\mathbf{k}}=\frac{1}{1+e^{\beta\xi_k}}
\end{equation}
be the maximizer of non-interacting $P_{\mathrm{BCS}}$. $(\gamma_{0,\mathbf{k}},0)$ is also the maximizer of non-interacting $P_{\mathrm{BCS}}^L$ when $\mathbf{k}\in(2\pi/L)\mathbb{Z}^3\times\mathcal{S}_{2}$. We let $g_0$ be the corresponding quasi-free state on $\mathcal{L}(\mathcal{F}_L)$\footnote{Here the subscript $L$ indicates the $L$ dependence.} (as already discussed in Section \ref{discrete sec}).

\par For $\mathbf{k}=(k,\sigma)\in(2\pi/L)\mathbb{Z}^3\times\mathcal{S}_{2}$, we define
\begin{equation}\label{test discrete}
  \tilde{\gamma}_{\mathbf{k}}=\gamma_{\mathbf{k}},\quad
  \tilde{\alpha}_{\mathbf{k}}=\alpha_{\mathbf{k}}.
\end{equation}
By the proof of Lemma \ref{check tr gamma lemma}, it is easy to check
\begin{equation}\label{tr finite discrete}
  \sum_{\sigma}\sum_{k\in(2\pi/L)\mathbb{Z}^3}\tilde{\gamma}_{\mathbf{k}}<\infty.
\end{equation}
Then (\ref{define alpha gamma function assumption}) generates two functions $\gamma(\mathbf{x}),\alpha(\mathbf{x})$ on $\mathbf{x}\in\Lambda_L\times\mathcal{S}_{2}$, and two operators $\gamma,\alpha$ on $L^2\big(\Lambda_L\times\mathcal{S}_{2}\big)$ with integral kernel given by (\ref{rewritten alpha gamma function}). By Proposition \ref{1to1correspondencequasifree1-pdm}, $(\tilde{\gamma},\tilde{\alpha})$ generates a quasi-free state $\tilde{g}$ on $\mathcal{L}(\mathcal{F}_L)$. Testing $P^L_{\mathrm{BCS}}$ with $\tilde{g}$, we have
\begin{equation}\label{test g tilde discrete}
  \begin{aligned}
  -P^L_{\mathrm{BCS}}[\tilde{g}]=&\frac{1}{L^3}\sum_{k\in(2\pi/L)\mathbb{Z}^3
  ,\sigma}\big(Q(ik)-\mu\big)\gamma_{\mathbf{k}}+\frac{1}{\beta L^3}\sum_{k\in(2\pi/L)\mathbb{Z}^3
  ,\sigma}\tr_{\mathbb{C}^2}M(\mathbf{k})\ln M(\mathbf{k})\\
  &+\frac{1}{2L^6}\sum_{k,k^\prime\in(2\pi/L)\mathbb{Z}^3}\sum_{\sigma}
  \hat{v}(k-k^\prime)\alpha_{k^\prime,\sigma}\bar{\alpha}_{k,\sigma}
  \geq\min-P^L_{\mathrm{BCS}}.
  \end{aligned}
\end{equation}
Since $\Delta$ is uniformly bounded by Lemma \ref{coro super}, we can apply dominated convergence theorem to get to
\begin{equation}\label{limit upper bound}
  \min-P_{\mathrm{BCS}}\geq\limsup_{L\to\infty}\big(\min-P^L_{\mathrm{BCS}}\big).
\end{equation}
Notice that for $0\leq T<T_c(v,\mu,Q,\infty)$,
\begin{equation}\label{limite non interacting}
  \lim_{L\to\infty}-P^L_{\mathrm{BCS}}[g_0]=
  -P_{\mathrm{BCS}}[\gamma_0,0]>\min-P_{\mathrm{BCS}}.
\end{equation}
Thus there exists a constant $C^\prime(v,\mu,Q)>C(v,\mu,Q)$, such that for $L>C^\prime(v,\mu,Q)$,
\begin{equation}\label{limite non interacting2}
  -P^L_{\mathrm{BCS}}[g_0]>\min-P^L_{\mathrm{BCS}}.
\end{equation}
This implies when $L>C^\prime(v,\mu,Q)$, equation (\ref{Gap eqn intro}), or (\ref{Gap eqn intro T=0}) when $T=0$, has a non-vanishing solution. By Theorem \ref{main1}, $T<T_c(v,\mu,Q,L)$, thus $T_c(v,\mu,Q,\infty)\leq T_c(v,\mu,Q,L)$.

\par On the other hand, for $L>C^\prime(v,\mu,Q)$ and $0\leq T<T_c(v,\mu,Q,\infty)\leq T_c(v,\mu,Q,L)$, by Corollary \ref{uniqueness maximizer intro}, we can choose the unique maximizer, for $\mathbf{k}\in(2\pi/L)\mathbb{Z}^3\times\mathcal{S}_{2}$,
\begin{align}
  \gamma^{L}_{k,\sigma}=&\frac{1}{2}\Big(1-\frac{\xi_k}{E_{k,\sigma}}\tanh\frac{\beta}{2}
  E_{k,\sigma}\Big),\label{Gap gamma limit2}\\
  \alpha^{L}_{k,\sigma}=&\frac{\Delta^{L}_{k,\sigma}}
  {2E_{k,\sigma}}\tanh\frac{\beta}{2}E_{k,\sigma},
  \label{Gap alpha limit2}
  \end{align}
where $\Delta^{L}_{k,\uparrow}>0$, $\Delta^{L}_{k,\downarrow}<0$ solves (\ref{Gap eqn intro}) when $T>0$ or (\ref{Gap eqn intro T=0}) when $T=0$. By Lemma \ref{check tr gamma lemma} and Proposition \ref{1to1correspondencequasifree1-pdm}, we let $g^L$ denote the quasi-free state on $\mathcal{L}(\mathcal{F}_L)$ induced by $(\gamma^{L},\alpha^{L})$.

\par We let
\begin{equation}\label{define C_k^L proof}
    C_k^L\coloneqq [-\pi/L,\pi/L)^3+k,\quad k\in(2\pi/L)\mathbb{Z}^3,
  \end{equation}
  such that $\mathbb{R}^3=\bigcup_k C_k^L$. Then we define for $(k,\sigma)\in\mathbb{R}^3\times\mathcal{S}_{2}$,
  \begin{equation}\label{test continuous}
    \tilde{\gamma}^L_{k,\sigma}=\sum_{k^\prime\in(2\pi/L)\mathbb{Z}^3}
    \gamma^{L}_{k,\sigma}\chi_{ C_{k^\prime}^L}(k),\quad
    \tilde{\alpha}^L_{k,\sigma}=\sum_{k^\prime\in(2\pi/L)\mathbb{Z}^3}
    \alpha^{L}_{k,\sigma}\chi_{ C_{k^\prime}^L}(k).
  \end{equation}
By Lemma \ref{check tr gamma lemma}, we can check that in fact $(\tilde{\gamma}^L,\tilde{\alpha}^L)\in\mathcal{D}$. Therefore, by direct calculation, we have
\begin{equation}\label{test gamma^L alpha^L continuous}
  -P_{\mathrm{BCS}}[\tilde{\gamma}^L,\tilde{\alpha}^L]=-P^L_{\mathrm{BCS}}[g^L]
  =\min-P^L_{\mathrm{BCS}}\geq \min-P_{\mathrm{BCS}}.
\end{equation}
Therefore,
\begin{equation}\label{limit lower bound}
  \liminf_{L\to\infty}\big(\min-P^L_{\mathrm{BCS}}\big)\geq\min-P_{\mathrm{BCS}}.
\end{equation}
Combining (\ref{limit upper bound}) and (\ref{limit lower bound}), we reach (\ref{limit model}).

\end{proof}

\subsection{Convergence of the gap equation and the critical temperature}
\
\par We first prove (\ref{limit delta}) and split the proof into several lemmas below. The proof of (\ref{limit critical temperature}) also follows based on the ideas in those lemmas.

\par  Recall that for $L>C(v,\mu,Q)$ and any $T\geq0$, $\tilde{\Delta}^L$ is defined by
\begin{equation}\label{define tildeDelta^L proof}
    \tilde{\Delta}^L_{k,\sigma}\coloneqq-\frac{1}{2L^3}\sum_{k^\prime\in(2\pi/L)\mathbb{Z}^3}
    \hat{v}(k-k^\prime)\frac{\Delta_{k^\prime,\sigma}^L}{E_{k^\prime,\sigma}}
    \tanh\frac{\beta}{2}E_{k^\prime,\sigma},\quad k\in\mathbb{R}^3.
  \end{equation}
Since $\Delta^L$ satisfies (\ref{Gap eqn intro}), or (\ref{Gap eqn intro T=0}) when $T=0$, $\tilde{\Delta}^L_{k,\sigma}={\Delta}^L_{k,\sigma}$ when $k\in(2\pi/L)\mathbb{Z}^3$. Also, since $\hat{v}\leq0$ and $\hat{v}(0)<0$, we can check that when $0\leq T<T_c(v,\mu,Q,L)$, $\tilde{\Delta}^L_{k,\uparrow}>0$ and $\tilde{\Delta}^L_{k,\downarrow}<0$.

\begin{lemma}\label{precompact}
  $\{\tilde{\Delta}^L\}$ is precompact in $L^\infty$, and hence there exists a subsequence $\tilde{\Delta}^{L_j}$ converges to $\tilde{\Delta}$ in $L^\infty$, which solves the gap equation (\ref{gap eqn intro continuous}) and $\tilde{\Delta}_{k,\uparrow}\geq0$, $\tilde{\Delta}_{k,\downarrow}\leq0$.
\end{lemma}
\begin{proof}
  \par By Assumptions \ref{assumption Q} and \ref{assumption v}, we can bound,
\begin{equation}\label{precompact 2}
\begin{aligned}
  \sup_{L>C(v,\mu,Q)}&\Vert\tilde{\Delta}^L_{k+h,\sigma}-\tilde{\Delta}^L_{k,\sigma}
  \Vert_{L_{k,\sigma}^\infty(\mathbb{R}^3\times
  \mathcal{S}_{2})}\\
  &\leq C_1(v,\mu,Q)\Vert\hat{v}(k+h)-\hat{v}(k)\Vert_{L_k^s(\mathbb{R}^3)}\stackrel{h\to0}
  {\longrightarrow}0,
\end{aligned}
\end{equation}
where
\begin{equation}\label{choose s}
  s=\left\{\begin{aligned}
  &\frac{3+\varepsilon}{3-d+\varepsilon},\quad 0<d\leq3,\\
  &\infty,\quad d>3,\\
  \end{aligned}\right.
\end{equation}
By (\ref{precompact 2}) and Lemma \ref{coro super}, $\tilde{\Delta}^L$ is bounded uniformly (independent of $L$):
\begin{equation}\label{precompact 1}
\begin{aligned}
  \sup_{L>C(v,\mu,Q)}\Vert\tilde{\Delta}^L\Vert_{L^\infty(\mathbb{R}^3\times
  \mathcal{S}_{2})}\leq&
  C_1(v,\mu,Q)\sup_{h\leq 2\pi/C(v,\mu,Q)}
  \Vert\hat{v}(k+h)-\hat{v}(k)\Vert_{L_k^s(\mathbb{R}^3)}\\
  &+\delta_0(v,\mu,Q)\\
  \leq& C_2(v,\mu,Q).
\end{aligned}
\end{equation}
Also, by Lemma \ref{decay Delta}, we have
\begin{equation}\label{precompact 3}
  \lim_{R\to\infty}\sup_{L>C(v,\mu,Q)}\Vert\tilde{\Delta}^L_{k,\sigma}\chi_{\vert k\vert>R}\Vert_{L_{k,\sigma}^\infty(\mathbb{R}^3\times
  \mathcal{S}_{2})}=0.
\end{equation}
Combining (\ref{precompact 2}), (\ref{precompact 1}) and (\ref{precompact 3}), we can find a subsequence of $\tilde{\Delta}^L$ converges to some $\tilde{\Delta}$ in $L^\infty$ (see for example \cite[Theorem IV.5.6]{dunford1988linear}), i.e. $\{\tilde{\Delta}^L\}$ is precompact in $L^\infty$. By dominated convergence theorem, $\tilde{\Delta}$ solves the gap equation (\ref{gap eqn intro continuous}) (the legitimacy of applying the dominated convergence theorem when $T=0$ is shown in (\ref{lower bound T=0 tilde}) below). Moreover, it is easy to check $\tilde{\Delta}_{k,\uparrow}\geq0$, $\tilde{\Delta}_{k,\downarrow}\leq0$.
\end{proof}

\begin{lemma}\label{convergence T>=T_c lemma}
  If $T\geq T_c(v,\mu,Q,\infty)$, then
  \begin{equation}\label{convergence T>=T_c}
    \lim_{L\to\infty}\Vert\tilde{\Delta}^L\Vert_{L^\infty}=0.
  \end{equation}
\end{lemma}

\begin{proof}
  \par  Notice that, when $T\geq T_c(v,\mu,Q,\infty)$, by Theorem \ref{main1}, we have $\tilde{\Delta}=\Delta\equiv0$, where $\tilde{\Delta}$ is given in Lemma \ref{precompact}. If (\ref{convergence T>=T_c}) is not true, we may find another subsequence $\{\Delta^{L_j}\}_j$, such that
\begin{equation}\label{contradiction T>=T_c}
  \Vert\Delta^{L_j}\Vert_{L^\infty}>c>0
\end{equation}
for some constant $c$ independent of $L_j$. However, Lemma \ref{precompact} and Theorem \ref{main1} implies that,
\begin{equation}\label{contradiction T>=T_c2}
  \lim_{k\to\infty}\Vert\Delta^{L_{j_k}}\Vert_{L^\infty}=0
\end{equation}
for some secondary subsequence, which leads to a contradiction. Hence (\ref{convergence T>=T_c}), i.e. (\ref{limit delta}) for $T\geq T_c(v,\mu,Q,\infty)$, must hold true .
\end{proof}

\begin{lemma}\label{convergence T=0 lemma}
  If $T=0$, then
  \begin{equation}\label{convergence T=0}
    \lim_{L\to\infty}\Vert\tilde{\Delta}^L-\Delta\Vert_{L^\infty}=0.
  \end{equation}
\end{lemma}

\begin{proof}

 \par When $T=0$, that is, $\beta=\infty$. We can then apply Corollary \ref{lower bound gap Delta} such that for any $k\in A^L$ ($A^L$ is defined in (\ref{annulus A})),
 \begin{equation}\label{lower bound T=0}
   \Delta^L_{k,\uparrow}>\varepsilon(v,\mu,Q)>0.
 \end{equation}
 By Lemma \ref{existence of subsolution} and the definition of $\tilde{\Delta}^L$, we also have for $k\in A$ ($A$ is defined in (\ref{annulus A continuous})),
  \begin{equation}\label{lower bound T=0 tilde}
   \tilde{\Delta}^L_{k,\uparrow}>\varepsilon(v,\mu,Q)>0.
 \end{equation}
 (\ref{lower bound T=0 tilde}) implies immediately $\tilde{\Delta}_{k,\uparrow}>0$. By the uniqueness of the positive solution given in Theorem \ref{main1}, we have $\tilde{\Delta}=\Delta$. Then (\ref{convergence T=0}), i.e. (\ref{limit delta}) for $T=0$, follows by using the uniqueness of positive solution in Theorem \ref{main1} and a reductio ad absurdum similar to that in Lemma \ref{convergence T>=T_c lemma}.
\end{proof}

\vspace{1em}

\par The proof of (\ref{limit delta}) for $0<T<T_c(v,\mu,Q,\infty)$ is slightly technical. We first introduce, for $0<T<\infty$, i.e. $0<\beta<\infty$, an operator $\mathcal{S}_{\beta}$ on the Banach space $C_0(\mathbb{R}^3)$ of continuous functions that vanish at infinity. Notice that the dual space of $C_0(\mathbb{R}^3)$ is the space $\mathcal{M}(\mathbb{R}^3)$ consisting of complex Radon measures on $\mathbb{R}^3$. We define for $g\in C_0(\mathbb{R}^3)$
\begin{equation}\label{define S}
  \big(\mathcal{S}_\beta g\big)(k)=-\frac{1}{2(2\pi)^3}\int_{\mathbb{R}^3}
    \hat{v}(k-k^\prime)\frac{g(k^\prime)}{\big\vert Q(ik^\prime)-\mu\big\vert}
    \tanh\frac{\beta}{2}\big\vert Q(ik^\prime)-\mu\big\vert dk^\prime,\quad k\in\mathbb{R}^3.
\end{equation}
For the properties of $\mathcal{S}_{\beta}$, we have:
\begin{lemma}\label{basic property of S_beta lemma}
  For $0<\beta<\infty$, $\mathcal{S}_\beta$ is a compact operator on $C_0(\mathbb{R}^3)$ with positive kernel, that is, for any $g\geq0$ not identically zero,
\begin{equation}\label{positive kernel}
  \mathcal{S}_\beta g>0.
\end{equation}
\end{lemma}
\begin{proof}
  \par Since $0<\beta<\infty$, it can be verified, similarly to proof of Lemma \ref{precompact}, that $\mathcal{S}_\beta$ is a compact operator on $C_0(\mathbb{R}^3)$. Notice that the positivity of $T$, i.e. finiteness of $\beta$ is crucial here, it guarantees the finiteness:
  \begin{equation*}
    \frac{\tanh\frac{\beta}{2}\big\vert Q(ik^\prime)-\mu\big\vert}{\big\vert Q(ik^\prime)-\mu\big\vert}\leq \frac{\beta}{2}.
  \end{equation*}

  \par For the verification of (\ref{positive kernel}), we just need to apply the observation at the beginning of Section \ref{provemain1}.
\end{proof}

\begin{lemma}\label{Krein-Rutman}
  If there exists a function $g\geq0$ in $C_0(\mathbb{R}^3)$ and not identically zero, and a positive constant $c>0$, such that $\mathcal{S}_\beta g=cg$, then $c=r(\mathcal{S}_\beta)$ is the largest and non-degenerated eigenvalue of $\mathcal{S}_\beta$ with $g$ being its eigenfunction. Here $r(\mathcal{S}_\beta)$ represents the spectral radius of $\mathcal{S}_\beta$. Moreover, $g>0$.
\end{lemma}
\begin{proof}
\par  By Lemma \ref{basic property of S_beta lemma} and Krein-Rutman theorem, $\lambda=r(\mathcal{S}_\beta)>0$ is the largest eigenvalue of $\mathcal{S}_\beta$, and it has a unique normalized eigenvector $h\in C_0(\mathbb{R}^3)$, which satisfies $h>0$. Similarly, by the definition of $\mathcal{S}_\beta$ (\ref{define S}) and Lemma \ref{basic property of S_beta lemma}, $\mathcal{S}_\beta^*$ is also a compact operator on $\mathcal{M}(\mathbb{R}^3)$ with positive kernel. We apply Krein-Rutman theorem to $\mathcal{S}_\beta^*$ and find its largest eigenvalue $\lambda=r(\mathcal{S}_\beta^*)=r(\mathcal{S}_\beta)$ and the corresponding unique eigenvector $\tau\in\mathcal{M}(\mathbb{R}^3)$ such that $\tau$ is a positive measure.

\par Since $\mathcal{S}_\beta g=g$, $g\geq 0$ and $g$ is not identically zero, then
\begin{equation}\label{lambda=1}
 0<\int_{\mathbb{R}^3}g(k)d\tau(k)=\frac{1}{\lambda}\int_{\mathbb{R}^3}
 g(k)d\big(S^*_\beta\tau\big)(k)=
 \frac{1}{\lambda}\int_{\mathbb{R}^3}
 S_\beta g d\tau(k)=\frac{c}{\lambda}\int_{\mathbb{R}^3}
  g(k)d\tau(k).
\end{equation}
Therefore $\lambda=c$. Since $\lambda$ is non-degenerated, we have $g=\kappa h$ for some $\kappa>0$ and hence $g>0$.
\end{proof}

\begin{lemma}\label{lamba>1}
  Let $\lambda=r(\mathcal{S}_\beta)>0$ be the largest and non-degenerated eigenvalue of $\mathcal{S}_\beta$. If there exists a function $g\geq0$ in $C_0(\mathbb{R}^3)$ and not identically zero, such that $\mathcal{S}_\beta g>g$, then $\lambda>1$.
\end{lemma}
\begin{proof}
  \par Let $h$ and $\tau$ be as in the proof of Lemma \ref{Krein-Rutman}, we have
  \begin{equation}\label{order lemma}
  \int_{\mathbb{R}^3}g(k)d\tau(k)=\frac{1}{\lambda}\int_{\mathbb{R}^3}
 g(k)d\big(S^*_\beta\tau\big)(k)=\frac{1}{\lambda}\int_{\mathbb{R}^3}
 S_\beta gd\tau(k)>\frac{1}{\lambda}\int_{\mathbb{R}^3}
  g(k)d\tau(k)>0.
\end{equation}
\end{proof}

\par With preparations above, we can now prove (\ref{limit delta}) for $0<T<T_c(v,\mu,Q,\infty)$.
\begin{lemma}\label{convergence 0<T<T_c lemma}
  If $0<T<T_c(v,\mu,Q,\infty)$, then
  \begin{equation}\label{convergence 0<T<T_c}
    \lim_{L\to\infty}\Vert\tilde{\Delta}^L-\Delta\Vert_{L^\infty}=0.
  \end{equation}
\end{lemma}

\begin{proof}
  \par When $0<T<T_c(v,\mu,Q,\infty)$, we first need to show $\tilde{\Delta}=\Delta$, i.e. $\tilde{\Delta}_{k,\uparrow}>0$. We argue by contradiction. Assuming $\tilde{\Delta}\neq\Delta$, then by Theorem \ref{main1}, we must have $\tilde{\Delta}\equiv0$. Since we have already proved below (\ref{limite non interacting2}) that, when $L>C^\prime(v,\mu,Q)$, $T_c(v,\mu,Q,\infty)\leq T_c(v,\mu,Q,L)$, we then define for $L>C^\prime(v,\mu,Q)$,
\begin{equation}\label{normalize Delta}
  f_{k,\sigma}^L=\frac{\tilde{\Delta}^L_{k,\sigma}}
  {\Vert\tilde{\Delta}_{k,\sigma}^L\Vert_{L^\infty}}.
\end{equation}
Obviously,
\begin{equation}\label{precompact f1}
  \sup_{L>C^\prime(v,\mu,Q)}\Vert f^L\Vert_{L^\infty(\mathbb{R}^3\times
  \mathcal{S}_{2})}=1<\infty.
\end{equation}
Since $f^L$ satisfies for $k\in\mathbb{R}^3$,
\begin{equation}\label{f^L eqn proof}
    f^L_{k,\sigma}=-\frac{1}{2L^3}\sum_{k^\prime\in(2\pi/L)\mathbb{Z}^3}
    \hat{v}(k-k^\prime)\frac{f_{k^\prime,\sigma}^L}{E_{k^\prime,\sigma}}
    \tanh\frac{\beta}{2}E_{k^\prime,\sigma},\quad k\in\mathbb{R}^3,
  \end{equation}
  where we still let
  \begin{equation}\label{E_k f^L}
    E_{k,\sigma}=\sqrt{\big(Q(ik)-\mu\big)^2+\vert\Delta^L_{k,\sigma}\vert^2},\quad
    k\in(2\pi/L)\mathbb{Z}^3,
  \end{equation}
we can bound similarly to (\ref{precompact 2}) that, for fixed temperature $T$,
\begin{equation}\label{precompact f2}
  \begin{aligned}
  \sup_{L>C^\prime(v,\mu,Q)}&\Vert f^L_{k+h,\sigma}-f^L_{k,\sigma}
  \Vert_{L_{k,\sigma}^\infty(\mathbb{R}^3\times
  \mathcal{S}_{2})}\\
  &\leq C_2(v,\mu,Q,T)\Vert\hat{v}(k+h)-\hat{v}(k)\Vert_{L_k^s(\mathbb{R}^3)}\stackrel{h\to0}
  {\longrightarrow}0,
\end{aligned}
\end{equation}
with $s$ chosen in (\ref{choose s}). By (\ref{f^L eqn proof}), we can deduce, in the way similar to Lemma \ref{decay Delta}, that
\begin{equation}\label{precompact f3}
  \lim_{R\to\infty}\sup_{L>C^\prime(v,\mu,Q)}\Vert f^L_{k,\sigma}\chi_{\vert k\vert>R}\Vert_{L_{k,\sigma}^\infty(\mathbb{R}^3\times
  \mathcal{S}_{2})}=0.
\end{equation}
(\ref{precompact f1})-(\ref{precompact f3}) together imply that there also exists a subsequence of $f^L$, which converges to some ${f}$ in $L^\infty$. It is easy to verify $f_{k,\uparrow}\geq0$ and $f_{k,\downarrow}\leq0$. We can w.l.g. assume (recall that we are now assuming $\tilde{\Delta}=0$)
\begin{equation}\label{w.l.g}
  \lim_{j\to\infty}\Vert\tilde{\Delta}^{L_j}\Vert_{L^\infty}=0,\quad
  \lim_{j\to\infty}\Vert f^{L_j}-f\Vert_{L^\infty}=0
\end{equation}
 By dominated convergence theorem, as $j$ tends to infinity, (\ref{f^L eqn proof}) becomes
\begin{equation}\label{f^L eqn proof L=infty}
    f_{k,\sigma}=-\frac{1}{2(2\pi)^3}\int_{\mathbb{R}^3}
    \hat{v}(k-k^\prime)\frac{f_{k^\prime,\sigma}}{\big\vert Q(ik^\prime)-\mu\big\vert}
    \tanh\frac{\beta}{2}\big\vert Q(ik^\prime)-\mu\big\vert dk^\prime,\quad k\in\mathbb{R}^3,
  \end{equation}

\par Let $f_{\sigma}(k)=f_{k,\sigma}$. Using (\ref{f^L eqn proof L=infty}), it is straightforward to check that $f_{\uparrow}\in C_0(\mathbb{R}^3)$, and $\mathcal{S}_\beta f_{\uparrow}=f_{\uparrow}$. Recall that $f_{\uparrow}\geq 0$ and $f_{\uparrow}\geq 0$ is not identically zero, then by Lemma \ref{Krein-Rutman}, $1=r(\mathcal{S}_\beta)$ is the largest and non-degenerated eigenvalue of $\mathcal{S}_\beta$.
%\begin{equation}\label{lambda=1 u}
% 0<\int_{\mathbb{R}^3}f_{\uparrow}(k)d\tau(k)=\frac{1}{\lambda}\int_{\mathbb{R}^3}
% f_{\uparrow}d\big(S^*_\beta\tau\big)(k)=
% \frac{1}{\lambda}\int_{\mathbb{R}^3}
% S_\beta f_{\uparrow}d\tau(k)=\frac{1}{\lambda}\int_{\mathbb{R}^3}
%  f_{\uparrow}d\tau(k).
%\end{equation}
\par However, since $T<T_c(v,\mu,Q,\infty)$, by Theorem \ref{main1}, we have $\Delta_{\uparrow}\in C_0(\mathbb{R}^3)$ solves (\ref{gap eqn intro continuous again}) and $\Delta_{\uparrow}>0$. By monotonicity, we have (recall that $\mathcal{M}$ is defined in (\ref{op gap eqn continuous})):
\begin{equation}\label{order}
  \mathcal{S}_\beta\Delta_{\uparrow}>\mathcal{M}[\Delta_{\uparrow}]=\Delta_{\uparrow}.
\end{equation}
By Lemma \ref{lamba>1}, we have $r(\mathcal{S}_\beta)>1$, which is absurd. Therefore, we must have $\tilde{\Delta}=\Delta$. A uniqueness argument leads to (\ref{convergence 0<T<T_c}) i.e.  (\ref{limit delta}) for $0<T<T_c(v,\mu,Q,\infty)$.

%Since $\lambda=1$, we get to
%\begin{equation}\label{conrtradiction complete}
%  \int_{\mathbb{R}^3}\Delta_{\uparrow}(k)d\tau(k)=\int_{\mathbb{R}^3}
% \Delta_{\uparrow}d\big(S^*_\beta\tau\big)(k)=\int_{\mathbb{R}^3}
% S_\beta \Delta_{\uparrow}d\tau(k)>\int_{\mathbb{R}^3}
%  \Delta_{\uparrow}d\tau(k)>0,
%\end{equation}

\end{proof}

\vspace{1em}

\par Now we can complete the proofs of (\ref{limit critical temperature}) and  (\ref{limit delta})

\begin{proof}[Proof of (\ref{limit critical temperature}) and  (\ref{limit delta})]
\

\par (\ref{limit delta}) follows directly from Lemmas \ref{convergence T>=T_c lemma}, \ref{convergence T=0 lemma} and \ref{convergence 0<T<T_c lemma}.

\par To show (\ref{limit critical temperature}), we argue by contradiction. Since we have already proved below (\ref{limite non interacting2}) that, when $L>C^\prime(v,\mu,Q)$, $T_c(v,\mu,Q,\infty)\leq T_c(v,\mu,Q,L)$, we can w.l.g. assume
\begin{equation}\label{contradiction assumption}
  \lim_{L\to\infty}T_c(v,\mu,Q,L)>T_c(v,\mu,Q,\infty).
\end{equation}
Then there exist two temperatures $T_1$ and $T_2$ such that, when $L>C^{\prime\prime}(v,\mu,Q)$, $T_c(v,\mu,Q,\infty)<T_1<T_2<T_c(v,\mu,Q,L)$. On the other hand, we have already proved that when $T\geq T_c(v,\mu,Q,\infty)$, $\tilde{\Delta}^L$ converges to $0$ in $L^\infty$ norm. Therefore we can also apply the argument in Lemma \ref{convergence 0<T<T_c lemma} to deduce that both $\mathcal{S}_{\beta_1}$ and $\mathcal{S}_{\beta_2}$ have largest and non-degenerated eigenvalue $1$. Let $h_i$, $i=1,2$ denote the positive eigenfunction of $\mathcal{S}_{\beta_i}$ associated with $1$, then by monotonicity and positivity,
\begin{equation}\label{order2}
  \mathcal{S}_{\beta_1}h_2>\mathcal{S}_{\beta_2}h_2=h_2.
\end{equation}
Let $\tau_i$ be the positive eigenfunction of $\mathcal{S}_{\beta_i}^*$ associated with $1$, using Lemma \ref{lamba>1}, we can deduce
\begin{equation}\label{order3}
  \int_{\mathbb{R}^3}h_2(k)d\tau_1(k)=\int_{\mathbb{R}^3}h_2(k)d\big(\mathcal{S}^*_{\beta_1}\tau_1\big)
  (k)=\int_{\mathbb{R}^3}\mathcal{S}_{\beta_1}h_2d\tau_1(k)>
  \int_{\mathbb{R}^3}h_2(k)d\tau_1(k)>0,
\end{equation}
a contradiction. Therefore, we must have
\begin{equation}\label{ydusgf}
    \lim_{L\to\infty}T_c(v,\mu,Q,L)=T_c(v,\mu,Q,\infty),
\end{equation}
i.e. (\ref{limit critical temperature}).

%Then (\ref{test gamma^L alpha^L continuous}) also holds true at this temperature $T$. On the other hand, we also have
%\begin{equation}\label{dufb}
%  -P^L_{\mathrm{BCS}}[g_0]>-P^L_{\mathrm{BCS}}[g^L].
%\end{equation}
%However, we can use (\ref{noninteract gamma continuous}) and Assumption \ref{assumption Q} to calculate directly, when $L\to\infty$,
%\begin{equation}\label{saufrbg}
%\begin{aligned}
%  -P^L_{\mathrm{BCS}}[g_0]=&-\frac{1}{\beta L^3}\sum_{k\in(2\pi/L)\mathbb{Z}^3,\sigma}
%  \ln\big(1+e^{-\beta\xi_k}\big)\\
%  &\searrow-\frac{1}{(2\pi)^3\beta}
%  \int_{\mathbb{R}^3\times\mathcal{S}_{2}}
%  \ln\big(1+e^{-\beta\xi_k}\big)dkd\sigma=-P_{\mathrm{BCS}}[\gamma_0,0].
%\end{aligned}
%\end{equation}
%Combining (\ref{dufb}) with (\ref{test gamma^L alpha^L continuous}), we have
%\begin{equation}\label{duffrvv}
%  \lim_{L\to\infty}-P^L_{\mathrm{BCS}}[g_0]=-P_{\mathrm{BCS}}[\gamma_0,0]
%  \geq\lim_{L\to\infty}-P^L_{\mathrm{BCS}}[g^L]
%\end{equation}

\end{proof}

\section*{Acknowledgments}

\par X. Chen was supported in part by NSF grant DMS-2406620, and Z. Zhang was partially supported by NSF of China under Grant 12288101. We appreciate the delightful discussions with Chushun Tian and Xiaoquan Yu, who have provided us many useful physical backgrounds and helped us enhance the understanding to the physical meanings of this problem. We also appreciate the discussions with Christian Brenecke and Reika Fukuizumi, who gave us many kind advices and comments. We also appreciate the discussions with Chengyang Wu, who provided us knowledge about the index theorems that help us complete the proof of Proposition \ref{1to1correspondencequasifree1-pdm} in this paper.

\section*{AI Disclosure}

\par This is an AI-free paper and we would like to keep it that way.

\appendix
\setcounter{equation}{0}

\renewcommand\theequation{A.\arabic{equation}} %（A1）
\section{Properties of $v^L$}\label{property of v^L}

\begin{proposition}\label{Lp v^L}
 Let $v^L$ defined by (\ref{v^L}) under Assumption \ref{assumption v}, then $v^L\in L^1$ and for any $k\in (2\pi/L)\mathbb{Z}^3$,
 \begin{equation}\label{v^L_k}
   \hat{v}(k)=\int_{\Lambda_L}v^L(x)e^{-ikx}dx.
 \end{equation}
 Moreover, if $0<d<3$ with $d$ defined in (\ref{deg Q}), we have $v^L\in L^{3/d}(\Lambda_L)$, and for $d=3$, $v^L\in L^{3/d^\prime}(\Lambda_L)$, where $d^\prime$ is given in (\ref{requirement for v 3}).
\end{proposition}
\begin{proof}
  \par Since $v\in L^1(\mathbb{R}^3)$, $v^L$ given in (\ref{v^L}) is a.e. finite and lies in $L^1(\Lambda_L)$. Then dominated convergence theorem guarantees (\ref{v^L_k}).

  \par We only need to show $v^L\in L^{3/d}$ for $0<d\leq 3$. This can be proved via
  \begin{equation}\label{eq1 appA}
    \Vert v^L\Vert_{L^{3/d}(\Lambda_L)}\leq \sum_{k\in\mathbb{Z}^3}\Vert v\Vert_
    {L^{3/d}(\Lambda_L+kL)},
  \end{equation}
  while (\ref{requirement for v 2}) ensures
  \begin{equation}\label{eq2 appA}
  \begin{aligned}
    &\sum_{k\in\mathbb{Z}^3}\Vert v\Vert_
    {L^{3/d}(\Lambda_L+kL)}\\\leq &
    \Big(\sum_{k\in\mathbb{Z}^3}\langle kL\rangle^{-3(1+\varepsilon)}\Big)^{\frac{3-d}{3}}
    \Big(\sum_{k\in\mathbb{Z}^3}\int_{\Lambda_L+kL}
    \vert v(x)\vert^{\frac{3}{d}}\langle kL\rangle^{\frac{3(3-d)}{d}(1+\varepsilon)}dx\Big)^{\frac{d}{3}}\\
    \leq& C\int_{\mathbb{R}^3} \vert v(x)\vert^{\frac{3}{d}}\langle x\rangle^{\frac{3(3-d)}{d}(1+\varepsilon)}dx<\infty
  \end{aligned}
  \end{equation}
  for some universal constant $C>0$. Similarly, using (\ref{requirement for v 3}), we can deduce $v^L\in L^{3/d^\prime}$.
\end{proof}

\bibliographystyle{abbrv}
\bibliography{refHF}
\end{document}